\documentclass[preprint,3p,authoryear,a4paper]{elsarticle}

\usepackage{booktabs}
\usepackage{graphicx}
\usepackage{epstopdf}
\usepackage{diagbox}
\usepackage{color}
\usepackage{dsfont}
\usepackage{amsmath}
\usepackage{amssymb}
\usepackage{amsfonts}
\usepackage{amsbsy}
\usepackage{caption}
\usepackage{subcaption}
\usepackage{amsthm}
\usepackage{array}
\usepackage{booktabs} 
\usepackage{verbatim}
\usepackage[english]{babel}
\usepackage{natbib}
\usepackage{ulem}
\usepackage{url}
\usepackage{multirow}
\usepackage{subcaption}
\usepackage{float}
\usepackage{dcolumn}
\usepackage[breaklinks=true]{hyperref}
\usepackage{breakcites}
\usepackage{soul}
\usepackage{tablefootnote}
\usepackage{bbm}
\usepackage{longtable}
\usepackage{environ}
\usepackage{threeparttable}
\usepackage{threeparttablex}
\usepackage{afterpage}
\usepackage{lscape}
\usepackage{caption}
\usepackage{subcaption}

\newcounter{tablerowno}
\makeatletter
\newenvironment{refltable}[2]{%
    \setcounter{tablerowno}{0}
    \footnotesize
    \begin{longtable}{>{\stepcounter{tablerowno}\raisebox{\f@size pt}{\phantomsection}\label{#2\thetablerowno}}#1}}%
  {%
    \end{longtable}%
}
\makeatother
\renewcommand\appendix{\par
\setcounter{section}{0}
\setcounter{subsection}{0}
\setcounter{table}{0}
\setcounter{figure}{0}
\gdef\thetable{\Alph{table}}
\gdef\thefigure{\Alph{figure}}
\gdef\thesection{\Alph{section}}
\setcounter{section}{0}} 

\begin{document}

\begin{frontmatter}

\title{On the evolution of data breach reporting patterns and frequency in the United States: a cross-state analysis}

 \author[UMelb]{Benjamin Avanzi}
 \ead{b.avanzi@unimelb.edu.au}

 \author[UMelb]{Xingyun Tan\corref{cor}}
 \ead{xingyunt@student.unimelb.edu.au}

 \author[UNSW]{Greg Taylor}
 \ead{gregory.taylor@unsw.edu.au}

 \author[UNSW]{Bernard Wong}
 \ead{bernard.wong@unsw.edu.au}

 \cortext[cor]{Corresponding author. }

 \address[UMelb]{Centre for Actuarial Studies, Department of Economics, University of Melbourne VIC 3010, Australia}
 \address[UNSW]{School of Risk and Actuarial Studies, UNSW Australia Business School, UNSW Sydney NSW 2052, Australia}

\begin{abstract}

Understanding the emergence of data breaches is crucial for cyber insurance and risk management. However, analyses of data breach frequency trends in the current literature lead to contradictory conclusions. We put forward that those discrepancies may be (at least partially) due to inconsistent data collection standards, as well as reporting patterns, over time and space. We set out to carefully control both.

In this paper, we conduct a joint analysis of state Attorneys General's publications on data breaches across eight states (namely, California, Delaware, Indiana, Maine, Montana, North Dakota, Oregon, and Washington), all of which are subject to established data collection standards—namely, state data breach (mandatory) notification laws. Thanks to our explicit recognition of these notification laws, we are capable of modelling frequency of breaches in a consistent and comparable way over time. Hence, we are able to isolate and capture the complexities of reporting patterns, adequately estimate IBNRs, and yield a highly reliable assessment of historical frequency trends in data breaches.

Our analysis also provides a comprehensive comparison of data breach frequency across the eight U.S. states, extending knowledge on state-specific differences in cyber risk, which has not been extensively discussed in the current literature. We thus illustrate how each state's unique regulations, market dynamics, demographic profiles, and risk factors can significantly impact insurance products and pricing. Furthermore, we uncover novel features not previously discussed in the literature, such as differences in cyber risk frequency trends between large and small data breaches.

Overall, we find that the reporting delays are lengthening. We also elicit commonalities and heterogeneities in reporting patterns across states, severity levels, and time periods. After adequately estimating IBNRs, we find that frequency is relatively stable before 2020 and increasing after 2020. This is consistent across states. Implications of our findings for cyber insurance reserving, pricing, underwriting, and experience monitoring are discussed.

\end{abstract}


\end{frontmatter}
{\centering \large}

%\newpage
\section{Introduction}\label{sec:intro}

\subsection{Background}

As the Internet and other digital networks are becoming increasingly vital to the functioning of the global economy, the threat posed by cybercriminals has risen in prominence \citep{OECD}. Cybercrime will result in an estimated economic loss of $\$$8 trillion USD in 2023, with the figure expected to rise to $\$$10.5 trillion annually by 2025 \citep{CybersecurityVentures}. Cyber is identified as the greatest risk faced by organisations globally, consecutively for 2021 and 2022 \citep{Allianz22,Allianz23}, and the most significant threat to both U.S. financial institutions and the overall financial system \citep{Aon21}.

In addition to adopting effective cyber hygiene practices, businesses are turning to cyber insurance policies for financial coverage and expert guidance in preventing and managing cyber incidents \citep{Deloitte20}. Cyber insurance direct written premiums in the U.S. in 2021 were around $\$$6.5 billion USD, an increase of 61$\%$ from 2020 \citep{NAIC22}. The global estimated gross direct premiums written in 2022 reached approximately $\$$14 billion USD, with the U.S. contributing more than half of the total \citep{InsuranceBusiness}. A survey conducted by Marsh and Microsoft found that 61$\%$ of organisations purchase some type of cyber insurance \citep{MarshMicrosoft22}. A major component of cyber risk is data breaches, which are the focus of the paper. 
According to \citet{NAAG24}, ``A data breach is the illegal and unauthorized access to personal information that jeopardizes its security, confidentiality, or integrity''.

Generally speaking, insurers operating across several states in the U.S. must accurately account for differences across those jurisdictions, such as regulation, market dynamics, and generally any risk factor that can significantly impact insurance products and pricing. Examples of insurance types that vary significantly across states include health insurance, workers' compensation, auto insurance, and homeowners' and renters' insurance \citep{AAA09,nasi22,AutoInsurance,III}. Cyber insurance is no different \citep[see, e.g.,][]{ChHaDiJiDoZhGuGa23}.

\subsection{Data breach reporting patterns and frequency}

To more accurately price cyber insurance, one needs to understand the statistical properties of various types of cyber incidents and model their frequency and/or severity. Unfortunately, we believe that the current literature's understanding of the evolution of data breaches is lacking. While there are several rigorous analyses of frequency trends, their conclusions do not agree \citep[see, e.g.,][]{MaSo10,EdHoFo16,XuScBaXu18,Rom16,WhHoSo21,Jun21,ElIbNi23}. Where could those discrepancies come from?

It is important to note that conclusions on cyber frequency trends are generally based on three main cyber databases: Data Breach Chronology provided by Privacy Rights Clearinghouse \footnote{Here we are referring to the public data - Data Breach Chronology Archive (2005-2018)} \citep{prc}, Cyber Loss Data by Advisen \citep{Advisen}, and SAS OpRisk Global Data by SAS \citep{SAS}. For the rest of this paper, we will refer to these datasets as follows: the PRC dataset, the Advisen dataset, and the SAS dataset. Unfortunately, the exact data collection standards of these datasets are unknown, as they are secondary data sources which collect data from multiple sources (e.g., media, state attorneys general, company websites). Since all of the three datasets are subject to unknown data collection standards, it is challenging to judge the reliability of any particular conclusion. 

Some research suggests an increase in cyber event counts over time \citep[e.g.,][]{ElIbNi23}. Without questioning the worth of those studies (whose main focus is generally elsewhere), it is unclear whether the observed increase is due to an actual increase of risk frequency or not. In particular, factors that enhance data collection capacities over time could drive increases in event counts, even if the level of risks remains unchanged. They include: 

\begin{enumerate}
    \item The introduction of various reporting mandates at different times may have led to sudden increases in the number of events reported. Businesses were forced to report more incidents, subsequently inflating the dataset \citep{AlGaGiLe22,Jun21}. Both the PRC and the Advisen datasets collect data from data breach notification laws of various states, which are introduced at different points in time \citep{prc,Advisen}.
    \item Increasing media attention in cybersecurity \citep{HaGa23} may have led to an increase in the number of events collected by the dataset from media sources. The Advisen the SAS datasets source part of their data from the media \citep{MaPeShTrJaSo22,WeLiZh18}.
    \item Increasing number of data sources used by data maintainers over time may have led to increases in the number of events collected by the dataset over time \citep{PaGuSh20}. This issue is likely to be present in all three datasets - the PRC, the Advisen, and the SAS datasets.
\end{enumerate}

Overall, to reliably assess frequency trends of cyber risks, we should analyse data that follow established and consistent data collection standards over time and space. This approach serves two critical purposes: firstly, it allows us to precisely delineate the scope of our conclusions; secondly, it helps mitigate the influence of above biases on event counts, thereby enabling a more nuanced understanding of the evolution of cyber risks.

\subsection{State Attorney General data breach information}

In contrast to the three databases mentioned in the previous section, the publications of data breaches by state Attorneys General are not influenced by the issue of enhanced data collection capacities. This allows for a more reliable assessment of frequency trends in cyber risks or cyber insurance risks. Details can be found in \citet{CAAG,DEAG,HIAG,INAG,IAAG,MEAG,MDAG,MTAG,NHAG,NJAG,NDAG,OKAG,ORAG,TXAG,VTAG,WAAG,WIAG}, and are developed in Section \ref{sec:data}. Salient properties of this set of data include:
\begin{enumerate}
    \item State Attorneys General's data are collected under state data breach notification laws in the United States \citep{NCSL}. They are less affected by the issue of increasing reporting incentives, as reporting standards have remained consistent over time at the state level. For example, from January 1 2012, California requires notification of the California Attorney General regarding breaches that affect more than 500 California residents. 
    \item The data sources consist of data breach reports submitted by business entities, rather than being derived from media sources. Consequently, they remain unaffected by the bias associated with escalating media attention.
    \item The only data sources are data breach reports submitted by business entities, thereby insulating them from the bias stemming from the introduction of new data sources.
\end{enumerate}

Crucially, state Attorneys General's data provide information on reporting delays (i.e., the lag between the date of breach occurrence and the date of breach notification), which allows for the analysis of the number of IBNR data breaches. Furthermore, in state Attorneys General's data, each business affected by cyber events is considered as a separate event, which aligns with the perspective of insurers.

In summary, datasets such as the Advisen dataset offer rich information on specific individual cyber events (``depth''). However, their coverage, in terms of how many of the occurred events are included, remains limited and ambiguously defined. In contrast, data from state Attorneys General provide less information on individual events, yet they are comprehensive and offer clearly defined coverage (``breadth''). This important observation enables the generation of results in this paper that are distinct and original compared to the existing literature, as we are the first ones to extract and analyse the state Attorneys General's data breach data to our stated purposes.

\subsection{Statement of contributions} \label{sec:intro_sub4}

In this paper, we provide a rigorous examination of data breach reporting patterns and frequency trends in the information published by state Attorneys General, by applying Generalised Additive Models (GAMs) to Over-dispersed Poisson (ODP) observations based on run-off triangles \citep[e.g.,][]{Tay12}. By ``development profile/pattern'' or ``reporting pattern'', we refer to the pattern or trend over time of event notifications following the occurrence of a cyber event. 

We first generate a quarterly estimation of the number of Incurred But Not Reported (IBNR) data breaches. Beyond providing useful insights, this step is essential for reliably assessing the frequency trends of data breaches within the jurisdictions included in our set of data. It is also a contribution in itself. While \citet{ElIbNi23} account for reporting delay and provide estimates of the number of cyber IBNRs using the Advisen dataset (which already presents potential issues due to its nature as explained above), their estimation is opaque. The frequency model utilised in \citet{ElIbNi23} includes some parameters to allow for changes in reporting delay, but unfortunately, the algebraic form of this quantity is not visible. Therefore, it is unclear whether and how changes in reporting delays (over time) are accounted for in their study. In contrast, we uncover specific characteristics of the IBNR data breach counts. By recognising changes of these characteristics over time, we can generate an accurate estimation of the number of IBNR breaches for different quarters of origin. 

Once IBNRs are reliably determined, the emergence of total data breaches (whether reported or not) can be analysed in detail. As our modelling attends to the fine details of the data, we are able to extract features from the frequency data that are material, but have not been mentioned in the prior literature. An understanding of these features, such as the commonalities among states and severity levels measured by the number of affected state residents, is of considerable value to pricing, reserving, and general appreciation of the evolution of cyber experience. Our analysis may serve as an example of the kind of analysis that could be performed on cyber data elsewhere.

Besides, our modelling jointly analyses the state Attorneys General's publications of data breaches across the eight states which explicitly include reporting delay information (i.e., California, Delaware, Indiana, Maine, Montana, North Dakota, Oregon, and Washington). This yields the most comprehensive comparison of cyber risk frequency across states in the literature to the extent of our knowledge. While the current academic literature analyses cyber events within the U.S. at a broad, aggregate level \citep{WhMaSo16,EdHoFo16,ElLo17,ElJu18,XuScBaXu18,SuXuZh21,BeBoHi21,WhHoSo21,LuZhZh23,FaLoTh21,LiMa23,LiLiDa22,PoCaMcPiCo20,Jun21,ElWi19,Rom16,KeZh20,Str19,MaPeShTrJaSo22,PaGuSh20,ShJaMaPeSoTr23,ElJuSh22,FaLoTh21}, it lacks a nuanced examination of these events at a more granular, state level. Although some academic research and industry reports explore global variations in cyber risks \citep[e.g.,][]{BrLuKaPhVa24,Deloitte23,Verizon24}, there is a notable absence of insights into state-level differences within the U.S.

Moreover, we further enrich our analysis with a comparative analysis of data breaches across different severity levels and states. This approach yields valuable insights into cyber risk frequency variations across severity levels. Indeed, a proper understanding of the disparities in trends between large and small data breaches is crucial for insurers to effectively manage risks, allocate resources, and optimize their operations to remain competitive in the insurance market. 

An important observation is that the frequency of data breaches remains relatively stable prior to 2020 but shows an upward trend post-2020 across severity levels and states. This lends support to the hypothesis that the frequency of cyber events changed after the onset of COVID-19 \citep[see, e.g.,][]{CyberInsuranceAcademy,USGAO}.

Finally, we summarise our results into an extensive and holistic discussion of their implications on the pricing, reserving, underwriting, capital needs, and experience monitoring of cyber insurance.

\subsection{Structure of the paper}

The paper is organised as follows. Section \ref{sec:data} distinguishes state Attorneys General's datasets from the PRC dataset and discusses data analysis considerations for cyber data. Section \ref{sec:data2} outlines the selection and processing of state Attorneys General's data before they are modelled. Section \ref{sec:methodology} describes the construction of the model used in this paper. Section \ref{sec:tech} provides details of the model output, including the discovery of data features not mentioned in prior literature that could provide valuable insights into cyber risks. Section \ref{sec:discussion} highlights key research findings of this paper for academic researchers and cyber insurers, including a detailed discussion of the insurance implications of the main model output found in Section \ref{sec:tech}. Section \ref{sec:conclusion} concludes. 

\section{State Attorneys General's publications of data breaches}\label{sec:data}

This section contains two parts. First, we introduce the datasets of state Attorneys General in Section \ref{sec:data_sub1}. As they have not yet been sufficiently acknowledged by the literature, we offer some background for cyber risk researchers to understand the differences among the datasets of individual state Attorneys General. State Attorneys General's publications should serve as important data sources in the current climate of scarce public data on cyber events, as they constitute one of the primary sources of major existing datasets and provide comprehensive and valuable information. 

Second, in Section \ref{sec:data_sub2}, we compare state Attorneys General's datasets with the most commonly used public dataset of data breaches (the PRC dataset), in order for academic researchers to better utilise these two data sources to gain insights into data breach risks. From the perspective of understanding and modelling the risks of data breaches over time, datasets of state Attorneys General have some advantages over the PRC dataset. State Attorneys General's data contain more extensive and detailed information of data breaches (see Section \ref{sec:data_sub2_1}). Furthermore, they provide a more detailed description of data breaches that affect multiple businesses due to the use of common services and providers, revealing the resulting interdependence of businesses (see Section \ref{sec:data_sub2_2}). This has significant value in understanding the consequences of dependent cyber policies, which is the major impediment to the market's expansion. In addition, as we could more confidently assume a constant reporting propensity for this set of data, the frequency trend of reported breaches appears to be a more accurate depiction of the actual breaches that have occurred (see Section \ref{sec:data_sub2_3}). The \textbf{reporting propensity} refers to the ratio of the number of breaches reported in a period to those actually occurring.

\subsection{Introduction of public datasets underrepresented in the current literature} \label{sec:data_sub1}

In this paper, we use data breaches published by individual state Attorneys General in the United States to investigate changes in data breach reporting patterns and frequency trends. These data breaches are subject to state reporting guidelines, and they are publicly accessible on the websites of state Attorneys General (see the references below). 

As of June 2023, 17 state Attorneys General publicly publish data breaches that they collect. See \citet{CAAG,DEAG,HIAG,INAG,IAAG,MEAG,MDAG,MTAG,NHAG,NJAG,NDAG,OKAG,ORAG,TXAG,VTAG,WAAG,WIAG}.

\subsubsection{Inconsistent data breach notification laws across states and over time in the United States}

At this time, the protection of private information in the United States is provided by a mixture of sector-specific federal statutes (i.e., covering financial services, healthcare, telecommunication, and education) and state laws, which differ in their scope and jurisdiction \citep{ICLG}. The National Conference of State Legislatures (NCSL) publishes state data breach notification laws in the United States \citep{NCSL}. As of June 2023, data breach notification laws have been implemented in all 50 states, the District of Columbia, Guam, Puerto Rico, and the Virgin Islands.

States in the United States imposed their own data breach notification laws at different times, each of which protects the privacy of its residents \citep{NCSL}. For instance, California has had a data breach notification law in effect since 2002, Mississippi since 2010, and South Dakota and Alabama in 2018, the last two states to enact such legislation. 

While most state data breach notification laws have similar elements, there are still variations. Key disparities include definition of what constitutes personally identifiable information, whom entities must notify, the number of affected state residents above which notification to the state Attorney General becomes mandatory, and when the notification must be made once an obligation is triggered. Additionally, the content of breach notice, whether the state publishes breach data publicly, and any exemptions from reporting also vary across states \citep{prcdiff}. For example, Table \ref{tab:defs} shows the varying definitions of reportable data breaches to state Attorneys General in the U.S. in 2021 \citep{iapp}. The same breach might require notification of multiple state Attorneys General, if it affects residents from multiple states.

\begin{table}[H]
\centering
\caption{Definitions of reportable data breaches in the U.S. in 2021}
\label{tab:defs}
\begin{threeparttable}
\begin{tabular}{@{}cclll@{}}
\toprule
Notification to state Attorney General & Number of states \tnote{a} \\ \midrule
No obligations                          & 17               \\
Yes                                     & 14               \\
Yes if more than 250 state residents    & 4                \\
Yes if more than 500 state residents    & 8                \\
Yes if more than 1000 state residents   & 7                \\
Others                                  & 4                \\ \bottomrule
\end{tabular}
\begin{tablenotes}
    \item[a] including 50 states, District of Columbia, Guam, Puerto Rico, and the Virgin Islands
\end{tablenotes}
\end{threeparttable}

\end{table}

Additionally, state laws are amended on a regular basis. Common trends include expanding the number of data items that constitute personally identifiable information, reducing reporting timeframe, and requiring notification of the state Attorney General. According to \citet{ME_amend}, an additional requirement of the state Attorney General effective from September 19, 2019 is to report breaches no later than 30 days after their discovery. Table \ref{tab:eight} presents the dates when some states' statutes began to require notification of the Attorney General.

\subsubsection{Differing fields contained in the datasets of individual state Attorneys General}

Table \ref{tab:notif1} and \ref{tab:notif2} distinguish the information provided by individual state Attorneys General on data breaches, as of June 2023. Attorneys General of New Hampshire, New Jersey, Vermont, and Wisconsin also publish breach information, but the breaches are presented only by individual notice letters. The second column of Table \ref{tab:notif1} lists the notification requirement of the state Attorney General. The fourth and fifth columns of Table \ref{tab:notif2} present the requirements of maximum notification timeframes from the discovery of a breach. The last column of Table \ref{tab:notif2} references state statutes. 

\begin{table}[htb]
\caption{Summary of datasets from state Attorneys General - 1}
\label{tab:notif1}
\resizebox{\columnwidth}{!}{%
\begin{tabular}{@{}ccccccccc@{}}
\toprule
State         & \begin{tabular}[c]{@{}c@{}}Notification to \\ State Attorney General\end{tabular} & \begin{tabular}[c]{@{}c@{}}Breach \\ notice\end{tabular} & \begin{tabular}[c]{@{}c@{}}Organisation \\ Name\end{tabular} & \begin{tabular}[c]{@{}c@{}}Start of \\ breach date\end{tabular} & \begin{tabular}[c]{@{}c@{}}End of \\ breach date\end{tabular} & Reported date & \begin{tabular}[c]{@{}c@{}}Number of persons \\ affected (state)\end{tabular} & \begin{tabular}[c]{@{}c@{}}Number of persons \\ affected (total)\end{tabular} \\ \midrule
California    & Yes if more than 500 California residents                                         & Y                                                        & Y                                                            & Y                                                               & Y                                                             & Y             &                                                                               &                                                                               \\
Delaware      & Yes if 500 Delaware residents                                                     & Y                                                        & Y                                                            & Y                                                               & Y                                                             & Y             & Y                                                                             &                                                                               \\
Hawaii        & Yes if 1000 Hawaii residents                                                      & Y                                                        & Y                                                            & Y                                                               & Y                                                             & Y             & Y                                                                             &                                                                               \\
Indiana       & Yes                                                                               &                                                          & Y                                                            & Y                                                               &                                                               & Y             & Y                                                                             & Y                                                                             \\
Iowa          & Yes if 500 Iowa residents                                                         & Y                                                        & Y                                                            &                                                                 &                                                               & Y             &                                                                               &                                                                               \\
Maine         & Yes                                                                               & Y after 2020                                             & Y                                                            & Y                                                               & Y                                                             & Y             & Y                                                                             & Y after 2018                                                                  \\
Maryland      & Yes                                                                               & Y                                                        & Y                                                            &                                                                 &                                                               & Y             & Y                                                                             &                                                                               \\
Massachusetts & Yes                                                                               &                                                          & Y                                                            &                                                                 &                                                               & Y             & Y                                                                             &                                                                               \\
Montana       & Yes                                                                               & Y                                                        & Y                                                            & Y                                                               & Y                                                             & Y             & Y                                                                             &                                                                               \\
North Dakota  & Yes if 250 North Dakota residents                                                 & Y                                                        & Y                                                            & Y                                                               & Y                                                             & Y             & Y                                                                             &                                                                               \\
Oregon        & Yes if 250 Oregon residents                                                       &                                                          & Y                                                            & Y                                                               & Y                                                             & Y             &                                                                               &                                                                               \\
Texas         & Yes if 250 Texas residents                                                        &                                                          & Y                                                            &                                                                 &                                                               & Y             & Y                                                                             &                                                                               \\
Washington    & Yes if 500 Washington residents                                                   & Y                                                        & Y                                                            & Y                                                               & Y                                                             & Y             & Y                                                                             &                                                                               \\ \bottomrule
\end{tabular}%
}
\end{table}

\begin{table}[htb]
\caption{Summary of datasets from state Attorneys General - 2}
\label{tab:notif2}
\begin{threeparttable}
\resizebox{\columnwidth}{!}{%
\begin{tabular}{@{}cccccc@{}}
\toprule
State & \begin{tabular}[c]{@{}c@{}}Cause of \\ breach\end{tabular} & \begin{tabular}[c]{@{}c@{}}Information \\ breached\end{tabular} & \begin{tabular}[c]{@{}c@{}}Timeframe of Notification to Individuals \\ (law enforcement exceptions)\end{tabular} & \begin{tabular}[c]{@{}c@{}}Timeframe of Notification to \\ State Attorney General\end{tabular} & Breach Notification Statutes \\ \midrule
California &  &  & \begin{tabular}[c]{@{}c@{}}as expediently as possible (AEAP) \\ and ``without unreasonable delay" \tnote{a}\end{tabular} &  & Cal. Civ. Code 1798.82 et seq. \\
Delaware &  &  & \begin{tabular}[c]{@{}c@{}}``without unreasonable delay but \\ no later than 60 days after determination \\ of the breach of security" \tnote{b}\end{tabular} & \begin{tabular}[c]{@{}c@{}}``no later than the time when notice \\ is provided to the resident" \tnote{b}\end{tabular} & Del. Code Ann. tit. 6 § 12B-101 et seq. \\
Hawaii & Y &  & ``without unreasonable delay" \tnote{c} & ``without unreasonable delay" \tnote{c} & Haw. Rev. Stat. § 487N-1 et seq. \\
Indiana &  &  & ``without unreasonable delay" \tnote{d} & ``without unreasonable delay" \tnote{d} & Ind. Code § 24-4.9-1-1 et seq. \\
Iowa &  &  & AEAP and ``without unreasonable delay" \tnote{e} & \begin{tabular}[c]{@{}c@{}}``within 5 business days after giving notice \\ of the breach of security to consumers" \tnote{e}\end{tabular} & Iowa Code § 715C.1 -- 2 \\
Maine &  & Y & \begin{tabular}[c]{@{}c@{}}AEAP and ``without unreasonable delay, \\ no more than 30 days after awareness of a breach \\ of security and identification of its scope" \tnote{f}\end{tabular} & ``without unreasonable delay" \tnote{f} & Me. Rev. Stat. tit. 10 § 1346 et seq. \\
Maryland & Y & Y & \begin{tabular}[c]{@{}c@{}}``as soon as reasonably practicable, \\ but no later than 45 days after the business \\ discovers or is notified of the breach of \\ the security of a system" \tnote{g}\end{tabular} & ``prior to notification to individuals" \tnote{g} & Md. Code Com. Law § 14-3504 et seq. \\
Massachusetts &  & Y & \begin{tabular}[c]{@{}c@{}}``as soon as practicable and \\ without unreasonable delay" \tnote{h}\end{tabular} & ``without unreasonable delay" \tnote{h} & Mass. Gen. Laws 93H § 1 et seq. \\
Montana &  &  & AEAP and ``without unreasonable delay" \tnote{i} & ``simultaneous with notification to individual" \tnote{i} & Mont. Code § 30-14-1701 et seq. \\
North Dakota &  &  & AEAP and ``without unreasonable delay" \tnote{j} & ``without unreasonable delay" \tnote{j} & N.D. Cent. Code § 51-30-01 et seq. \\
Oregon &  &  & \begin{tabular}[c]{@{}c@{}}AEAP and ``without unreasonable delay, \\ but no later than 45 days after discovering \\ or receiving notification of the breach of security" \tnote{k}\end{tabular} & \begin{tabular}[c]{@{}c@{}}``without unreasonable delay, but\\ no later than 45 days after ..." \tnote{k}\end{tabular} & Or. Rev. Stat. §§ 646A.600 - 646A.604 \\
Texas &  & Y & \begin{tabular}[c]{@{}c@{}}``without unreasonable delay and in each case \\ no later than the 60th day after the date on which \\ the person determines that the breach occurred" \tnote{l}\end{tabular} & ``no later than the 60th day after ..." \tnote{l} & Tex. Bus. $\&$ Com. Code § 521.053 \\
Washington &  & Y & \begin{tabular}[c]{@{}c@{}}AEAP and ``without unreasonable delay, and \\ no more than 30 calendar days \\ after the breach was discovered" \tnote{m}\end{tabular} & \begin{tabular}[c]{@{}c@{}}``no more than 30 days \\ after the breach was discovered" \tnote{m}\end{tabular} & Wash. Rev. Code § 19.255.010 et seq. \\ \bottomrule
\end{tabular}%
}
\begin{tablenotes}[para]
    Sources:
    \item[a] \citet{CA_law};
    \item[b] \citet{DE_law};
    \item[c] \citet{HI_law};\\
    \item[d] \citet{IN_law};
    \item[e] \citet{IA_law};
    \item[f] \citet{ME_law};\\
    \item[g] \citet{MD_law};
    \item[h] \citet{MA_law};
    \item[i] \citet{MT_law};\\
    \item[j] \citet{ND_law};
    \item[k] \citet{OR_law};
    \item[l] \citet{TX_law};\\
    \item[m] \citet{WA_law}
\end{tablenotes}
\end{threeparttable}
\end{table}

\subsection{Advantages of datasets from state Attorneys General over the PRC database on frequency modelling} \label{sec:data_sub2}

The main dataset used in data breach frequency and severity modelling is the PRC dataset. The PRC dataset obtains most of its data from state Attorneys General and the U.S. Department of Health and Human Services; the former are the data sources of this paper and the latter collects breaches of protected health information under a federal regulation (i.e., The Health Insurance Portability and Accountability Act of 1996 (HIPAA)), which is also publicly available \citep{USDHHS}. 

Data from state Attorneys General are more suitable for the purpose of this study than the PRC dataset due to the considerations below. 

\subsubsection{Additional information provided by state Attorneys General} \label{sec:data_sub2_1}

\paragraph{\\~Dates of occurrence}
As major sources of the PRC dataset, data breaches publicised by state Attorneys General contain the information necessary for assessing reporting delay - date of breach incidence - that is absent from the PRC dataset. Shown in Table \ref{tab:notif1}, among all states that publicly publish data breaches, 8 explicitly present information regarding dates of occurrence (i.e., California, Delaware, Indiana, Maine, Montana, North Dakota, Oregon, and Washington). Maine provides two additional dates, including date of discovery and date of consumer notification. 

\paragraph{Data breach notification letters}
Some state Attorneys General provide access to data breach notification letters that are submitted by organisations as part of regulatory requirements. These letters typically contain comprehensive descriptions of the breaches that have occurred. 

\paragraph{Up-to-date breaches}
State Attorneys General are updating their own database daily to include the newest reported data breaches, making the examination of breaches occurred in 2020 and 2021 possible. However, the PRC dataset contains few data breaches which are reported after 2018 and none after 2019, and thus makes it difficult to see the most recent changes, including those affected by the COVID-19 pandemic. \citet{WhHoSo21} find that the number of data breaches in 2018-2019 present in the PRC dataset is far less than those in previous years, suspecting incomplete data. \cite{LiMa23} suggest that the PRC dataset is only reliable until 2017. 

\subsubsection{Richer description of data breaches} \label{sec:data_sub2_2}

\paragraph{\\~Two different event definitions could be used} ~\\

One of the peculiarities of cyber risks is that event definition can be complicated by interdependencies among certain security incidents \citep{WhHoSo21}, namely third-party cyber events. A \textbf{third-party data breach} refers to a data breach that occurs at a service provider, vendor, or other third-party organisation that has access to other companies' data \citep{Prevalent22}. It can be considered as 1) a single event occurred at the third-party provider or one of its affected client firms, or 2) a series of correlated events at the provider and all of its affected client firms. The choice of event definition will impact on the derived frequency trend, as the latter will result in a greater number of occurred data breaches. 

In the case of a third-party breach, the PRC dataset follows the former event definition \citep{Ben21} and datasets of state Attorneys General follows the latter. For example, In 2017, Sabre, a travel company, experienced a data breach in its Hospitality Solutions system that affected its business partners who used its central reservations booking engine \citep{Fortra}. In the PRC dataset, this data breach is recorded as a single event under Sabre. However, state Attorneys General's datasets classify it as multiple breaches involving various organisations, including Sabre. This is because the affected companies that outsourced services to Sabre were obligated to report the breach to relevant state Attorneys General individually.

\paragraph{Which event definition should be used} ~\\

According to \citet{WhHoSo21}, taking into account all affected organisations in the case of a third-party cyber event provides a richer description of the event. Also, both from an economic and a cyber insurer's point of view, a third-party data breach should be regarded as a series of events rather than a single occurrence, which will be explained below.

To avoid underestimating the risk of a data breach, a third-party data breach should be viewed as a series of events occurred at both the third-party provider and all of its affected client firms. It should not be considered as a single event occurred at the third-party provider or one of its affected client firms. If not, we would underestimate the total number of businesses that are affected by such a breach, resulting in an overall underestimation of the frequency rate. Second, we would underestimate the economic impact of the third-party data breach, by failing to account for the impact on each of all affected businesses. Third, we would underestimate the dependencies across organisations because the data do not capture the dependence resulting from the utilisation of common services and providers.

From the point of view of cyber insurance pricing, considering all affected businesses in the case of a third-party data breach 
has significant value in understanding the consequences of dependent cyber policies, which is the major impediment to the market's expansion. A cyber liability policy covers financial loss in the event of a data breach, irrespective of who was accountable for the loss of data \citep{ws20}. Therefore, when the insurance company insures a third-party provider and its client firms, a breach occurred at the provider may result in multiple claims to the insurer, rather than a single claim, which can be an aggregation problem. 

When the data of an organisation that has purchased cyber insurance is compromised within a third party's system, the insurer will incur two kinds of costs. First, the organisation is responsible for the costs related to the data breach, including regulatory compliance, potential litigation, and related costs. When the contract with the third-party vendor is not enough to cover these costs, the insurer is responsible for the rest \citep{ws20}. Second, the insurer may take the lead in executing the organisation's contractual rights with the service provider directly accountable for the breach. This is known as subrogation \citep{ws20}.

Third-party data breaches require particular attention, as the ``cloud'' has become a ubiquitous part of corporate IT networks, and most breaches are found to be caused by third-party vendors. In 2022, 94$\%$ of businesses use cloud services in some capacity to hold and process their data \citep{flexera22}.  \citet{Ponemon22} found that among 1162 cybersecurity experts surveyed, a majority of the them, accounting for 59 percent, acknowledged that their organisations had encountered a data breach originating from third parties.

In addition, the current literature has not investigated trends in data breach frequency that attempt to take into account all affected client firms of the third party vendor in the case of a third-party data breach. Therefore, we use datasets of state Attorneys General in our frequency analysis. 

\subsubsection{A more constant reporting propensity} \label{sec:data_sub2_3}

\paragraph{\\~ Some assumption about the reporting propensity is required to assess risks over time} ~\\

The ultimate goal of modelling the frequency and severity of cyber risks in the context of cyber insurance is to estimate the loss distribution of insurance claims resulting from occurred cyber incidents. Trend analysis of occurred events can shed light on the risks associated with insurance claims over time. However, the limitation of any datasets that collect real events is that they can only record those that are publicly disclosed; not all events that have occurred are known.

Nonetheless, we can learn about the trend of actual events by analysing reported ones, so long as we can make a valid assumption about the reporting propensity (see the definition in Section \ref{sec:data}). For example, if the reporting propensity remains constant over occurrence periods, then counts of reported events will vary proportionately with those incurred, and the former will validly reflect any trend in the latter.

The data breach reporting propensity in the United States may have shifted over time, complicating efforts to identify actual frequency trends with reported incidents. From this perspective, individual state Attorneys General's datasets are more appropriate for frequency analysis as they provide a more reliable basis for assuming a constant reporting propensity compared to the PRC dataset. Detailed discussion is presented below. 

\paragraph{A likely change in the reporting propensity in the U.S.} \label{sec:changes_rp} ~\\

The reporting propensity of data breaches is heavily influenced by legal requirements, as organisations are reluctant to disclose their security incidents unless necessary, in the fear that they could tarnish the reputation of the brand and instill mistrust among customers. As a result, notification laws play a crucial role in the disclosure of security incidents, and the enactment and amendment of such mandates could have a substantial effect on the events that come to light. For example, In Australia, the total number of data breach notifications increased by 712 percent under the mandatory reporting scheme (i.e., the Notifiable Data Breaches (NDB) scheme) compared to the previous year under the voluntary scheme \citep{oaic}.

The propensity to report data breaches with varying severities in the U.S. may have changed over time, as a result of differing state laws governing data breach notification obligations and the evolution of such requirements over time within individual states. We have identified two key disparities in state laws that materially affect the reporting propensity. 

First, over time, data breach notification regulations have evolved to require breached organisations to not only notify affected customers, but also government bodies, which may have resulted in a greater number of breaches being made public following this change. For example, from January 1 2012, California requires notification of the California Attorney General regarding breaches that affect more than 500 California residents. Considering that a sizable percentage of the PRC dataset comes from state Attorneys General, including the California Attorney General, this requirement may have led to more breaches being collected by the California Attorney General and subsequently by the PRC dataset. In addition, notification of the respective state Attorney General regarding certain breaches is enacted at different times across different states' statutes (see Table \ref{tab:eight}). For example, Washington state added this requirement at a much later date than California, on July 24 2015. Such variation in reporting requirements over time across different states could potentially lead to changes in the reporting propensity of data breaches in general at the national level.

Second, the varying definitions of reportable data breaches across states may have also resulted in differing reporting propensity of data breaches with different severities. The number of affected state residents above which notification to the respective state Attorney General become mandatory varies by state (see Table \ref{tab:defs}). For example, the Washington Attorney General requires the reporting of data breaches that affect more than 500 Washington residents, whereas the Indiana Attorney General requires the reporting of all breaches that affect Indiana residents (see Table \ref{tab:notif1}). As organisations are generally reluctant to disclose data breaches unless they are legally required to do so, the breaches reported to the Washington Attorney General will be larger in size than those reported to the Indiana Attorney General. This is reflected in the actual data: $7\%$ of data breaches published by the Indiana Attorney General affect more than 500 state residents, compared to $90\%$ of data breaches by the Washington Attorney General.

\paragraph{The reporting propensity of the PRC dataset} ~\\

The PRC dataset relies heavily on information provided by individual state Attorneys General. Hence, any change in the propensity to report to individual state Attorneys General would likely cause a shift in the reporting of the PRC dataset. It is possible that the reporting propensity of individual state Attorneys General may have varied throughout the PRC dataset's duration (2005-2018) due to differences in when states implemented the reporting requirement to their respective Attorney General. This variability could result in the reporting propensity of the PRC dataset being less constant and we could not make a valid assumption about its reporting propensity over time. For example, organisations became mandatory to notify the California Attorney General of data breaches affecting more than 500 California residents at the beginning of 2012. As a result, more of such breaches might have been captured by the California Attorney General and subsequently by the PRC dataset. The California Attorney General alone is the source of almost $10\%$ of breaches in the PRC dataset, and thus the change in the reporting propensity of California can materially affect that of the PRC dataset. 

The PRC dataset may not be suitable for performing national trend analysis due to variations in the definitions of reportable data breaches among its data sources, which may have resulted in differing reporting propensity for breaches with varying severities. Since various state Attorneys General, one of the primary data sources for the PRC dataset, require the reporting of breaches with different severities, it can be challenging to derive meaningful insights from aggregated analysis across all states.

Additionally, as noted by \citet{LiMa23}, concerns about data collection reliability have led to suspicions that the PRC dataset's reporting propensity may have been subject to changes. In particular, there has been a noticeable decline in the number of incidents reported for non-medical institutions after 2012, which seems inconsistent with the growing prevalence of e-commerce and increasing awareness of cyber risks.

In conclusion, we cannot reasonably conclude that the reporting propensity of data breaches in the PRC dataset is constant. While the PRC dataset is useful for tracking the earliest and biggest data breaches, the evolution of data breach risks can be difficult to unravel, given that we cannot make a reliable assumption about its reporting propensity over time.

\paragraph{The reporting propensity of datasets provided by state Attorneys General}~\\

The datasets from individual state Attorneys General, when analysed individually, are likely to be subject to a more constant reporting propensity than the PRC dataset. First, these are data breaches that were reported following mandatory notification of individual state Attorneys General (i.e., after the major shift in the reporting propensity within individual states). Second, the definition of reportable data breaches remains consistent over time in each state. Although the reporting propensity in any state could still change due to changes in legal environments, for instance, stronger penalties could increase the reporting propensity, we could assume with greater confidence that the reporting propensity of individual state Attorneys General's datasets is constant than we could for the PRC dataset.

\section{Data selection and processing} \label{sec:data2}

In this section, we cover the specific details of data selection, aggregation, and processing related to the state Attorneys General's data. This includes the identification of data segments for analysis, the definition of reporting delay, the selection of time periods for investigation, the choice of frequency aggregation, and any required data cleaning. These measures are necessary to eliminate potential biases and make the conclusions of this paper more relevant to cyber insurers. A summary of the data manipulations can be found in Online Appendix \ref{app:data_processing}.

\subsection{Differentiating data breaches by state and severity} 

In the previous section, we saw that to control for the reporting propensity, we should analyse datasets from individual state Attorneys General separately. When comparing across states, we should compare breaches under the same definition, as different states may not share the same definition of reportable breaches.

Therefore, we categorise data breaches by state and definition. All cases of comparison are shown in Table \ref{tab:notif3}; 15 data segments are investigated, consisting of 8 states and 4 severities. First, we study reporting delays of all eight states which provide information on both `Reported date/Date of notification' and `Date of breach (occurrence date)'. We exclude breaches with an unknown date of occurrence from the analysis. Second, we differentiate among data breaches with various severities (i.e., the number of state residents affected), as these states collect data breaches that are subject to different severities.

\begin{table}[htp]
\centering
\caption{Comparison by state and severity}
\label{tab:notif3}
\resizebox{\textwidth}{!}{%
\begin{tabular}{@{}cc@{}}
\toprule
The   number of state residents affected & State                                 \\ \midrule
0-249                                    & Indiana, Montana, Maine               \\
250-499                                  & Indiana, Montana, Maine, North Dakota \\
$\ge$ 250                                & Oregon                                \\
$\ge$ 500 & Indiana, Montana, Maine, North Dakota, Washington, Delaware, California \\ \bottomrule
\end{tabular}%
}
\end{table}

\subsection{Definition of reporting delay} 

The reporting delay consists of the time lag between breach occurrence and discovery and the time lag between discovery and notification of relevant parties. Both lags are of significant interest, but we cannot study them separately since only Maine has three years of breaches with dates of discovery. Therefore, we study the lag between breach occurrence and reporting. We consider the occurrence date to be the earliest possible date when breach might have occurred, as this determines coverage or not for most insurance contracts. The date of notification is also defined as the earliest date, as this approximates when insurers receive claims. Therefore, when multiple dates are present, only the earliest date is retained. 

\subsection{Selection of time periods for investigation} 

We analyse the time periods for which complete and unbiased data are available (see Table \ref{tab:eight}) to ensure we do not underestimate the number of breaches that occurred in earlier years. First, we exclude breaches that occurred before notification of state Attorney General was made mandatory. If we had included them, we may have significantly underestimate the number of breaches that occurred prior to the mandatory notification requirement, and we may have wrongly identified an increasing frequency trend (see Section \ref{sec:changes_rp}). 

Second, we also exclude breaches that occurred after the mandatory notification requirement but before the earliest reported date of all breaches in the database. For example, the mandatory notification requirement of the North Dakota Attorney General is effective from April 13, 2015. However, all breaches in the database are found to be reported after January 2, 2019. If we had included breaches that occurred prior to 2019, we would be again at risk of underestimating the number of data breaches.

\begin{table}[htp]
\centering
\caption{Eight states with recorded dates of breach occurrence}
\label{tab:eight}
\resizebox{\textwidth}{!}{%
\begin{tabular}{@{}ccccc@{}}
\toprule
State             & \begin{tabular}[c]{@{}c@{}}Notification to State \\ Attorney General \\ (Effective date)\end{tabular} & Earliest reported date                                                              & \begin{tabular}[c]{@{}c@{}}Period of analysis \\ (Date of occurrence)\end{tabular} & Accident quarters \\ \midrule
California (CA)   & January 1, 2012                                                                                       & January 20, 2012                                                                    & \begin{tabular}[c]{@{}c@{}}January 1, 2012 to \\ December 31, 2021\end{tabular}    & 2012Q1 - 2021Q4   \\
Delaware (DE)     & April 14, 2018                                                                                        & April 11, 2018                                                                      & \begin{tabular}[c]{@{}c@{}}April 1, 2018 to \\ December 31, 2021\end{tabular}      & 2018Q2 - 2021Q4   \\
Indiana (IN)      & \begin{tabular}[c]{@{}c@{}}2006 \\ (exact date unknown)\end{tabular}                                  & \begin{tabular}[c]{@{}c@{}}January 2, 2014 \\ (only a few in 2013)\end{tabular}     & \begin{tabular}[c]{@{}c@{}}January 1, 2014 to \\ July 31, 2021\end{tabular}        & 2014Q1 - 2021Q2   \\
Maine (ME)        & \begin{tabular}[c]{@{}c@{}}2005 \\ (exact date unknown)\end{tabular}                                  & \begin{tabular}[c]{@{}c@{}}January 2, 2013 \\ (only a few before then)\end{tabular} & \begin{tabular}[c]{@{}c@{}}January 1, 2013 to \\ July 31, 2020\end{tabular}        & 2013Q1 - 2020Q2   \\
Montana (MT)      & October 1, 2015                                                                                       & \begin{tabular}[c]{@{}c@{}}October 1, 2015 \\ (only a few before then)\end{tabular} & \begin{tabular}[c]{@{}c@{}}October 1, 2015 to \\ December 31, 2021\end{tabular}    & 2015Q4 - 2021Q4   \\
North Dakota (ND) & April 13, 2015                                                                                        & January 2, 2019                                                                     & \begin{tabular}[c]{@{}c@{}}January 1, 2019 to\\  December 31, 2021\end{tabular}    & 2019Q1 - 2021Q4   \\
Oregon (OR)       & January 1, 2016                                                                                       & \begin{tabular}[c]{@{}c@{}}January 14, 2016 \\ (only two before then)\end{tabular}  & \begin{tabular}[c]{@{}c@{}}January 1, 2016 to \\ December 31, 2021\end{tabular}    & 2016Q1 - 2021Q4   \\
Washington (WA)   & July 24, 2015                                                                                         & August 11, 2015                                                                     & \begin{tabular}[c]{@{}c@{}}October 1, 2015 to \\ December 31, 2021\end{tabular}    & 2015Q4 - 2021Q4   \\ \bottomrule
\end{tabular}%
}
\end{table}

\subsection{Choice of frequency aggregation} \label{sec:fs_reporting}

A critical decision that has to be made is around frequency aggregation - annually, quarterly, or monthly? This depends on how often cyber insurers should monitor the reporting delay. As the U.S. holds the largest market size of cyber insurance, it might be worth looking at the regulations of Property and Casualty insurance industry in the U.S., which cyber insurance falls under. 

The regulation of the insurance industry in the United States is mainly executed by the respective states, with state insurance regulators being members of the National Association of Insurance Commissioners (NAIC). Regulatory filings of insurance companies consist of those required by the NAIC, which are identical for all, and those required by the state where the insurers are admitted to do business, to the \textit{NAIC Financial Data Repository} \citep{naic2}. Quarterly financial statements are one of the sets of financial statements that need to be completed in accordance with the NAIC \citep{naic1}. Given the need for cyber insurers to quantify their liabilities quarterly, monitoring reporting delay on a quarterly basis is necessary.

\subsection{Necessary data cleaning}

Two features of data breaches disclosed by state Attorneys General are worth noting: recording errors occur periodically, and sometimes officers record supplementary breach notices as separate entries to update the information contained in the original breach notice.

To fix the first, we retrieve correct dates from breach notices for breaches with a negative delay (i.e., occurrence dates later than discovery/notification, or discovery dates later than notification). If the breach notice does not contain the correct dates or is unavailable, we remove the breach. In the cases of errors that are not obvious (e.g., notification lag of 1 day), they have been accepted as correct as there is no obvious means of filtering them. 

Second, we should handle notices/entries related to the same breach with care to avoid double-counting. After transferring the updated information from the supplementary notices to the original notice, we delete entries related to supplementary notices. In addition, if a breach only contains dates when supplementary notices are submitted but not when the original notice is submitted (one breach in Maine met this criterion), we exclude it from the analysis to avoid overestimating the reporting delay. This is because supplementary notices are submitted after the original notice, resulting in a longer delay between the occurrence date and the reported date for supplementary notices compared to original notices.

\section{A model of data breach reporting patterns and frequency} \label{sec:methodology}

There has been little formal examination of quarterly data breach development patterns in the current cyber risk literature. Additionally, there is a lack of adequate accounting for potential changes in reporting delays when estimating the number of IBNR breaches. \citet{WhHoSo21} provide some summary statistics on data breach reporting delay prior to 2013 from a previously available public dataset (Open Security Foundation Dataloss Data Base). \citet{KaNa13} compute the mean delay of different types of breaches in the healthcare industry between 2009 and 2010 from the U.S. Department of Health and Human Services \citep{USDHHS}. In the computer science literature, \citet{SaDaWh22} estimate the reporting delay distributions of cyber events across four industries in the U.S. between 2010 and 2019 using a proprietary dataset created by integrating various cyber event datasets. These distributions differ across industries, and the business sector has the longest reporting delay in comparison to the other three industries studied. However, the analysis does not specify the types of cyber events that were included. It also does not explore the quarterly development patterns of data breaches. Moreover, it assumes that the reporting delay remains constant over time, which may not be an accurate depiction of reality. Similar issues are seen in \citep{ElIbNi23} (see details in Section \ref{sec:intro_sub4}. In this paper, we explore changes in quarterly data breach development patterns over time, which enables us to estimate development profiles, IBNRs, and changes in frequency. We find that the reporting delay may not be constant and ignoring these extra steps can cause an underestimation of IBNRs and consequently the frequency of data breaches.

In order to uncover the underlying changes in reporting patterns and frequency of data breaches, we apply parameter reduction techniques to Over-dispersed Poisson (ODP) cross-classified model, resulting in a GAM. We begin with a brief description of the ODP cross-classified model (see Section \ref{sec:methodology_sub1}), and then move on to descriptions of the model components in the GAM in general terms (see Section \ref{sec:methodology_sub2}).

The data used in the model are run-off triangles, which will be explained shortly and can be found in Online Appendix \ref{app:qtr_triangles}. The GAM itself is provided in Online Appendix \ref{app:glm}. Model diagnostics is provided in Online Appendix \ref{app:diagnostics}. 

\subsection{Preliminary: Over-dispersed Poisson (ODP) cross-classified model} \label{sec:methodology_sub1}

Chain Ladder and its extensions based on run-off trinagles are widely used in IBNR (Incurred But Not Reported) reserve estimation \citep[][]{Mac94,Mac93,ReVe98,Gri15,CoPiAt16,Kre82,Ver94,Ver00,EnVe98,EnVe01,EnVe02,Tay12,WuMe08,PeSeAeVe17,SiSh21,AnBe08,PiAnDe03,Shi17}. They estimate IBNRs by completing the lower triangle of a run-off triangle using information from the upper triangle, which represents the experience to date. 

A run-off triangle is a matrix of numbers that shows claim observations for each period, such as the number of claims filed, the amount paid out in claims, and the average cost of claims. The rows of the matrix represent the accident periods, the columns depict the development periods. Accident periods are the periods in which the claims occurred, while development periods are the periods in which the claims were reported, developed, and ultimately closed. The triangle has a third orientation, the diagonal, which is also known as the calendar period. Each diagonal represents claim experience during a particular calendar period. 

The ODP cross-classified model assumes that the claim observations $C_{i j}$ in row $i$ and column $j$, the incremental reported claim counts in our case, follow an over-dispersed Poisson distribution. Mean and variance are as follows:

$$
\mathrm{E}\left[C_{i j}\right]=\alpha_{i} \beta_{j} \quad \text { and } \quad \operatorname{Var}\left[C_{i j}\right]=\phi_{ij} \alpha_{i} \beta_{j}
$$

The parameters of the ODP cross-classified model include both row and column parameters, denoted as $\alpha_{i}$ and $\beta_{j}$ respectively. If the dispersion parameter $\phi_{i j}$ is identical for all cells (i.e., $\phi_{i j}=\phi$) in an ODP cross-classified model, the resulting Maximum Likelihood Estimators (MLEs) are equivalent to the conventional chain ladder estimators \citep[][]{EnVe02,TaMc16,ReVe98}.

$$
\begin{array}{c}
C_{i j} \sim O D P\left(\alpha_{i} \beta_{j}, \phi\right)=O D P\left(\mu_{i j}, \phi\right) \\
\end{array}
$$

This can be recognised as Generalised Linear Models (GLMs), with a log link:

\begin{equation} \label{eqn_1_1}
\ln \left(\mu_{i j}\right)= \ln (\alpha_{i})+ \ln (\beta_{j})
\end{equation}

The predictor structure in this case follows the chain-ladder model, with a separate parameter for each row and each column.

\subsection{Model construction}
\label{sec:methodology_sub2}

To decrease the number of parameters in the ODP cross-classified model, we utilise GAMs. Where appropriate, we replace the categorical variables in the GLM with parametric forms that are parameter-efficient to produce a GAM \citep{TaMc16}. We start with development period and accident period simplifications, adding calendar period effects. Then, we explore interactions between accident periods and development periods (i.e., shifts in development pattern between accident periods), followed by treatments of exceptional observations. 

We first formulate a GAM for each state and severity, and then fuse these various models into one by taking advantage of the commonalities across states and severities. The GAM for all states and severities is provided in Online Appendix \ref{app:glm}. 

In Section \ref{sec:methodology_sub2_1}, we provide an overview of the prototype model that serves as the foundation for this study. Then, we explain the basic construction of each of the aforementioned model components using examples, from Section \ref{sec:gen_dev_pro} to \ref{sec:exceptional_obs}. Each of these subsections commences with an equation that outlines several example covariates, which are the focus of the subsequent discussion. The subsection then proceeds to examine how these covariates are designed.

Our analysis focuses on quarterly patterns of reported breaches over time, which we evaluate using run-off triangles. For brevity, we refer to accident quarter, development quarter, and calendar quarter as AQ, DQ, and CQ, respectively.

\subsubsection{Prototype model}
\label{sec:methodology_sub2_1}

The reporting pattern of data breaches is modelled by modifying the Hoerl curve \citep{EnVe02}, also known as the gamma curve, which is the most popular parametric form to describe development patterns in general insurance. The Hoerl curve is produced by substituting the column parameters in equation \eqref{eqn_1_1} with:

\begin{equation} \label{eqn_1}
\ln \left(\mu_{i j}\right)=\alpha_{i}+\beta_{i} \cdot \ln (j)+\gamma_{i} \cdot j
\end{equation}

The development time $j$ is considered as a continuous variable, which allows extrapolation outside the observed range of development times. The development pattern adheres to a predetermined parametric structure, and it is allowed to be different for each accident period. The Hoerl curve has a general shape that resembles the typical development pattern of incremental claims, with a steep increase to a peak followed by an asymptotically exponential decline.

\subsubsection{General development profile} \label{sec:gen_dev_pro}

\begin{equation} \label{eqn_2}
\begin{split}
\beta_{1} \cdot \ln(j+1) \cdot 1_{[1,4]}(j)+\beta_{2} \cdot \ln(j+1)+\gamma_{1} \cdot 1_{1}(j)
\end{split}
\end{equation}

Let $1_A(x)$ be the indicator function for a set $A$. Then:
\[ 1_A(x) = \begin{cases} 1 & x \in A \\ 0 & \text{otherwise} \end{cases} \]

Equation \eqref{eqn_2} presents one of the general development profiles utilised in this paper. The terms that only depend on development periods represent a contribution to the overall runoff pattern across all accident periods, serving the same function as column effects in the basic Chain Ladder. DQs are numbered starting from 1. 

The first two terms in equation \eqref{eqn_2} yield a change in gradient at $j=4$:

\[\begin{cases} (\beta_{1}+\beta_2) \cdot \ln(j+1) & \text{for} \ 1\le j \le4  \\ \beta_2 \cdot \ln(j+1) & \text{for} \ j>4 \end{cases}\]

And similarly for the last term, which adds a specific constant at $j=1$.

\subsubsection{Accident period trend}

\begin{equation} \label{eqn_3}
\begin{split}
\alpha_{0} + \alpha_{1} \cdot i + \alpha_{2} \cdot i^2 +\alpha_{3} \cdot 1_{i \ge i_1}(i)
+ \alpha_{4} \cdot \max(0,i-i_2) \quad \quad \quad \quad (i_2>i_1)
\end{split}
\end{equation}

After characterizing development pattern, we are now concerned with parameter reduction in row effects $\alpha_{i}$. A quadratic serves as the baseline trend, and all the other terms fix the area of poor fit introduced by the quadratic. For example, $1_{i \ge i_1}(i)$ indicates that a break-point occurs at AQ $i_1$; the gradient of the AQ trend is increased by $\alpha_3$ afterwards. Similarly, the gradient is increased by a further $\alpha_4$ after $i=i_2$. 

\subsubsection{Calendar period effects}

\begin{equation} \label{eqn_4}
\phi \cdot 1_{c^*}(c)
\end{equation}

The calendar period $c$ of a claim observation is calculated as $c=i+j-1$. The addition of calendar period effects to a model requires caution. It is important to note that accident, development, and calendar periods are not independent of one another, so one needs sufficient simplicity over all three time dimensions to avoid problems of multi-collinearity. In this paper, simple indicator functions are sufficient to recognise special calendar periods. 

\subsubsection{Interactions between development periods and accident periods}

\begin{equation} \label{eqn_5}
\begin{split}
\beta_{1} \cdot \ln(j+1) \cdot 1_{[1,4]}(j) \cdot \max(0,i^*-i) + \beta_{2} \cdot \ln(j+1) \cdot1_{[1,4]}(j) \cdot j \cdot \max(0,i-i^*)
\end{split}
\end{equation}

\begin{equation} \label{eqn_6}
\begin{split}
\gamma_{1} \cdot j \cdot \max(0,i-i^*) +\gamma_{2} \cdot \max(0,j-6) \cdot \max(0,i-i^*)
\end{split}
\end{equation}

Sometimes, the development profile is not constant, and changes with accident periods. In this case, in addition to the terms in equation \eqref{eqn_2}, we have interaction terms in equations \eqref{eqn_5} and/or \eqref{eqn_6}, which depend on both development periods and accident periods. The ramp function $\max(0,i-i^*)$ is zero up to $i=i^*$ and then linearly increases, and $\max(0,i^*-i)$ linearly decreases down to $i=i^*$ and becomes zero afterwards. The interactions containing $\max(0,i-i^*)$ allow for a constant development pattern up to $\text{AQ}=i^*$, and a smoothly changing one thereafter, and vice versa for interactions containing $\max(0,i^*-i)$.

\subsubsection{Exceptional observations}
\label{sec:exceptional_obs}

\begin{equation} \label{eqn_7}
\phi_{1} \cdot 1_{i^*}(i) +  \phi_{2} \cdot 1_{i^*}(i) \cdot 1_{j^*}(j)
\end{equation}

After main effects and interactions are adequately modelled above, areas of poor fit may come from exceptional accident periods and irregular experience that is different from those preceding and those succeeding. To address these, in equation \eqref{eqn_7}, we propose indicator functions, which are specific to a particular accident period or cell.

Outliers which are not representative of anything need to be handled carefully, as they can contaminate the development profile and are not useful for prediction. Sometimes we need to assign zero weight to all observations in a specific accident period, if the development pattern related to this accident period is exceptional. This prevents data from the accident period in question from influencing the model fit, and results in a poor fit for the exceptional accident period but a better fit in all other periods. 

\section{Analysis of model output}\label{sec:tech}

In this section, we present the output of the GAM model built for the 15 data segments (i.e., 8 states and 4 severities in Table \ref{tab:notif3}). First, from Section \ref{sec:tech_sub1} to \ref{sec:tech_sub5}, we present the most important results and their interpretation in the same order as Section \ref{sec:methodology}. Second, we summarise residual model effects that consist of interesting model features without a clear interpretation (see Section \ref{sec:tech_sub6}). Finally, we present the limitations of our analysis in Section \ref{sec:tech_sub7}. The GAM itself is provided in Online Appendix \ref{app:glm}, and model diagnostics is provided in Online Appendix \ref{app:diagnostics}. 

Data segments are represented by abbreviations such as IN(0-249), where the state is abbreviated (see Table \ref{tab:eight}) and the number inside the parentheses corresponds to the range of individuals affected by the breaches being investigated. As an example, the data segment abbreviation IN(0-249) refers to breaches that affect between 0 and 249 Indiana residents. Henceforth, larger breaches are referred to as those that affect more than 500 state residents, and smaller breaches are referred to as those that affect between 0 and 249 state residents. In addition, the numbering of AQs follows YYYYQQ format, provided in the last column of Table \ref{tab:eight}.

The analysis focuses on the four largest data segments IN(0-249), MT(0-249), ME(0-249), and CA($>$499), with WA($>$499) and OR($>$249) following behind. The remaining data segments deserve less emphasis due to their low average number of notifications per cell in their respective quarterly run-off triangle, which is less than 5 (see Online Appendix \ref{app:qtr_triangles}). This creates challenges in deriving meaningful insights from the data. 

IBNR breaches are projected by extrapolating historical development profiles to the future to complete the lower triangle; the number of ultimate incurred breaches is calculated by adding actual counts in the upper triangle to the forecasts of IBNRs in the lower, which then reveals frequency trends. 

\subsection{General development profile}
\label{sec:tech_sub1}

\subsubsection{Cyber insurance is a short-tailed business}

The data breach notification component of cyber insurance is a short-tailed business. At least 80$\%$ of breaches are reported within a year of occurrence, and 90$\%$ within a year and a half. 

\subsubsection{Larger breaches have longer delay between occurrence and notification than smaller breaches}

\begin{figure}[htp]
     \centering
     \begin{subfigure}[b]{0.49\textwidth}
         \centering
         \includegraphics[width=\textwidth]{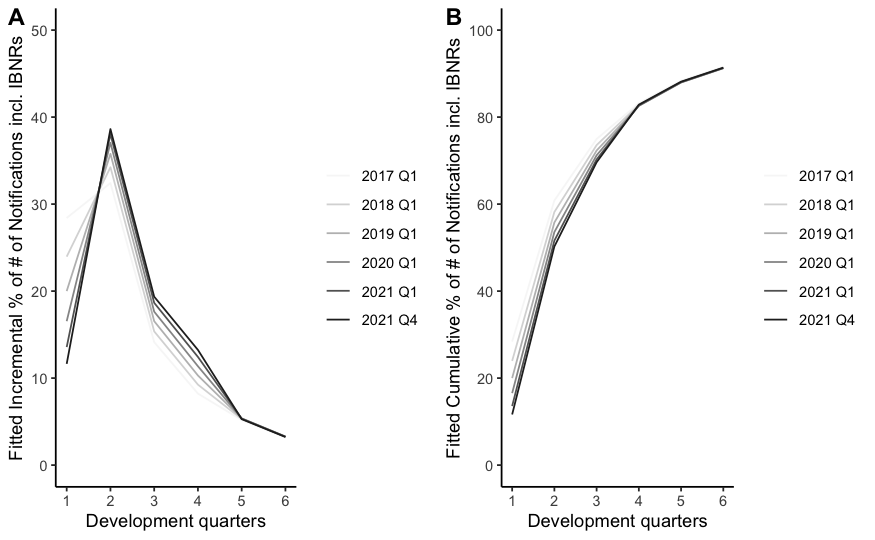}
         \caption{AQ 2017Q1 - AQ 2021Q4, CA($>$499)}
         \label{fig:CA_DP3}
     \end{subfigure}
     \hfill
     \begin{subfigure}[b]{0.49\textwidth}
         \centering
         \includegraphics[width=\textwidth]{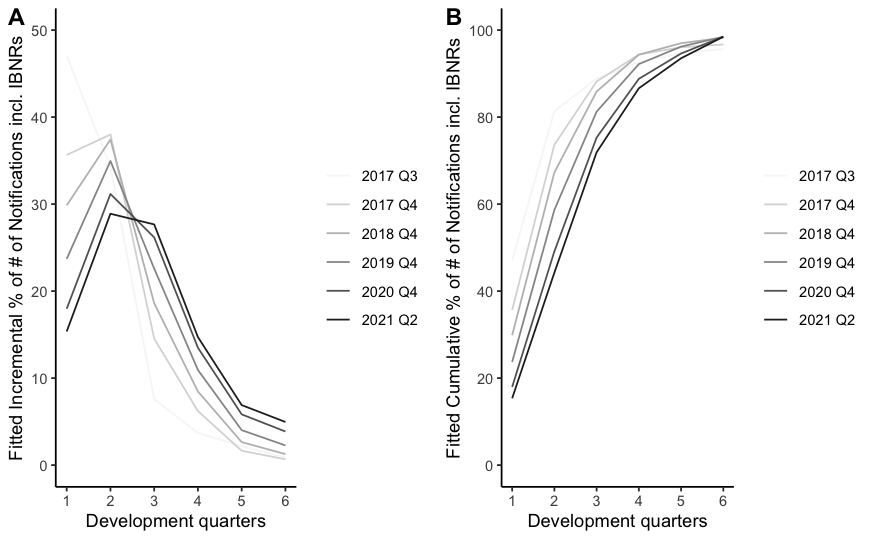}
         \caption{AQ 2014Q1 - AQ 2021Q2, IN (0-249)}
         \label{fig:IN_DP}
     \end{subfigure}
        \caption{Development pattern trends}
\end{figure}

On average, 80$\%$ of larger breaches are reported within a year of occurrence, and 90$\%$ within one year and a half (see Panel B of Figure \ref{fig:CA_DP3}). 90$\%$ of smaller breaches are disclosed within a year of occurrence, and almost all breaches are reported within one year and a half (see Panel B of Figure \ref{fig:IN_DP}).

\subsection{Accident period trend}
\label{sec:tech_sub2}

\subsubsection{The inclusion of IBNRs reveals escalating frequency}

\begin{figure}[htp]
     \centering
     \begin{subfigure}[b]{0.49\textwidth}
         \centering
         \includegraphics[width=\textwidth]{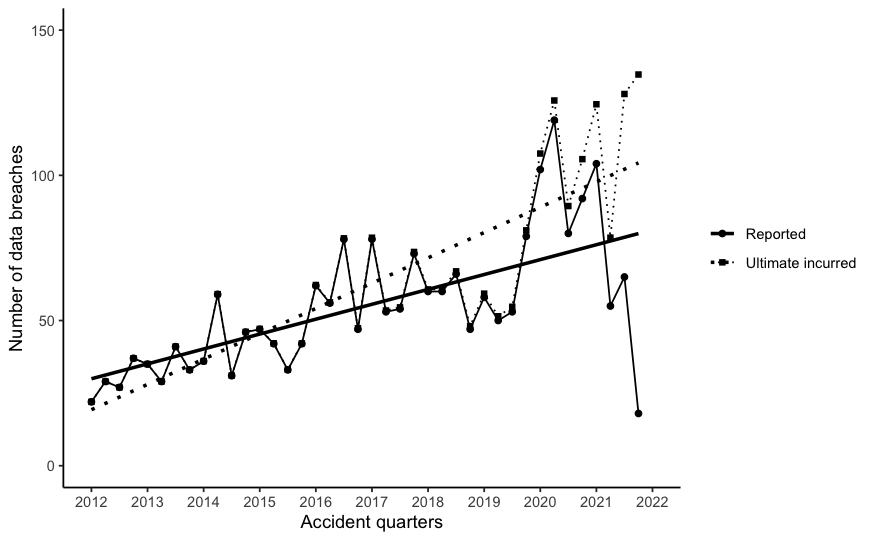}
         \caption{AQ 2012Q1 - AQ 2021Q4, CA($>$499)}
         \label{fig:CA_freq}
     \end{subfigure}
     \hfill
     \begin{subfigure}[b]{0.49\textwidth}
         \centering
         \includegraphics[width=\textwidth]{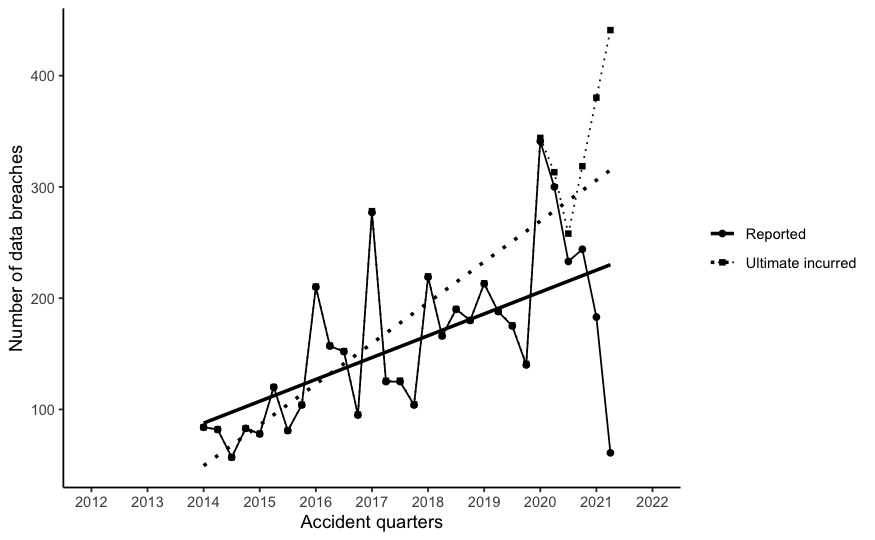}
         \caption{AQ 2014Q1 - AQ 2021Q2, IN (0-249)}
         \label{fig:IN_freq}
     \end{subfigure}
        \caption{Reported versus Ultimate incurred breaches}
\end{figure}

The inclusion of projected IBNRs leads to a notable rise in claim frequency across all states and severities beyond 2020, as opposed to the decrease in frequency that would be seen if actual counts were the only consideration. See Figure \ref{fig:CA_freq} for CA($>$499) and Figure \ref{fig:IN_freq} for IN(0-249), and all other major cases in Online Appendix \ref{app:freq}. 

\subsubsection{CA, IN, MT, and ME exhibit highly similar frequency trends}

\begin{figure}[htp]
    \centering
    \includegraphics[width=10cm]{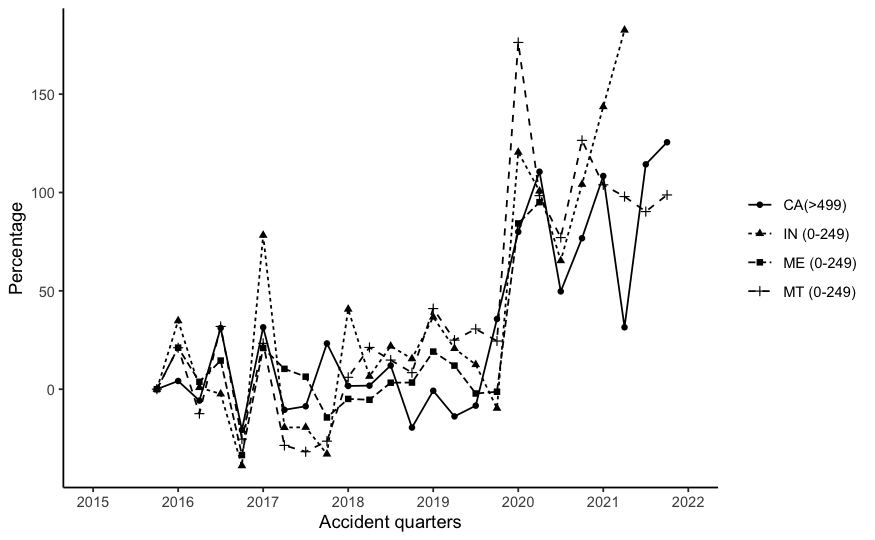}
    \caption{Ultimate incurred breaches by AQ, expressed as \\ a percentage of the average over AQs 2015Q4 to 2016Q3}
    \label{fig:all_freq}
\end{figure}

The growth of quarterly ultimate incurred breaches between 2016Q1 and 2021Q4 is similar across all four states when compared to the average quarterly number of breaches between 2015Q4 and 2016Q3. See Figure \ref{fig:all_freq}.

\begin{figure}[htp]
    \centering
    \includegraphics[width=10cm]{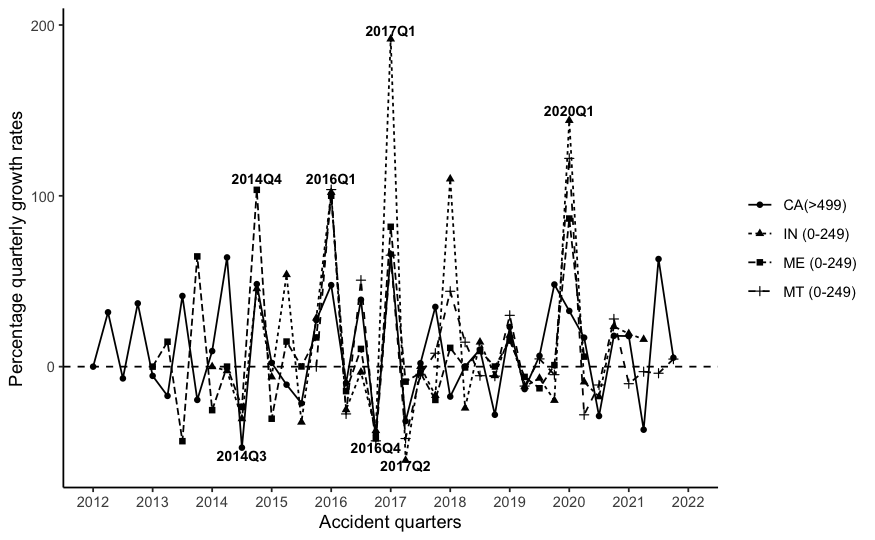}
    \caption{Quarterly growth rates of ultimate incurred breaches (AQ 2012Q1 - 2021Q4)}
    \label{fig:Qtr_growth}
\end{figure}

Starting from 2014Q2, the growth rates of four states often share the same sign for the same period. Furthermore, for the periods with the highest or lowest growth rates, except for 2018Q1, the growth rates consistently maintain both sign and magnitude across states. See Figure \ref{fig:Qtr_growth}. 

\subsubsection{2020Q1 marks a break-point in frequency trends for all states}

Equation \eqref{eqn_3_1} shows the forms of model tested for frequency trends in individual cases, with log of the number of notifications being the dependent variable. $\alpha_3$ is statistically significant in most states and severities, meaning that 2020Q1 is the break-point in their frequency trends. Below, we will discuss the differences in the frequency trends across individual cases. 

\begin{equation} \label{eqn_3_1}
\begin{split}
\alpha_{0} + \alpha_{1} \cdot i + \alpha_{2} \cdot i^2 +\alpha_{3} \cdot 1_{i \ge 2020Q1}(i)+ \alpha_{4} \cdot 1_{2020Q1}(i)
+ \alpha_{5} \cdot 1_{2020Q2}(i)+ \alpha_{6} \cdot \max(0,i-2020Q2)
\end{split}
\end{equation}

\begin{itemize}
    \item \textbf{CA($>$499), IN(0-249):} $\alpha_4$ and $\alpha_5$ are insignificant, all else significant in CA($>$499). $\alpha_4$ is insignificant in IN(0-249). There is a sudden increase in the quarterly number of breaches in 2020Q1, followed by a continuous upward trend.
    \item \textbf{MT(0-249), ME(0-249), WA($>$499), OR($>$250), IN($>$499), MT(250-499), ME(250-499), ND(250-499):} $\alpha_4$, $\alpha_5$, $\alpha_6$ are insignificant. Following the spike in 2020Q1, the quarterly number of breaches stabilises at a relatively higher level.
    \item \textbf{ND($>$499), IN(250-499):} $\alpha_5$, $\alpha_6$ are insignificant. After the sharp increase in 2020Q1, the quarterly count of breaches declines to a level below that of 2020Q1, but stays elevated compared to pre-2020Q1 periods.
\end{itemize}

\subsubsection{Smaller breaches show faster growth and greater volatility compared to larger breaches}

IN(0-249) exhibits the highest and lowest growth rates among the four major data segments (see Figure \ref{fig:Qtr_growth}).

To ensure meaningful comparisons among states with varying data availability, we calculate the mean and standard deviation of percentage quarterly growth rates based on the largest common time periods observed in two or more states. The resulting values are presented in Table \ref{tab:mean_sd_gr}. On average, IN(0-249) has the highest and most volatile growth rates, followed by ME(0-249) and MT(0-249), and then CA($>$499).

\begin{table}[htp]
\centering
\caption{Mean and standard deviation of percentage quarterly growth rates}
\label{tab:mean_sd_gr}
\begin{tabular}{@{}ccccc@{}}
\toprule
              & CA(\textgreater{}499) & ME(0-249)    & IN(0-249)    & MT(0-249)    \\ \midrule
2013Q2-2014Q1 & 3.47(28.40)           & 2.51(47.97)  & NA           & NA           \\
2014Q2-2015Q4 & 8.90(39.66)           & 11.62(44.28) & 8.09(35.03)  & NA           \\
2016Q1-2020Q2 & 10.41(30.41)          & 11.89(38.27) & 20.36(68.48) & 14.97(46.64) \\
2020Q3-2021Q2 & -7.46(29.56)          & NA           & 10.29(18.86) & 1.07(18.24)  \\
2021Q3-2021Q4 & 34.12(40.85)          & NA           & NA           & 0.29(5.87)   \\ \bottomrule
\end{tabular}%
\end{table}

\subsection{Interactions between development periods and accident periods}
\label{sec:tech_sub3}

\subsubsection{All states experience shifts in their reporting patterns, and the reporting patterns vary based on breach severity}

Equations \eqref{eqn_5_1} and \eqref{eqn_5_2} show a subset of terms to model reporting patterns across accident periods in different states, with log of the number of notifications being the dependent variable.

\begin{equation} \label{eqn_5_1}
\begin{split}
\beta_{1} \cdot \ln(j+1) + \beta_{2} \cdot \ln(j+1) \cdot 1_{[1,4]}(j) 
+ \beta_3 \cdot \ln(j+1) \cdot 1_{[1,4]}(j) \cdot \max(0,i_1-i)
\end{split}
\end{equation}

\begin{equation} \label{eqn_5_2}
\begin{split}
\gamma_{1}\cdot j + \gamma_{2}\cdot \max(0,j-6)
+ \gamma_{3}\cdot 1_{\{1,2,5\}}(j)+ \gamma_{4}\cdot 1_{\{2,5\}}(j) + \gamma_{5}\cdot 1_{\{5\}}(j) + \gamma_{6} \cdot \max(0,j-6) \cdot \max(0,i-i_2)
\end{split}
\end{equation}

For larger breaches (see equation \eqref{eqn_5_1}), in the case of CA($>$499), the number of notifications declines as a power function of DQ, with different exponents before and after DQ 4. In addition, the before-DQ-4 exponent varies with AQ, indicating that the short-term notification profile has varied with AQ. See Figure \ref{fig:CA_DP3} for the trends in the notification profile between 2017Q1 and 2021Q4. For detailed descriptions of the changes in trends across all AQs, see Online Appendix \ref{app:CA_DP}. 

For smaller breaches (see equation \eqref{eqn_5_2}), IN(0-249), MT(0-249), and ME(0-249) share another type of reporting pattern. The number of notifications decreases exponentially with change in decay factor at DQ 6, and corrections at DQ 1, 2, and 5. The decay factors used to model the notification profile also vary with AQ. See Figure \ref{fig:IN_DP} for the trend in the notification profile of IN(0-249). See Online Appendix \ref{app:IN_DP} to \ref{app:ME_DP} for descriptions of the changes in trends of IN(0-249), MT(0-249), and ME(0-249). 

\subsubsection{Around 2017, all states observe a shift in their reporting patterns}

After some point in 2017, all states observe a different trend in their reporting patterns. One of the two change-points in the reporting pattern of CA($>$499) is 2017Q1 (see Online Appendix \ref{app:CA_DP}). IN(0-249), MT(0-249), and ME(0-249) experience relatively constant reporting patterns before and including 2017Q3, and shift away from them since 2017Q4 (see Online Appendix \ref{app:IN_DP} to \ref{app:ME_DP}). 

\subsubsection{The reporting delay has been getting longer in most states} \label{sec:AD}

\begin{figure}[htp]
     \centering
     \begin{subfigure}[b]{0.49\textwidth}
         \centering
         \includegraphics[width=\textwidth]{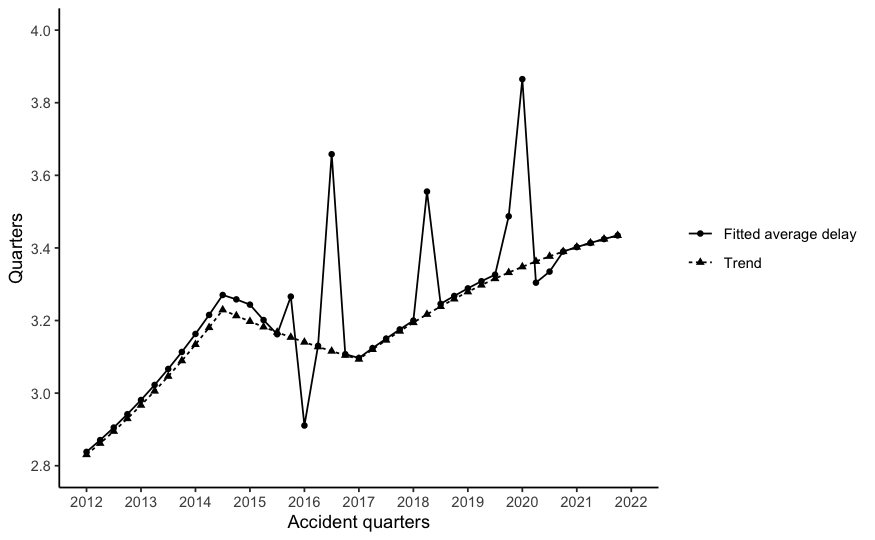}
         \caption{CA($>$499)}
         \label{fig:CA_AD}
     \end{subfigure}
     \hfill
     \begin{subfigure}[b]{0.49\textwidth}
         \centering
         \includegraphics[width=\textwidth]{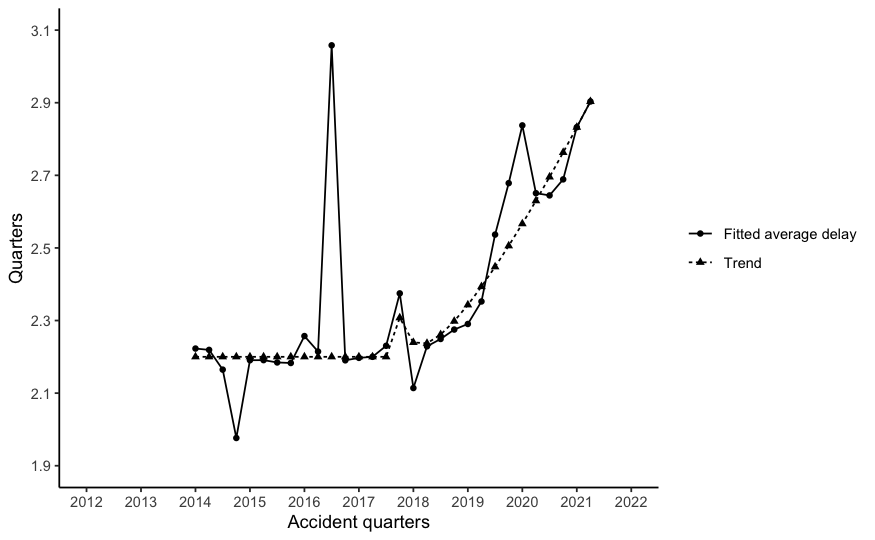}
         \caption{IN(0-249)}
         \label{fig:IN_AD}
     \end{subfigure}
        \caption{Fitted average delay and its trend}
\end{figure}

Figure \ref{fig:CA_AD} - \ref{fig:IN_AD}, and Online Appendix \ref{app:AD} compare the fitted average delay and its trend (i.e., assume no calendar period effects and exceptional observations) for the four major cases. For periods that have been assigned zero weight (see notes at the bottom of Table \ref{tab:glm_mean}), the figure compares the actual delay observed in the data and the trend. 

The average delays of CA($>$499), IN(0-249), and ME(0-249) have lengthened to different extents after 2017 compared to prior periods. The average delay of MT(0-249) has decreased slightly, but not significantly compared to previous periods. 

\paragraph{CA($>$499)}

Shown in Figure \ref{fig:CA_AD}, the average time to report data breaches is 2.8 quarters in AQ 2012Q1, increases to 3.2 quarters in AQ 2014Q3, decreases slightly to 3.1 quarters in AQ 2017Q1, and increases again to 3.4 quarters in AQ 2021Q4. It takes 2 quarters to reach 65$\%$ of reporting in AQ 2012Q1, but takes 3 quarters in AQ 2021Q4 (see Online Appendix \ref{app:CA_DP}).

\paragraph{IN(0-249)}

Shown in Figure \ref{fig:IN_AD}, the average time to report data breaches is 2.2 quarters between AQ 2014Q1 and AQ 2017Q3, and increases to 2.9 quarters in AQ 2021Q2. It takes 2 quarters to reach 80$\%$ of reporting between AQ 2014Q1 and AQ 2017Q3, but takes 4 quarters in AQ 2021Q2 (see Online Appendix \ref{app:IN_DP}).

\subsubsection{The reporting of data breaches shifts away from the first quarter of occurrence across all states}

The percentage of breaches that are reported within the quarter of their occurrence has decreased across all states, regardless of whether the average delay has improved or gotten worse (see Online Appendix \ref{app:DP}). On average, it has decreased from more than 30$\%$ in 2017 to less than 15$\%$ in 2021. 

\begin{figure}[htp]
     \centering
     \begin{subfigure}[b]{0.5\textwidth}
         \centering
         \includegraphics[width=\textwidth]{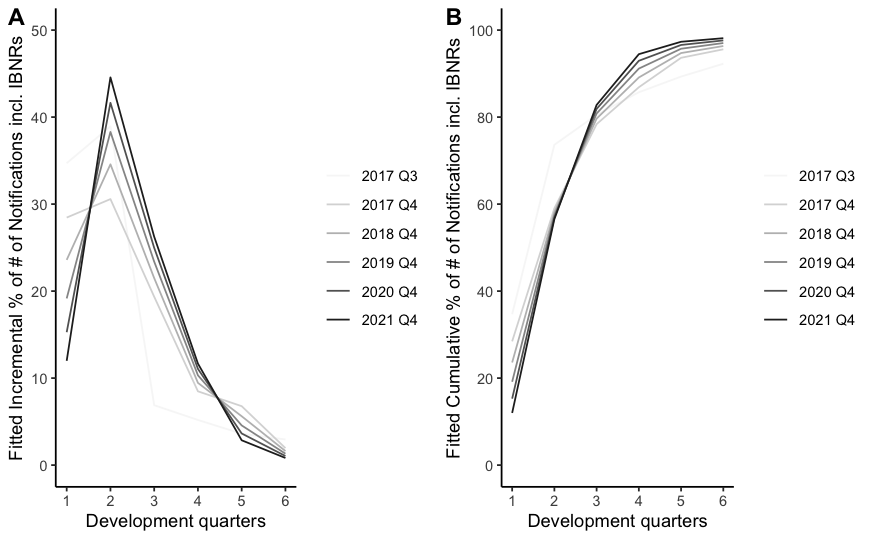}
         \caption{Development pattern trend}
         \label{fig:MT_DP}
     \end{subfigure}
     \hfill
     \begin{subfigure}[b]{0.48\textwidth}
         \centering
         \includegraphics[width=\textwidth]{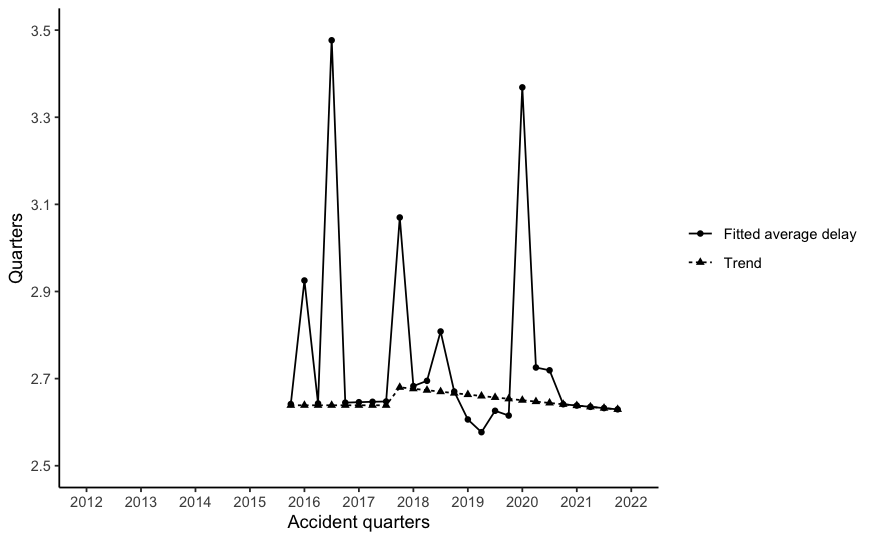}
         \caption{Fitted average delay and its trend}
         \label{fig:MT_AD}
     \end{subfigure}
        \caption{AQ 2015Q4 - AQ 2021Q4, MT (0-249)}
\end{figure}

For example, MT(0-249) is the only state that has seen a slight improvement in the average reporting delay after 2017 (see Figure \ref{fig:MT_AD}). However, just like the other states with deteriorating reporting delay, the percentage of notifications received in the quarter of breach occurrence has decreased from 35$\%$ in 2017Q3 to 12$\%$ in 2021Q4, shown in Figure \ref{fig:MT_DP}. 

\subsection{Exceptional observations}
\label{sec:tech_sub4}

\subsubsection{Some periods of erratically long reporting delay are shared across all states}

During three periods (AQ 2016Q1, AQ 2016Q3, and AQ 2020Q1), all states have experienced significant delays in reporting breaches. These are isolated changes that are significantly off-trend. In addition to the four major cases, we also observe them in WA($>$499), OR($>$249) and IN($>$499).

We see the most drastic change, in the average delay of breaches occurred for each of these periods, in MT(0-249) (see Table \ref{tab:exceptions}). Table \ref{tab:exceptions} shares the same statistics with Figure \ref{fig:CA_AD} - \ref{fig:IN_AD}, and Online Appendix \ref{app:AD}. See Section \ref{sec:AD} for how we compute these statistics. Although there is no evidence of sustained change of the reporting pattern in MT(0-249), MT(0-249) is subject to the greatest volatility in the average reporting delay across accident periods (see Online Appendix \ref{app:MT_DP}). 

In AQ 2016Q3 and AQ 2020Q1, all states are subject to extended reporting delays. For example, in AQ 2020Q1, the average delay in MT(0-249) increases to 3.4 quarters, which would otherwise be 2.6 quarters. IN(0-249) is up by 0.3 quarters to 2.9, from 2.6 quarters. For larger breaches, in CA($>$499), the average delay increases to 3.9 quarters, which would have been 3.4 quarters. The delay experience in AQ 2020Q1 is exceptional compared to other AQs for IN(0-249), MT(0-249), CA($>$499), and WA($>$499); to avoid the distorting effects of such experience, we assign zero weight to it in the modelling. 

In AQ 2016Q1, almost all states are subject to longer-than-usual reporting delays, except that CA($>$499) has shorter-than-usual delay. CA($>$499) experiences extra-long average delay in AQ 2015Q4. It is also worth noting that WA($>$499) and OR($>$249) have also experienced longer delay (see Online Appendix \ref{app:qtr_triangles}). We assign zero weight to the entire AQ 2016Q1 in the modelling of these two states. 

CA($>$499) and IN(0-249), both of which experience non-trivial lengthening reporting delays after 2017, also see longer-than-usual delays in AQ 2019Q4. See Figure \ref{fig:CA_AD} and \ref{fig:IN_AD}. 

\begin{table}[htp]
\centering
\caption{Average delay in quarters - pre and aft accounting for exceptions}
\label{tab:exceptions}
\begin{tabular}{@{}cccc@{}}
\toprule
          & 2016Q1                & 2016Q3                & 2020Q1                \\ \midrule
CA(\textgreater{}499) & 3.1 $\rightarrow$ 2.9 & 3.1 $\rightarrow$ 3.7 & 3.4 $\rightarrow$ 3.9 \\
IN(0-249) & 2.2 $\rightarrow$ 2.3 & 2.2 $\rightarrow$ 3.1 & 2.6 $\rightarrow$ 2.9 \\
MT(0-249) & 2.6 $\rightarrow$ 2.9 & 2.6 $\rightarrow$ 3.5 & 2.6 $\rightarrow$ 3.4 \\
ME(0-249) & 2.4 $\rightarrow$ 2.5 & 2.4 $\rightarrow$ 3.2 & NA                    \\ \bottomrule
\end{tabular}%
\end{table}

\subsubsection{An increase in notifications at DQ 4 or DQ 5 is noted during periods of extended reporting delay}

During periods of extended reporting delay, most of the time, we observe a spike in notifications at DQ 4 or DQ 5, which is at least higher than the number of breaches reported at DQ 3.

Online Appendix \ref{app:qtr_triangles} show the actual quarterly triangles in all data segments. Among breaches occurred in 2016Q3, the number of notifications at DQ 5, in all four major cases and also IN($>$499), is equivalent to or even higher than any other DQs in the same AQ. Among breaches occurred in 2015Q4 of CA($>$499), the number of notifications at DQ 4 is close to the highest. Among breaches occurred in 2016Q1, IN(0-249), MT(0-249), ME(0-249), and IN($>$499) see a spike of notifications at DQ 5. 

\subsection{Calendar period effects}
\label{sec:tech_sub5}

\subsubsection{Opposite calendar period effects in adjacent periods}

A higher/lower total number of notifications received by the state Attorney General in a CQ is compensated by a lower/higher number in the next CQ, observed in IN(0-249) and ME(0-249). This often results in a period of longer-than-usual delay followed by shorter-than-usual delay, when speaking to accident periods, and vice versa. 

\paragraph{IN(0-249)} The number of notifications received in CQ 2017Q4 is 24$\%$ lower than what is modelled in the absence of CQ effects, whereas the number of notifications in CQ 2018Q1 is 31$\%$ higher. Consequently, breaches occurred in 2017Q4 has longer-than-usual delay, followed by 2018Q1 shorter-than-usual delay (see Figure \ref{fig:IN_AD}).

\paragraph{ME(0-249)} The number of notifications received in CQ 2018Q2 is 27$\%$ higher than what is modelled in the absence of CQ effects, whereas the number of notifications in CQ 2018Q3 is 21$\%$ lower. The number of notifications received in CQ 2018Q4 is 29$\%$ lower than what is modelled in the absence of CQ effects, whereas the number of notifications in CQ 2019Q1 is 41$\%$ higher. As a result, breaches occurred in 2018Q1 has shorter-than-usual delay, followed by 2018Q3 longer-than-usual delay, and 2019Q1 shorter-than-usual delay.

\subsection{Residual model effects}
\label{sec:tech_sub6}
~\\
\textit{Item 1:} For the 11 less numerous cases (e.g., IN($>$499)), we find some similarities between their development patterns. The general structure of their development patterns is represented by equation \eqref{eqn_8}, where $\alpha_i$ represents the accident period effects (abbreviated as such), and interactions are not explicitly shown in order to maintain focus. 

\begin{equation} \label{eqn_8}
\begin{split}
\ln(u_{ij})&=\alpha_{i}+\gamma_{1}\cdot j + \gamma_{2}\cdot \max(0,j-6)
+ \gamma_{3}\cdot 1_{\{1,2,5\}}(j)+ \gamma_{4}\cdot 1_{\{2,5\}}(j) + \gamma_{5}\cdot 1_{\{5\}}(j)
\end{split}
\end{equation}

The development pattern amounts to an exponential curve with change in decay factor at DQ 6, and corrections at DQ 1, 2, 5. Some might share a common set of coefficients, while others might have identical coefficients for the exponential curves (i.e., $\gamma_{1}$, $\gamma_{2}$) or the corrections (i.e., $\gamma_{3}$...$\gamma_{5}$). See the former presented by Term \hyperref[r10]{8} - \hyperref[r12]{10} in Table \ref{tab:glm_mean}, and the later presented by Term \hyperref[r3]{1} - \hyperref[r9]{7}. 

When two data segments share the same $\gamma_{3}$, it indicates that the DQ coefficient at $j=1$ in the two segments may be significantly different, but the difference between that coefficient and the coefficients of their respective exponential curves may not be. A common $\gamma_{4}$ suggests that the difference between the coefficients at $j=2$ and $j=1$ is identical, and a common $\gamma_{5}$ suggests that the difference between the coefficients at $j=5$ and $j=2$ is identical. 

~\\
\textit{Item 2:} To characterise accident period effects, we have trend terms and indicator functions to correct for certain anomalies. For example, in equation \eqref{eqn_9} where $\beta_{j}$ is abbreviated for development period effects, there is a quadratic trend and two indicator functions to address poor fit of two periods. 

In some instances, we find that the deviation from the general accident period trend across periods is similar in a single data segment (i.e., similar $\alpha_{3}$ and $\alpha_{4}$). See Term \hyperref[r21]{19} - \hyperref[r24]{22}. 

Sometimes, different data segments may show a similar deviation from their respective accident period trends in the same period or different periods. This means some segments might have a common $\alpha_{3}$ or the $\alpha_{3}$ in one segment is similar to the $\alpha_{4}$ in another segment. See Term \hyperref[r13]{11} - \hyperref[r15]{13}. However, it is worth noting that these data segments may not necessarily have the same accident period trend.

\begin{equation} \label{eqn_9}
\begin{split}
\ln(u_{ij})&=\beta_{j} + \alpha_{1} \cdot i + \alpha_{2} \cdot i^2 +\alpha_{3} \cdot 1_{i_{1}}(i) +\alpha_{4} \cdot 1_{i_{2}}(i)  
\end{split}
\end{equation}

~\\
\textit{Item 3:} We observe similar calendar period effects across different periods in a single data segment and in the same period across adjacent data segments. In equation \eqref{eqn_10}, $\phi_{1}$ and $\phi_{2}$ are similar in WA($>$499) for CQs 2017Q2 and 2018Q4. A common $\phi_{2}$ is shared across WA($>$499) and OR($>$249) for CQ 2018Q4, although they do not have the same accident period trend and development pattern. See Term \hyperref[r16]{14}. 

\begin{equation} \label{eqn_10}
\begin{split}
\ln(u_{ij})&=\alpha_{i} + \beta_{j} + \phi_{1} \cdot 1_{c_{1}}(c)  + \phi_{2} \cdot 1_{c_{2}}(c) 
\end{split}
\end{equation}

~\\
\textit{Item 4:} When correcting for exceptional cells, we find that sometimes the correction is similar across multiple exceptional cells in a single data segment (i.e., similar $\phi_{1}$ and $\phi_{2}$ in equation \eqref{eqn_11}). Sometimes, the correction for the same exceptional cell is identical across multiple data segments, indicated by a common $\phi_{1}$. For example, IN($>$499) has the same correction for (DQ 5, AQ 2016Q1), (DQ 5, AQ 2016Q3), and (DQ 5, AQ 2017Q4); the correction for (DQ 5, AQ 2016Q3) is similar among IN($>$499), WA($>$499), and OR($>$249). See Term \hyperref[r25]{23} and \hyperref[r26]{24}.

\begin{equation} \label{eqn_11}
\begin{split}
\ln(u_{ij})&=\alpha_{i} + \beta_{j} +  \phi_{1} \cdot 1_{i_1}(i) \cdot 1_{j_1}(j) +  \phi_{2} \cdot 1_{i_2}(i) \cdot 1_{j_2}(j)
\end{split}
\end{equation}

\subsection{Limitations}
\label{sec:tech_sub7}

Our analysis presents some limitations. First, despite the fact that we attempted to employ an event definition that takes into account both the third-party provider and all of its affected client firms in the event of a third-party data breach, we only included businesses that met state notification requirements due to data limitations. Second, the conclusions on data breach notifications and frequency trends were drawn based on a subset of state Attorneys General's data that explicitly provides dates of occurrence and dates of notification. Data breaches in other states may display distinct features. Third, we did not incorporate a national analysis, due to the lack of data. When more data are available, it would be interesting to perform multi-state analysis on breaches with the same definition, after filtering out common breaches across states. Fourth, it is unclear whether the lengthening delay between occurrence and reporting is the result of a lengthening delay between occurrence and discovery, between discovery and reporting, or both. This ambiguity arises because most states do not explicitly provide dates of discovery. By extracting discovery dates from breach notices, it may be possible to conduct additional research into the causes of lengthening delay. Fifth, it is unknown whether similarities among states in breach experience are the result of a high proportion of third-party data breaches or of breaches affecting residents of multiple states. This could be investigated further by identifying third-party breaches via breach notice letters and identifying breaches that affect residents of multiple states by identifying common breaches across states. 

\section{Insights and implications} \label{sec:discussion}

This section summarises the main research findings, with a particular emphasis on examining the insurance implications of the main model output presented in Section \ref{sec:tech}. Through this analysis, we offer valuable insights into a variety of insurance-related issues, including pricing, reserving, underwriting, and capital needs.

First, we present an initial attempt to more accurately assess the frequency of both historical and recent data breach risks in the United States, by taking into account some peculiarities when analysing actual event data. Section \ref{sec:dt_analysis} highlights key data considerations for informing data breach risk evolution from incident data that are discussed in Section \ref{sec:data}. 

Second, we analyse the delay between the occurrence of a data breach and the reporting to state regulators, and examine the frequency of data breaches. By focusing on the intricate details of the data in the model, we unearth characteristics that have not been previously discussed in the literature. Having knowledge of these features, including similarities in breaches with the same severity across different states, is extremely valuable when it comes to pricing, reserving, and gaining a general understanding of how cyber incidents are evolving. As cyber risks evolve quickly, we should carefully evaluate and consider the implications of these characteristics to make informed decisions. Section \ref{sec:P3} commences with a concise overview of the data usage and methodology utilised in this study. The section then highlights the key results from the modelling that were uncovered in Section \ref{sec:tech}, followed by an explanation of their potential implications for insurance. Table \ref{tab:summary} summarises key results and their implications. 

\subsection{Data analysis considerations} \label{sec:dt_analysis}

\subsubsection{A different event definition to assess the actual impact of data breaches} \label{sec:P1}

One of the peculiarities of cyber risks is that event definition can be complicated by interdependencies among certain security incidents \citep{WhHoSo21}, namely third-party cyber events. A third-party cyber event can be considered as 1) a single event occurred at the third-party provider or one of its affected client firms, or 2) a series of correlated events at the provider and all of its affected client firms. In this paper, we adopt the latter definition for third-party data breaches, which is necessary from both economic and cyber insurance perspectives. Reasons are briefly explained below (more details in Section \ref{sec:data_sub2_2}). 

To avoid underestimating the risk of a data breach, a third-party data breach should be viewed as a series of events (the second definition defined above). If not, we would underestimate the total number of businesses that are affected by such a breach, resulting in an overall underestimation of the frequency rate. Second, we would underestimate the economic impact of the third-party data breach, by failing to account for the impact on each of all affected businesses. Third, we would underestimate the dependencies across organisations because the data do not capture the dependence resulting from the utilisation of common services and providers.

Cyber insurers should also consider a third-party data breach as a series of events to assess its impact on a cyber portfolio. In the event of a third-party data breach, when a cyber insurer insures both the third-party provider and its client firms, the insurer may face multiple claims. Viewing this breach as a single event may lead to underestimation of the frequency and severity of such breaches, as well as the dependencies between insureds in the portfolio, for the same reasons as stated previously. As a result, the insurer would undervalue the aggregate pricing of the portfolio. Therefore, in order to reflect the actual position of the insurer, this data breach should be regarded as a group of dependent events. 

As this type of data breach becomes more prevalent than ever before, the definition and modelling of its frequency and severity will greatly affect the overall estimation of data breach risk, both at the societal level and from the perspective of a cyber insurer. Therefore, they should be carefully considered. The IBM \textit{Cost of a Data Breach Report 2022} states that 19$\%$ of breaches occurred because of a compromise at a business partner \citep{IBM}. 

\subsubsection{Frequency trend uncovered by controlling for changes in the reporting propensity} \label{sec:P2}

We uncover the frequency trend following an attempt to circumvent the limitation of changing propensity of organisations to report data breaches across periods, hence aiding in the estimation of the true frequency distribution of actual breach occurrences. 

Constant reporting means that the reporting propensity of occurred data breaches does not vary between time periods (see the definition of reporting propensity in Section \ref{sec:data}). If the reporting propensity remains constant over occurrence periods, then counts of reported events will vary proportionately with those incurred, and the former will validly reflect any trend in the latter. 

There have likely been changes in the propensity to report data breaches that are of varying severities in the U.S., as a result of differing state laws governing data breach notification obligations and the evolution of such requirements over time within individual states (more details in Section \ref{sec:changes_rp}). Therefore, to unravel the frequency trend, we segregate data breaches by the state laws to which they are subject, and only analyse breaches of selected time periods, whereas most current literature that studies data breaches in the U.S. performs national analysis across all available time periods. Then, we have no reason to believe that the reporting propensity of the data breaches under investigation has varied, and assume it constant for the purpose of this paper. Below, we provide a brief summary of the two main reasons why the reporting propensity might have materially changed and the selection of data breaches due to this consideration. 

First, the number of entities that must be notified of a breach has expanded, to include government bodies such as state Attorneys General, which may have led to a greater number of breaches being made public after this change. To overcome the change in the reporting propensity caused by this, we analyse data breaches made available directly by individual state Attorneys General, which are those occurred after mandatory notification of the respective Attorney General. 

In addition, states enact the requirement to notify the state Attorney General of certain breaches at different times, which creates challenges for a sound national analysis due to changing reporting propensity over time at the national level. Therefore, to determine the evolution of data breach risk over time, we perform an analysis over states that provide relevant data and differentiate states within the model to the extent that is needed. 

Second, states have different definitions of what constitutes a data breach that must be reported to government authorities. In particular, the number of affected state residents above which notifying the respective state Attorney General is required varies by state. This may have resulted in state-by-state variations in the reporting propensity of data breaches with varying severities. Consequently, when comparing across states, we should compare data breaches that are subject to the same definition, as is done in this paper.

\subsection{Key insights from numerical analysis and their implications} \label{sec:P3}

To unravel the evolution of data breach risk over time, we investigate data breaches at state levels which are made available directly by state Attorneys General, and compare data breaches that are subject to similar definitions, as explained above. As many cyber insurers need to quantify their liabilities quarterly (see Section \ref{sec:fs_reporting}), we aggregate data breaches quarterly for analysis. 

In particular, we investigate eight state Attorneys General publications of data breaches, out of 17. This is because they report breach incidence dates, which allows investigation of data breach reporting patterns and thus helps yield a more complete picture of recent data breach trends after reasonable estimation of IBNR breaches. They are California, Delaware, Indiana, Maine, Montana, North Dakota, Oregon, and Washington state. 

We conduct two-dimensional analyses of data breach frequency data, namely the change of reporting delay and the trend of data breaches between 2012 and 2021 simultaneously by applying GAMs to ODP observations based on run-off triangles. To create a comprehensive model, we first develop a GAM for each state and severity. Subsequently, we utilise the common characteristics among states and severities to merge these individual models into a single model. Projecting IBNR breaches involves extrapolating historical development profiles to the future, and the projected IBNR breaches are then added to the reported breaches to reveal the frequency trends. 

Our findings and their implications are explained in the following subsections, and summarised in  Table \ref{tab:summary} at the end of this section.

\subsubsection{Lengthening delay between data breach occurrence and reporting} \label{sec:D1}

\textbf{Result 1: The average delay of data breaches, between the first possible date of breach occurrence and the date reported to government bodies, has lengthened to different extents after 2017 compared to prior periods in California, Indiana, and Maine.} For example, for breaches that affect more than 500 California residents, it takes two quarters to reach a reporting rate of $65\%$ for breaches that occurred in 2012Q1, but it takes three quarters for breaches that occurred in 2021Q4. \\

\ul{\textit{Insight 1.1: If the lengthening delay between occurrence and reporting is the result of a lengthening delay between discovery and reporting,
cyber attacks that result in data breaches are becoming more effective and costly.}}

In cyber security terminology, the time to detect a breach is referred to as dwell time, the period of time between a cyberattacker's entry and removal from a system \citep{SecurityBoulevard}. Cybercriminals are surreptitious and persistent, frequently operating invisibly on a network for weeks or even months at a time. Long dwell times provide cybercriminals more time to understand the network architecture, the access they have through stolen credentials, and the location of sensitive data \citep{SecurityBrief21}. Consequently, they are given more opportunity to access personally identifiable information, compromise financial accounts, and insert malicious malware through newer and more sophisticated attacks. 

Longer intruder dwell times continue to associate with greater potential impact of a data breach, a finding noted by IBM since 2016 \citep{IBM}. In 2022, the average cost of a data breach with an average time to identify and contain of fewer than 200 days was USD 3.74 million, while the cost of breaches with an average time of more than 200 days was USD 4.86 million. For breaches with a shorter than 200-day lifespan, this difference provides an average cost reduction of USD 1.12 million, or 26.5$\%$.\\

\ul{\textit{Insight 1.2: Cyber insurers are less certain of the risk assumed at the underwriting stage, in terms of whether the potential insured has already been compromised. }}

Cyber insurance policy can be written on an \textbf{occurrence} basis or on a \textbf{discovery} basis. On an occurrence basis, the policy will cover insured events that occur during the policy period. On a discovery basis, the policy will cover events that are discovered during the policy period. 

The increase in dwell times suggests that enterprises are having a more difficult time spotting threats within their increasingly complex and hybrid networks \citep{Microsoft}. For covers that are provided on a discovery basis, such as those on the market, this means that when insurers write the policy, there is greater uncertainty regarding whether or not attackers have already lingered on the network. If this is the case, the organisation's likelihood of being attacked increases, thereby increasing the insurer's potential liability. \\

\ul{\textit{Insight 1.3: For policies that are provided on a discovery basis, cyber insurers should more thoroughly assess the historical attack probability of the insured.}}

For policies on a discovery basis, longer dwell time means that cyber insurers might be liable for hacks that are dated back longer, if they could not be detected until after policy commencement. For example, the insurer is responsible for a data breach that occurred a year ago, provided the breach is discovered after the policy takes effect. Therefore, at the underwriting stage, cyber insurers should conduct a more comprehensive assessment of a potential policyholder's historical security position, to ensure the premiums charged reflect the underwritten risk. For example, an organisation that implemented multi-factor authentication one month ago should be rated differently than one that did so one year ago, due to the increased likelihood of incurring an attack that remains undetected.\\

\ul{\textit{Insight 1.4: Cyber insurers should direct more efforts towards forecasting the financial coverage of incurred but not reported (IBNR) data breach claims.}}

A greater proportion of data breach claims is expected to be reported in later quarters. As cyber insurers are required to report their financial condition on a quarterly basis, the need to estimate data breach claims that have been incurred but not reported (IBNR) has increased.

\subsubsection{The reporting patterns of data breaches shift over time and differ by size of the breach} \label{sec:D2}

\textbf{Result 2: Larger breaches have longer delay between occurrence and notification than smaller breaches.} This is in line with the findings of the IBM \textit{Cost of a Data Breach Report 2022} \citep{IBM}. \\

\ul{\textit{Insight 2.1: Loss reserving should differentiate among breaches with varying severities.}}

As data breach reporting patterns significantly differ by severity of the breach, the severity of breaches should be factored into reserving. \\

\textbf{Result 3: The reporting patterns of data breaches with various severities have shifted over time. } The percentage of breaches reported in each of development quarters 1-6 varies by time periods. For example, an important finding is that the percentage of breaches that are reported within the quarter of their occurrence has decreased from more than 30$\%$ in 2017 to less than 15$\%$ in 2021, across all states. \\

\ul{\textit{Insight 3.1: Basic Chain Ladder method of reserving should not be used to predict IBNR breaches, as it will lead to inaccurate forecasting.}}

There is substantial evidence of shifting reporting patterns across all investigated states and breaches with different severities, which contradicts the assumption of the basic Chain Ladder method of reserving. Using basic Chain Ladder will lead to inaccurate forecasting. \\

\textbf{Result 4: Sometimes, the reporting delay is unusually lengthy.} In such instances, we observe a spike in the reporting of breaches in the fourth and fifth quarters after their occurrence, which can represent as much as 25$\%$ of the total number of breaches of the relevant accident quarter. \\

\ul{\textit{Insight 4.1: Cyber insurers need more capital for policies that are provided on a discovery basis, due to increased loss reserve uncertainty resulting from greater variations of reporting delay.}}

Due to increased variations of reporting delay across periods, suggested by \textbf{Result 3} and \textbf{4}, the number of eligible data breach claims may deviate from the estimated to a greater extent across different accident periods, for policies on a discovery basis. This increases the need for capital, which would occur naturally in any system where capital requirements depend on estimated forecast uncertainty (e.g., U.S., Europe, Australia).\\

\ul{\textit{Insight 4.2: Cyber insurers may have an estimation of the number of data breaches that will occur in each time period. If the received is lower than expected in early development quarters, cyber insurers should not assume a favourable position, as there is a possibility that the number of reported data breaches will spike in later development quarters. }}

Occasionally, a significant proportion of data breaches that are attributed to the same accident period might be reported in later development quarters. Consequently, even if insurers observe lower-than-usual number of claims in the first three development quarters, they are still uncertain about the ultimate liability, and have to wait until later. 

\subsubsection{Shared data breach experience among states} \label{sec:D3}

\textbf{Result 5: States observe similarities in breach frequency trends, the timing of change in reporting patterns, and trends in the average delay between occurrence and reporting.} Different states follow highly similar frequency trends, relatively stationary up to accident quarter 2020Q1 and increasing subsequently. The historical change in reporting pattern commences at a similar point in time across states, around 2017. All states have shown no sign of decreasing average delay between occurrence and reporting, and most have experienced lengthening delay.\\

\ul{\textit{Insight 5.1: There are three potential causes of commonalities among states, and none of them are encouraging.}}

\textit{Reason 1:} One possible reason might be that a significant proportion of breaches are those that impact residents of multiple states. State Attorneys General are concerned about breaches that affect residents of their state. As a result, if a breach affects people in more than one state, the affected organisation must submit a separate report to each state Attorney General of the affected states regarding this breach. When this kind of breach accounts for a sizable fraction of the total, there is considerable overlap of breaches across states. Then, we would anticipate similarities among states. This means that a substantial percentage of breaches incurs the cost of complying with regulations in multiple jurisdictions, which is more expensive than complying with regulations in a single jurisdiction.

\textit{Reason 2:} Another reason might be that a sizable portion of breaches are third-party breaches. In the event of a third-party data breach, as defined in Section \ref{sec:data_sub2_2}, both the third-party vendor and its affected client firms are obligated to separately report such a breach to the Attorneys General of the affected states, in the name of the affected organisation. As these reports are considered as breaches in the same occurrence period, we can expect positive state correlations, which lead to shared experience across states. A third-party data breach creates an aggregation problem for the insurer, as it will lead to multiple claims at the same time, when more than one insured are affected by such a breach, whether they be the vendor or its affected client firms.

\textit{Reason 3:} If we cannot attribute the similarities among states to either of the first two, the only explanation is that attackers have been more active during certain periods than others to launch attacks regardless of geographic location. Alternatively, attackers have become more successful in launching attacks that could compromise a variety of businesses. For instance, attackers can successfully compromise multiple organisations at the same time by exploiting a common vulnerability shared by them. As these breaches occur at roughly the same time, we can also anticipate positive state correlations. Consequently, positive interdependencies exist among organisations, which is undesirable for cyber insurers.

Regardless of the possible causes mentioned above, the fact that states exhibit similar characteristics in breach experience could have far-reaching implications for cyber insurers, as discussed above. Furthermore, if \textit{Reason 2} and \textit{Reason 3} explain most of the shared experience, there are positive dependencies among organisations, which brings about two further implications. \\

\ul{\textit{Insight 5.2: The aggregate premiums and reserves for a cyber portfolio should be higher than those that assume independence of policies.}}

Positive dependencies across organisations reduce diversification benefits over insureds from the portfolio perspective, compared to more independent policies. As a result, cyber insurers should collect more premiums and need more reserves.  

Whether the dependencies stem from the monoculture of software, hardware, or outsourcing services, there is a non-trivial possibility of a single catastrophic event that could affect the majority of the portfolio, which would lead to a significant loss for insurers. \\

\ul{\textit{Insight 5.3: When compared to more independent policies, cyber insurance is expected to have greater variations in actual experience across quarters.}}

The number of breaches may fluctuate widely from quarter to quarter at the national level, as states experience the ups and downs of breach frequency together.

\begin{table}[H]
\caption{Summary of results and insights}
\label{tab:summary}
\resizebox{\textwidth}{!}{%
\begin{tabular}{ll}
\hline
\multicolumn{1}{c}{Results} & \multicolumn{1}{c}{Insights} \\ \hline
\begin{tabular}[c]{@{}l@{}}\\\\ Result 1: Lengthening delay between data breach occurrence \\ and reporting after 2017 in California, Indiana, and Maine\\ \\\end{tabular} & \begin{tabular}[c]{@{}l@{}}\\\\ Insight 1.1: Longer time to identify a breach $\rightarrow$ increased cost of data breaches\\ \\ For policies provided on a discovery basis, \\ \hspace{0.5cm} $\bullet$ Insight 1.2: cyber insurers less certain of the actual level of risk assumed\\  \hspace{0.5cm}   $\bullet$ Insight 1.3: a greater assessment of historical attack probability \\ \hspace{0.5cm} of the insured needed when underwritting\\ \\ Insight 1.4: More efforts towards forecasting incurred but not reported (IBNR) data breach claims\\ \\\end{tabular} \\ \hline
\begin{tabular}[c]{@{}l@{}}\\\\ Result 2: Longer delay between occurrence and notification\\ observed in larger breaches than smaller breaches\\ \\\end{tabular} & \begin{tabular}[c]{@{}l@{}}\\\\ Insight 2.1: differentiating between breaches with varying severities in loss reserving\\ \\\end{tabular} \\ \hline
\begin{tabular}[c]{@{}l@{}}\\\\ Result 3: Shifting reporting patterns of data breaches \\ with various severities\\ \\\end{tabular} & \begin{tabular}[c]{@{}l@{}}\\\\ Insight 3.1: The basic Chain Ladder method of reserving inaccurate to forecast IBNRs\\ \\\end{tabular} \\ \hline
\begin{tabular}[c]{@{}l@{}}\\\\ Result 4: Occasional extremely lengthy reporting delay\\ \\\end{tabular} & \begin{tabular}[c]{@{}l@{}}\\\\ Insight 4.1: For policies provided on a discovery basis, more capital required due to \\ greater variations of eligible data breach claims (Result 3 and 4)\\ Insight 4.2: Cyber insurers should not assume a favourable position too early\\ \\\end{tabular} \\ \hline
\begin{tabular}[c]{@{}l@{}}\\\\ Result 5: States sharing experience in \\ breach frequency trends, \\ the timing of change in reporting patterns, and \\ trends in the average delay between occurrence and reporting\\ \\\end{tabular} & \begin{tabular}[c]{@{}l@{}}\\\\ Insight 5.1: three potential causes\\ \hspace{0.5cm} $\bullet$ \textit{Reason 1}: data breaches impacting residents of multiple states\\ \hspace{0.5cm} $\bullet$ \textit{Reason 2}: third-party data breaches\\ \hspace{0.5cm} $\bullet$ \textit{Reason 3}: attackers being more active during certain periods or becoming more successful\\ \hspace{0.5cm} in launching widespread attacks\\ \\ If \textit{Reason 2} and \textit{Reason 3} explain $\rightarrow$ positive dependencies among organisations: \\ \hspace{0.5cm} $\bullet$ Insight 5.2: reduced diversification benefits from the portfolio perspective and\\ \hspace{0.5cm} accumulation risks $\rightarrow$ higher aggregate premiums and reserves than assuming independence\\ \hspace{0.5cm} $\bullet$ Insight 5.3: expect actual frequency experience at the national level to vary widely across quarters\\ \\\end{tabular} \\ \hline
\end{tabular}%
}
\end{table}

\section{Conclusion} \label{sec:conclusion}

This paper sheds new light on data breach frequency and reporting patterns, by utilising an underrecognised set of public data provided by U.S. state Attorneys General. This set of data was collected based on mandatory state data breach notification laws, and provides comprehensive descriptions of data breaches. 

Some of our major takeaways include the following. First, the average reporting delay of data breaches has lengthened after 2017. In light of this finding, cyber insurers may expect a higher cost of data breaches, and should direct more effort towards forecasting the financial coverage of incurred but not reported (IBNR) data breach claims. The underwriting of policies on a discovery basis should incorporate a greater assessment of historical attack probability of the insured. Second, the reporting profile of events varies significantly across different accident periods and breach sizes. This means that any loss reserving would require more sophistication than a vanilla Chain Ladder technique, and would benefit from differentiating across severity levels. More capital may be required for policies provided on a discovery basis. Third, the frequency of data breaches remains relatively stable before 2020 but shows an upward trend across severity levels and states after 2020. This supports the hypothesis that the frequency of cyber events was affected by the pandemic \citep[see, e.g.,][]{CyberInsuranceAcademy,USGAO}. Fourth, states share similar experience in breach frequency trends, the timing of change in reporting patterns, and trends in the average delay between occurrence and reporting. Even though such similarities are arguably due to third-party breaches or other common vulnerabilities shared by organisations, this suggests that diversification benefits across states may be less than otherwise anticipated.

Compared to the existing literature, our data and frequency analysis presents significant benefits, which are all of particular importance to cyber insurers. The consistency and completeness of the data collection in our set of data allows us to isolate frequency trends from reporting delays, other unrelated trends (such as media attention), as well as differing reporting propensities over time (such as breach notification legislation; see Section \ref{sec:P2}). Furthermore, our definition of ``event'' aligns with what is typical in an insurance portfolio (see Section \ref{sec:P1}). 

Eventually, our analysis and its methodology can help cyber insurers project IBNR reserves and gain a deeper insight into data breach claim frequency trends. It offers a fresh perspective on the actual magnitude of cyber insurance risks. Our extensive discussion on the implications of our findings is useful for cyber insurance pricing, reserving, underwriting, capital needs, and experience monitoring. Furthermore, commonalities and differences across eight different U.S. states were also highlighted and discussed.

\section*{Acknowledgements}

Avanzi and Wong acknowledge support under Australian Research Council's Discovery Project \linebreak (DP200101859) funding scheme. The views expressed herein are those of the authors and are not necessarily those of the supporting organisations.

\section*{References}

\bibliographystyle{elsarticle-harv}

\newpage

\section*{Supplemental Materials for Online Publication Only}
\renewcommand{\theHsection}{A\arabic{section}} \appendix

\section{A summary of data manipulations} \label{app:data_processing}

With some manual work and the use of excel and R, the following steps are repeated for the raw data downloaded from the websites of state Attorneys General: 

\begin{itemize}
    \item Extract the earliest possible date of breach occurrence.
    \item Select breaches within the time periods for which complete and unbiased data are available.
    \item Examine breaches with negative reporting delay and correct the misrecorded dates. 
    \item Delete observations with no date of breach/organisation name/reported date/unknown number of state residents affected/ineligible breaches (i.e., the number of state residents affected is below the threshold that triggers the notification obligation). 
    \item Examine duplicate entries regarding the same breach. Update the original notice by the information provided by supplementary notices, and then delete supplementary notices. 
    
\end{itemize}

Assumptions are made for some special entries. For example, the occurrence date recorded as `mid Dec. 2019' is assumed to be December 15 2019. Upon submission, all details will be provided for replication purposes.

\section{Quarterly Run-off Triangles} \label{app:qtr_triangles}

AQ and DQ will be used as abbreviations for accident quarter and development quarter. Most numerous cases are displayed first. 

%CA500
\begin{landscape}
\begin{table}
\caption{Quarterly Run-off Triangle of the number of data breaches that affect more than 500 California residents}
\resizebox{\columnwidth}{!}{%
\begin{tabular}{@{}lrrrrrrrrrrrrrrrrrrrrrrrrrrrrrrrrrrrrrrrrr@{}}
\toprule
 & DQ1 & DQ2 & DQ3 & DQ4 & DQ5 & DQ6 & DQ7 & DQ8 & DQ9 & DQ10 & DQ11 & DQ12 & DQ13 & DQ14 & DQ15 & DQ16 & DQ17 & DQ18 & DQ19 & DQ20 & DQ21 & DQ22 & DQ23 & DQ24 & DQ25 & DQ26 & DQ27 & DQ28 & DQ29 & DQ30 & DQ31 & DQ32 & DQ33 & DQ34 & DQ35 & DQ36 & DQ37 & DQ38 & DQ39 & DQ40 & Sum \\ \midrule
AQ2012Q1 & 7 & 10 & 3 & 1 & 0 & 0 & 1 & 0 & 0 & 0 & 0 & 0 & 0 & 0 & 0 & 0 & 0 & 0 & 0 & 0 & 0 & 0 & 0 & 0 & 0 & 0 & 0 & 0 & 0 & 0 & 0 & 0 & 0 & 0 & 0 & 0 & 0 & 0 & 0 & 0 & 22 \\
AQ2012Q2 & 9 & 13 & 2 & 0 & 3 & 0 & 0 & 1 & 1 & 0 & 0 & 0 & 0 & 0 & 0 & 0 & 0 & 0 & 0 & 0 & 0 & 0 & 0 & 0 & 0 & 0 & 0 & 0 & 0 & 0 & 0 & 0 & 0 & 0 & 0 & 0 & 0 & 0 & 0 & 0 & 29 \\
AQ2012Q3 & 11 & 11 & 3 & 0 & 0 & 0 & 0 & 0 & 0 & 0 & 0 & 1 & 0 & 0 & 0 & 0 & 1 & 0 & 0 & 0 & 0 & 0 & 0 & 0 & 0 & 0 & 0 & 0 & 0 & 0 & 0 & 0 & 0 & 0 & 0 & 0 & 0 & 0 & 0 & 0 & 27 \\
AQ2012Q4 & 15 & 15 & 3 & 1 & 0 & 2 & 0 & 0 & 0 & 0 & 0 & 0 & 0 & 0 & 0 & 0 & 0 & 0 & 0 & 0 & 0 & 0 & 0 & 1 & 0 & 0 & 0 & 0 & 0 & 0 & 0 & 0 & 0 & 0 & 0 & 0 & 0 & 0 & 0 & 0 & 37 \\
AQ2013Q1 & 9 & 7 & 7 & 4 & 0 & 3 & 2 & 1 & 0 & 1 & 0 & 0 & 1 & 0 & 0 & 0 & 0 & 0 & 0 & 0 & 0 & 0 & 0 & 0 & 0 & 0 & 0 & 0 & 0 & 0 & 0 & 0 & 0 & 0 & 0 & 0 & 0 & 0 & 0 & 0 & 35 \\
AQ2013Q2 & 11 & 9 & 6 & 1 & 0 & 0 & 1 & 0 & 0 & 1 & 0 & 0 & 0 & 0 & 0 & 0 & 0 & 0 & 0 & 0 & 0 & 0 & 0 & 0 & 0 & 0 & 0 & 0 & 0 & 0 & 0 & 0 & 0 & 0 & 0 & 0 & 0 & 0 & 0 & 0 & 29 \\
AQ2013Q3 & 13 & 16 & 1 & 5 & 2 & 0 & 0 & 1 & 1 & 0 & 0 & 1 & 0 & 0 & 0 & 0 & 0 & 0 & 0 & 0 & 0 & 0 & 0 & 0 & 0 & 0 & 0 & 0 & 0 & 0 & 0 & 0 & 1 & 0 & 0 & 0 & 0 & 0 & 0 & 0 & 41 \\
\bottomrule
\end{tabular}%
}
\end{table}
\end{landscape}

%IN1
\begin{landscape}
\begin{table}
\caption{Quarterly Run-off Triangle of the number of data breaches that affect between 0 and 249 Indiana residents}
\resizebox{\columnwidth}{!}{%
\begin{tabular}{@{}lrrrrrrrrrrrrrrrrrrrrrrrrrrrrrrr@{}}
\toprule
 & DQ1 & DQ2 & DQ3 & DQ4 & DQ5 & DQ6 & DQ7 & DQ8 & DQ9 & DQ10 & DQ11 & DQ12 & DQ13 & DQ14 & DQ15 & DQ16 & DQ17 & DQ18 & DQ19 & DQ20 & DQ21 & DQ22 & DQ23 & DQ24 & DQ25 & DQ26 & DQ27 & DQ28 & DQ29 & DQ30 & Sum \\ \midrule
AQ2014Q1 & 39 & 26 & 8 & 4 & 4 & 0 & 1 & 0 & 0 & 2 & 0 & 0 & 0 & 0 & 0 & 0 & 0 & 0 & 0 & 0 & 0 & 0 & 0 & 0 & 0 & 0 & 0 & 0 & 0 & 0 & 84 \\
AQ2014Q2 & 37 & 32 & 5 & 2 & 2 & 2 & 0 & 0 & 0 & 0 & 0 & 0 & 0 & 0 & 1 & 0 & 0 & 0 & 0 & 0 & 0 & 0 & 0 & 0 & 0 & 0 & 0 & 1 & 0 & 0 & 82 \\
AQ2014Q3 & 18 & 27 & 4 & 2 & 1 & 0 & 0 & 2 & 0 & 0 & 0 & 0 & 0 & 1 & 0 & 0 & 0 & 0 & 0 & 0 & 1 & 0 & 0 & 0 & 0 & 0 & 1 & 0 & 0 & 0 & 57 \\
AQ2014Q4 & 52 & 18 & 9 & 2 & 1 & 1 & 0 & 0 & 0 & 0 & 0 & 0 & 0 & 0 & 0 & 0 & 0 & 0 & 0 & 0 & 0 & 0 & 0 & 0 & 0 & 0 & 0 & 0 & 0 & 0 & 83 \\
AQ2015Q1 & 39 & 26 & 4 & 3 & 1 & 2 & 0 & 1 & 0 & 0 & 0 & 0 & 0 & 0 & 0 & 0 & 0 & 0 & 1 & 0 & 0 & 0 & 1 & 0 & 0 & 0 & 0 & 0 & 0 & 0 & 78 \\
AQ2015Q2 & 64 & 40 & 7 & 3 & 1 & 0 & 0 & 0 & 0 & 1 & 0 & 1 & 2 & 0 & 0 & 1 & 0 & 0 & 0 & 0 & 0 & 0 & 0 & 0 & 0 & 0 & 0 & 0 & 0 & 0 & 120 \\
AQ2015Q3 & 33 & 29 & 10 & 0 & 0 & 1 & 1 & 0 & 0 & 2 & 0 & 2 & 2 & 0 & 0 & 0 & 0 & 0 & 0 & 1 & 0 & 0 & 0 & 0 & 0 & 0 & 0 & 0 & 0 & 0 & 81 \\
\bottomrule
\end{tabular}%
}
\end{table}
\end{landscape}

%MT1
\begin{landscape}
\begin{table}
\caption{Quarterly Run-off Triangle of the number of data breaches that affect between 0 and 249 Montana residents}
\resizebox{\columnwidth}{!}{%
\begin{tabular}{@{}lrrrrrrrrrrrrrrrrrrrrrrrrrr@{}}
\toprule
 & DQ1 & DQ2 & DQ3 & DQ4 & DQ5 & DQ6 & DQ7 & DQ8 & DQ9 & DQ10 & DQ11 & DQ12 & DQ13 & DQ14 & DQ15 & DQ16 & DQ17 & DQ18 & DQ19 & DQ20 & DQ21 & DQ22 & DQ23 & DQ24 & DQ25 & Sum \\ \midrule
AQ2015Q4 & 16 & 20 & 7 & 5 & 1 & 1 & 1 & 1 & 1 & 0 & 0 & 0 & 0 & 0 & 0 & 1 & 0 & 0 & 0 & 0 & 1 & 0 & 0 & 0 & 0 & 55 \\
AQ2016Q1 & 38 & 36 & 3 & 2 & 17 & 6 & 4 & 1 & 2 & 0 & 0 & 0 & 2 & 0 & 0 & 0 & 0 & 0 & 1 & 0 & 0 & 0 & 0 & 0 & 0 & 112 \\
AQ2016Q2 & 26 & 30 & 6 & 4 & 5 & 5 & 2 & 2 & 1 & 0 & 0 & 0 & 0 & 0 & 0 & 0 & 0 & 0 & 0 & 0 & 0 & 0 & 0 & 0 & 0 & 81 \\
AQ2016Q3 & 25 & 36 & 4 & 4 & 39 & 7 & 1 & 1 & 0 & 0 & 1 & 1 & 0 & 0 & 0 & 0 & 1 & 0 & 0 & 0 & 2 & 0 & 0 & 0 & 0 & 122 \\
AQ2016Q4 & 24 & 24 & 7 & 6 & 3 & 2 & 0 & 0 & 2 & 0 & 0 & 0 & 1 & 0 & 0 & 0 & 0 & 0 & 0 & 0 & 0 & 0 & 0 & 0 & 0 & 69 \\
AQ2017Q1 & 49 & 41 & 6 & 5 & 4 & 0 & 2 & 6 & 1 & 0 & 0 & 0 & 0 & 0 & 0 & 0 & 0 & 0 & 0 & 0 & 0 & 0 & 0 & 0 & 0 & 114 \\
AQ2017Q2 & 25 & 25 & 6 & 2 & 3 & 0 & 1 & 1 & 1 & 0 & 0 & 0 & 0 & 0 & 0 & 1 & 1 & 0 & 0 & 0 & 0 & 0 & 0 & 0 & 0 & 66 \\
\bottomrule
\end{tabular}%
}
\end{table}
\end{landscape}

%ME1
\begin{landscape}
\begin{table}
\caption{Quarterly Run-off Triangle of the number of data breaches that affect between 0 and 249 Maine residents}
\resizebox{\columnwidth}{!}{%
\begin{tabular}{@{}lrrrrrrrrrrrrrrrrrrrrrrrrrrrrrrr@{}}
\toprule
 & DQ1 & DQ2 & DQ3 & DQ4 & DQ5 & DQ6 & DQ7 & DQ8 & DQ9 & DQ10 & DQ11 & DQ12 & DQ13 & DQ14 & DQ15 & DQ16 & DQ17 & DQ18 & DQ19 & DQ20 & DQ21 & DQ22 & DQ23 & DQ24 & DQ25 & DQ26 & DQ27 & DQ28 & DQ29 & DQ30 & Sum \\ \midrule
AQ2013Q1 & 26 & 15 & 1 & 0 & 0 & 2 & 0 & 1 & 0 & 3 & 0 & 0 & 0 & 0 & 0 & 0 & 0 & 0 & 0 & 0 & 0 & 0 & 0 & 0 & 0 & 0 & 0 & 0 & 0 & 0 & 48 \\
AQ2013Q2 & 30 & 18 & 3 & 3 & 0 & 0 & 0 & 0 & 0 & 0 & 0 & 0 & 0 & 1 & 0 & 0 & 0 & 0 & 0 & 0 & 0 & 0 & 0 & 0 & 0 & 0 & 0 & 0 & 0 & 0 & 55 \\
AQ2013Q3 & 16 & 7 & 3 & 2 & 1 & 0 & 1 & 1 & 0 & 0 & 0 & 0 & 0 & 0 & 0 & 0 & 0 & 0 & 0 & 0 & 0 & 0 & 0 & 0 & 0 & 0 & 0 & 0 & 0 & 0 & 31 \\
AQ2013Q4 & 31 & 11 & 2 & 2 & 2 & 0 & 0 & 1 & 0 & 0 & 0 & 0 & 0 & 1 & 0 & 0 & 1 & 0 & 0 & 0 & 0 & 0 & 0 & 0 & 0 & 0 & 0 & 0 & 0 & 0 & 51 \\
AQ2014Q1 & 13 & 16 & 4 & 2 & 2 & 0 & 0 & 1 & 0 & 0 & 0 & 0 & 0 & 0 & 0 & 0 & 0 & 0 & 0 & 0 & 0 & 0 & 0 & 0 & 0 & 0 & 0 & 0 & 0 & 0 & 38 \\
AQ2014Q2 & 12 & 11 & 3 & 3 & 2 & 2 & 0 & 2 & 1 & 0 & 0 & 0 & 0 & 1 & 1 & 0 & 0 & 0 & 0 & 0 & 0 & 0 & 0 & 0 & 0 & 0 & 0 & 0 & 0 & 0 & 38 \\
AQ2014Q3 & 8 & 12 & 2 & 1 & 1 & 0 & 0 & 1 & 0 & 0 & 0 & 1 & 0 & 0 & 0 & 0 & 1 & 0 & 0 & 0 & 2 & 0 & 0 & 0 & 0 & 0 & 0 & 0 & 0 & 0 & 29 \\
\bottomrule
\end{tabular}%
}
\end{table}
\end{landscape}

%WA500
\begin{landscape}
\begin{table}
\caption{Quarterly Run-off Triangle of the number of data breaches that affect more than 500 Washington residents}
\resizebox{\columnwidth}{!}{%
\begin{tabular}{@{}lrrrrrrrrrrrrrrrrrrrrrrrrrr@{}}
\toprule
 & DQ1 & DQ2 & DQ3 & DQ4 & DQ5 & DQ6 & DQ7 & DQ8 & DQ9 & DQ10 & DQ11 & DQ12 & DQ13 & DQ14 & DQ15 & DQ16 & DQ17 & DQ18 & DQ19 & DQ20 & DQ21 & DQ22 & DQ23 & DQ24 & DQ25 & Sum \\ \midrule
AQ2015Q4 & 1 & 3 & 1 & 1 & 1 & 0 & 0 & 0 & 0 & 0 & 0 & 0 & 0 & 0 & 0 & 0 & 0 & 0 & 0 & 0 & 0 & 0 & 0 & 0 & 0 & 7 \\
AQ2016Q1 & 2 & 1 & 3 & 1 & 12 & 4 & 0 & 0 & 1 & 0 & 0 & 0 & 0 & 0 & 0 & 0 & 0 & 0 & 1 & 0 & 0 & 0 & 0 & 0 & 0 & 25 \\
AQ2016Q2 & 2 & 3 & 1 & 2 & 0 & 0 & 1 & 1 & 0 & 1 & 0 & 0 & 0 & 0 & 0 & 0 & 0 & 0 & 0 & 0 & 0 & 0 & 0 & 0 & 0 & 11 \\
AQ2016Q3 & 3 & 1 & 0 & 4 & 5 & 0 & 0 & 0 & 0 & 0 & 0 & 1 & 0 & 0 & 0 & 0 & 0 & 0 & 0 & 0 & 0 & 0 & 0 & 0 & 0 & 14 \\
AQ2016Q4 & 1 & 4 & 4 & 2 & 2 & 0 & 0 & 0 & 0 & 0 & 0 & 0 & 0 & 0 & 1 & 0 & 0 & 0 & 0 & 0 & 0 & 0 & 0 & 0 & 0 & 14 \\
AQ2017Q1 & 4 & 4 & 0 & 2 & 1 & 0 & 0 & 0 & 0 & 0 & 0 & 1 & 0 & 0 & 0 & 0 & 0 & 0 & 0 & 0 & 0 & 0 & 0 & 0 & 0 & 12 \\
AQ2017Q2 & 1 & 4 & 4 & 2 & 0 & 0 & 0 & 0 & 0 & 0 & 0 & 0 & 0 & 0 & 0 & 0 & 0 & 0 & 0 & 0 & 0 & 0 & 0 & 0 & 0 & 11 \\
\bottomrule
\end{tabular}%
}
\end{table}
\end{landscape}

%OR250
\begin{landscape}
\begin{table}
\caption{Quarterly Run-off Triangle of the number of data breaches that affect more than 250 Oregon residents}
\resizebox{\columnwidth}{!}{%
\begin{tabular}{@{}lrrrrrrrrrrrrrrrrrrrrrrrrr@{}}
\toprule
 & DQ1 & DQ2 & DQ3 & DQ4 & DQ5 & DQ6 & DQ7 & DQ8 & DQ9 & DQ10 & DQ11 & DQ12 & DQ13 & DQ14 & DQ15 & DQ16 & DQ17 & DQ18 & DQ19 & DQ20 & DQ21 & DQ22 & DQ23 & DQ24 & Sum \\ \midrule
AQ2016Q1 & 0 & 9 & 4 & 2 & 12 & 6 & 0 & 1 & 0 & 0 & 0 & 0 & 0 & 0 & 0 & 0 & 0 & 0 & 1 & 0 & 0 & 0 & 0 & 0 & 35 \\
AQ2016Q2 & 3 & 7 & 1 & 1 & 1 & 1 & 1 & 0 & 1 & 1 & 1 & 0 & 0 & 0 & 0 & 0 & 0 & 0 & 0 & 0 & 0 & 0 & 0 & 0 & 18 \\
AQ2016Q3 & 2 & 7 & 1 & 1 & 5 & 1 & 0 & 0 & 0 & 0 & 0 & 1 & 0 & 0 & 0 & 0 & 0 & 0 & 0 & 0 & 0 & 0 & 0 & 0 & 18 \\
AQ2016Q4 & 6 & 3 & 1 & 2 & 2 & 0 & 0 & 0 & 1 & 0 & 0 & 0 & 0 & 0 & 1 & 0 & 1 & 0 & 0 & 0 & 0 & 0 & 0 & 0 & 17 \\
AQ2017Q1 & 5 & 5 & 3 & 3 & 1 & 1 & 1 & 0 & 0 & 0 & 0 & 0 & 0 & 0 & 0 & 0 & 0 & 0 & 0 & 0 & 0 & 0 & 0 & 0 & 19 \\
AQ2017Q2 & 6 & 8 & 4 & 2 & 0 & 0 & 0 & 0 & 1 & 0 & 0 & 0 & 0 & 0 & 0 & 0 & 0 & 1 & 0 & 0 & 0 & 0 & 0 & 0 & 22 \\
AQ2017Q3 & 3 & 4 & 3 & 4 & 1 & 2 & 0 & 0 & 0 & 0 & 0 & 0 & 0 & 0 & 0 & 0 & 0 & 0 & 0 & 0 & 0 & 0 & 0 & 0 & 17 \\
\bottomrule
\end{tabular}%
}
\end{table}
\end{landscape}

%IN500
\begin{landscape}
\begin{table}
\caption{Quarterly Run-off Triangle of the number of data breaches that affect more than 500 Indiana residents}
\resizebox{\columnwidth}{!}{%
\begin{tabular}{@{}lrrrrrrrrrrrrrrrrrrrrrrrrrrrrrrr@{}}
\toprule
 & DQ1 & DQ2 & DQ3 & DQ4 & DQ5 & DQ6 & DQ7 & DQ8 & DQ9 & DQ10 & DQ11 & DQ12 & DQ13 & DQ14 & DQ15 & DQ16 & DQ17 & DQ18 & DQ19 & DQ20 & DQ21 & DQ22 & DQ23 & DQ24 & DQ25 & DQ26 & DQ27 & DQ28 & DQ29 & DQ30 & Sum \\ \midrule
AQ2014Q1 & 0 & 2 & 0 & 0 & 1 & 1 & 0 & 0 & 0 & 1 & 0 & 0 & 0 & 0 & 0 & 0 & 0 & 0 & 0 & 0 & 0 & 0 & 0 & 0 & 0 & 0 & 0 & 0 & 0 & 0 & 5 \\
AQ2014Q2 & 2 & 2 & 1 & 2 & 0 & 2 & 0 & 0 & 0 & 0 & 0 & 0 & 0 & 0 & 0 & 0 & 0 & 0 & 0 & 0 & 0 & 0 & 0 & 0 & 0 & 0 & 0 & 0 & 0 & 0 & 9 \\
AQ2014Q3 & 0 & 2 & 0 & 1 & 3 & 0 & 0 & 0 & 0 & 0 & 0 & 0 & 0 & 0 & 0 & 0 & 0 & 0 & 0 & 0 & 0 & 0 & 0 & 0 & 0 & 0 & 0 & 0 & 0 & 0 & 6 \\
AQ2014Q4 & 2 & 2 & 2 & 0 & 0 & 0 & 0 & 0 & 0 & 0 & 0 & 0 & 0 & 0 & 0 & 0 & 0 & 0 & 0 & 0 & 0 & 0 & 0 & 0 & 0 & 0 & 0 & 0 & 0 & 0 & 6 \\
AQ2015Q1 & 4 & 1 & 1 & 0 & 0 & 0 & 0 & 0 & 0 & 0 & 0 & 0 & 0 & 0 & 0 & 0 & 0 & 0 & 0 & 0 & 0 & 0 & 0 & 0 & 0 & 0 & 0 & 0 & 0 & 0 & 6 \\
AQ2015Q2 & 2 & 1 & 0 & 1 & 0 & 1 & 0 & 0 & 0 & 0 & 0 & 0 & 0 & 0 & 0 & 0 & 0 & 0 & 1 & 0 & 0 & 0 & 0 & 0 & 0 & 0 & 0 & 0 & 0 & 0 & 6 \\
AQ2015Q3 & 0 & 3 & 1 & 0 & 0 & 0 & 0 & 0 & 0 & 0 & 0 & 0 & 0 & 0 & 0 & 0 & 0 & 0 & 0 & 0 & 0 & 0 & 0 & 0 & 0 & 0 & 0 & 0 & 0 & 0 & 4 \\
\bottomrule
\end{tabular}%
}
\end{table}
\end{landscape}

%IN250
\begin{landscape}
\begin{table}
\caption{Quarterly Run-off Triangle of the number of data breaches that affect between 250 and 499 Indiana residents}
\resizebox{\columnwidth}{!}{%
\begin{tabular}{@{}lrrrrrrrrrrrrrrrrrrrrrrrrrrrrrrr@{}}
\toprule
 & DQ1 & DQ2 & DQ3 & DQ4 & DQ5 & DQ6 & DQ7 & DQ8 & DQ9 & DQ10 & DQ11 & DQ12 & DQ13 & DQ14 & DQ15 & DQ16 & DQ17 & DQ18 & DQ19 & DQ20 & DQ21 & DQ22 & DQ23 & DQ24 & DQ25 & DQ26 & DQ27 & DQ28 & DQ29 & DQ30 & Sum \\ \midrule
AQ2014Q1 & 1 & 1 & 0 & 1 & 0 & 0 & 0 & 0 & 0 & 0 & 0 & 0 & 0 & 0 & 0 & 0 & 1 & 0 & 0 & 0 & 0 & 0 & 0 & 0 & 0 & 0 & 0 & 0 & 0 & 0 & 4 \\
AQ2014Q2 & 3 & 0 & 1 & 1 & 0 & 1 & 0 & 0 & 0 & 0 & 0 & 0 & 0 & 0 & 0 & 0 & 0 & 0 & 0 & 0 & 0 & 0 & 0 & 0 & 0 & 0 & 0 & 0 & 0 & 0 & 6 \\
AQ2014Q3 & 0 & 0 & 2 & 0 & 0 & 0 & 0 & 0 & 0 & 0 & 0 & 1 & 0 & 0 & 0 & 0 & 0 & 0 & 0 & 0 & 0 & 0 & 0 & 0 & 0 & 0 & 0 & 0 & 0 & 0 & 3 \\
AQ2014Q4 & 2 & 1 & 0 & 0 & 1 & 0 & 0 & 0 & 0 & 0 & 0 & 0 & 0 & 0 & 0 & 0 & 0 & 0 & 0 & 0 & 0 & 0 & 0 & 0 & 0 & 0 & 0 & 0 & 0 & 0 & 4 \\
AQ2015Q1 & 0 & 1 & 0 & 0 & 0 & 0 & 0 & 0 & 1 & 0 & 0 & 0 & 0 & 0 & 0 & 0 & 0 & 0 & 0 & 0 & 0 & 0 & 0 & 0 & 0 & 0 & 0 & 0 & 0 & 0 & 2 \\
AQ2015Q2 & 1 & 1 & 0 & 0 & 0 & 0 & 0 & 0 & 0 & 0 & 0 & 0 & 0 & 0 & 0 & 0 & 0 & 0 & 0 & 0 & 0 & 0 & 0 & 0 & 0 & 0 & 0 & 0 & 0 & 0 & 2 \\
AQ2015Q3 & 1 & 0 & 1 & 0 & 0 & 0 & 0 & 1 & 0 & 0 & 0 & 0 & 0 & 0 & 0 & 0 & 0 & 0 & 0 & 0 & 0 & 0 & 0 & 0 & 0 & 0 & 0 & 0 & 0 & 0 & 3 \\
\bottomrule
\end{tabular}%
}
\end{table}
\end{landscape}

%MT500
\begin{landscape}
\begin{table}
\caption{Quarterly Run-off Triangle of the number of data breaches that affect more than 500 Montana residents}
\resizebox{\columnwidth}{!}{%
\begin{tabular}{@{}lrrrrrrrrrrrrrrrrrrrrrrrrrr@{}}
\toprule
 & DQ1 & DQ2 & DQ3 & DQ4 & DQ5 & DQ6 & DQ7 & DQ8 & DQ9 & DQ10 & DQ11 & DQ12 & DQ13 & DQ14 & DQ15 & DQ16 & DQ17 & DQ18 & DQ19 & DQ20 & DQ21 & DQ22 & DQ23 & DQ24 & DQ25 & Sum \\ \midrule
AQ2015Q4 & 0 & 1 & 0 & 0 & 0 & 0 & 0 & 0 & 0 & 0 & 0 & 0 & 0 & 0 & 0 & 0 & 0 & 0 & 0 & 0 & 0 & 0 & 0 & 0 & 0 & 1 \\
AQ2016Q1 & 1 & 2 & 1 & 0 & 0 & 1 & 0 & 0 & 0 & 0 & 0 & 0 & 0 & 0 & 0 & 0 & 0 & 0 & 1 & 0 & 0 & 0 & 0 & 0 & 0 & 6 \\
AQ2016Q2 & 0 & 3 & 0 & 1 & 0 & 0 & 1 & 0 & 0 & 0 & 0 & 0 & 0 & 0 & 0 & 0 & 0 & 0 & 0 & 0 & 0 & 0 & 0 & 0 & 0 & 5 \\
AQ2016Q3 & 2 & 0 & 0 & 2 & 1 & 0 & 0 & 0 & 0 & 0 & 0 & 0 & 0 & 0 & 0 & 0 & 0 & 0 & 0 & 0 & 0 & 0 & 0 & 0 & 0 & 5 \\
AQ2016Q4 & 0 & 1 & 0 & 0 & 0 & 0 & 0 & 0 & 0 & 0 & 0 & 0 & 0 & 0 & 0 & 0 & 0 & 0 & 0 & 0 & 0 & 0 & 0 & 0 & 0 & 1 \\
AQ2017Q1 & 1 & 0 & 1 & 0 & 1 & 0 & 0 & 0 & 0 & 0 & 0 & 0 & 0 & 0 & 0 & 0 & 0 & 0 & 0 & 0 & 0 & 0 & 0 & 0 & 0 & 3 \\
AQ2017Q2 & 1 & 1 & 0 & 0 & 0 & 0 & 0 & 0 & 0 & 0 & 0 & 0 & 0 & 0 & 0 & 0 & 0 & 0 & 0 & 0 & 0 & 0 & 0 & 0 & 0 & 2 \\
\bottomrule
\end{tabular}%
}
\end{table}
\end{landscape}

%MT250
\begin{landscape}
\begin{table}
\caption{Quarterly Run-off Triangle of the number of data breaches that affect between 250 and 499 Montana residents}
\resizebox{\columnwidth}{!}{%
\begin{tabular}{@{}lrrrrrrrrrrrrrrrrrrrrrrrrrr@{}}
\toprule
 & DQ1 & DQ2 & DQ3 & DQ4 & DQ5 & DQ6 & DQ7 & DQ8 & DQ9 & DQ10 & DQ11 & DQ12 & DQ13 & DQ14 & DQ15 & DQ16 & DQ17 & DQ18 & DQ19 & DQ20 & DQ21 & DQ22 & DQ23 & DQ24 & DQ25 & Sum \\ \midrule
AQ2015Q4 & 0 & 2 & 0 & 0 & 0 & 0 & 0 & 0 & 0 & 0 & 0 & 0 & 0 & 0 & 0 & 0 & 0 & 0 & 0 & 0 & 0 & 0 & 0 & 0 & 0 & 2 \\
AQ2016Q1 & 3 & 1 & 1 & 0 & 2 & 1 & 0 & 0 & 0 & 0 & 0 & 0 & 0 & 0 & 0 & 0 & 0 & 0 & 0 & 0 & 0 & 0 & 0 & 0 & 0 & 8 \\
AQ2016Q2 & 0 & 1 & 0 & 0 & 0 & 0 & 0 & 0 & 0 & 0 & 0 & 0 & 0 & 0 & 0 & 0 & 0 & 0 & 0 & 0 & 0 & 0 & 0 & 0 & 0 & 1 \\
AQ2016Q3 & 0 & 0 & 1 & 0 & 0 & 0 & 0 & 0 & 0 & 0 & 0 & 0 & 0 & 0 & 0 & 0 & 0 & 0 & 0 & 0 & 0 & 0 & 0 & 0 & 0 & 1 \\
AQ2016Q4 & 1 & 1 & 0 & 0 & 1 & 0 & 0 & 0 & 0 & 0 & 0 & 0 & 0 & 0 & 1 & 0 & 0 & 0 & 0 & 0 & 0 & 0 & 0 & 0 & 0 & 4 \\
AQ2017Q1 & 1 & 0 & 0 & 0 & 0 & 0 & 0 & 0 & 0 & 0 & 0 & 0 & 0 & 0 & 0 & 0 & 0 & 0 & 0 & 0 & 0 & 0 & 0 & 0 & 0 & 1 \\
AQ2017Q2 & 0 & 2 & 0 & 0 & 0 & 0 & 0 & 0 & 0 & 0 & 0 & 0 & 0 & 0 & 0 & 0 & 0 & 0 & 0 & 0 & 0 & 0 & 0 & 0 & 0 & 2 \\
\bottomrule
\end{tabular}%
}
\end{table}
\end{landscape}

%ME500
\begin{landscape}
\begin{table}
\caption{Quarterly Run-off Triangle of the number of data breaches that affect more than 500 Maine residents}
\resizebox{\columnwidth}{!}{%
\begin{tabular}{@{}lrrrrrrrrrrrrrrrrrrrrrrrrrrrrrrr@{}}
\toprule
 & DQ1 & DQ2 & DQ3 & DQ4 & DQ5 & DQ6 & DQ7 & DQ8 & DQ9 & DQ10 & DQ11 & DQ12 & DQ13 & DQ14 & DQ15 & DQ16 & DQ17 & DQ18 & DQ19 & DQ20 & DQ21 & DQ22 & DQ23 & DQ24 & DQ25 & DQ26 & DQ27 & DQ28 & DQ29 & DQ30 & Sum \\ \midrule
AQ2013Q1 & 1 & 3 & 0 & 0 & 0 & 0 & 0 & 0 & 0 & 0 & 0 & 0 & 0 & 0 & 0 & 0 & 0 & 0 & 0 & 0 & 0 & 0 & 0 & 0 & 0 & 0 & 0 & 0 & 0 & 0 & 4 \\
AQ2013Q2 & 1 & 0 & 0 & 0 & 0 & 0 & 0 & 0 & 0 & 0 & 0 & 0 & 0 & 0 & 0 & 0 & 0 & 0 & 0 & 0 & 0 & 0 & 0 & 0 & 0 & 0 & 0 & 0 & 0 & 0 & 1 \\
AQ2013Q3 & 0 & 1 & 0 & 0 & 0 & 0 & 0 & 0 & 0 & 0 & 0 & 0 & 0 & 0 & 0 & 0 & 0 & 0 & 0 & 0 & 0 & 0 & 0 & 0 & 0 & 0 & 0 & 0 & 0 & 0 & 1 \\
AQ2013Q4 & 2 & 0 & 0 & 0 & 0 & 0 & 0 & 0 & 0 & 0 & 0 & 0 & 0 & 0 & 0 & 0 & 0 & 0 & 0 & 0 & 0 & 0 & 0 & 0 & 0 & 0 & 0 & 0 & 0 & 0 & 2 \\
AQ2014Q1 & 1 & 0 & 0 & 0 & 0 & 0 & 0 & 0 & 0 & 0 & 0 & 0 & 0 & 0 & 0 & 0 & 0 & 0 & 0 & 0 & 0 & 0 & 0 & 0 & 0 & 0 & 0 & 0 & 0 & 0 & 1 \\
AQ2014Q2 & 0 & 2 & 0 & 0 & 0 & 0 & 0 & 0 & 0 & 0 & 0 & 0 & 0 & 0 & 0 & 0 & 0 & 0 & 0 & 0 & 0 & 0 & 0 & 0 & 0 & 0 & 0 & 0 & 0 & 0 & 2 \\
AQ2014Q3 & 0 & 0 & 0 & 0 & 1 & 0 & 0 & 0 & 0 & 0 & 0 & 0 & 0 & 0 & 0 & 0 & 0 & 0 & 0 & 0 & 0 & 0 & 0 & 0 & 0 & 0 & 0 & 0 & 0 & 0 & 1 \\
\bottomrule
\end{tabular}%
}
\end{table}
\end{landscape}

%ME250
\begin{landscape}
\begin{table}
\caption{Quarterly Run-off Triangle of the number of data breaches that affect between 250 and 499 Maine residents}
\resizebox{\columnwidth}{!}{%
\begin{tabular}{@{}lrrrrrrrrrrrrrrrrrrrrrrrrrrrrrrr@{}}
\toprule
 & DQ1 & DQ2 & DQ3 & DQ4 & DQ5 & DQ6 & DQ7 & DQ8 & DQ9 & DQ10 & DQ11 & DQ12 & DQ13 & DQ14 & DQ15 & DQ16 & DQ17 & DQ18 & DQ19 & DQ20 & DQ21 & DQ22 & DQ23 & DQ24 & DQ25 & DQ26 & DQ27 & DQ28 & DQ29 & DQ30 & Sum \\ \midrule
AQ2013Q1 & 1 & 1 & 1 & 1 & 0 & 0 & 0 & 0 & 0 & 0 & 0 & 0 & 0 & 0 & 0 & 0 & 0 & 0 & 0 & 0 & 0 & 0 & 0 & 0 & 0 & 0 & 0 & 0 & 0 & 0 & 4 \\
AQ2013Q2 & 1 & 0 & 0 & 0 & 0 & 0 & 0 & 0 & 0 & 0 & 0 & 0 & 0 & 0 & 0 & 0 & 0 & 0 & 0 & 0 & 0 & 0 & 0 & 0 & 0 & 0 & 0 & 0 & 0 & 0 & 1 \\
AQ2013Q3 & 1 & 0 & 1 & 1 & 0 & 0 & 0 & 0 & 0 & 0 & 0 & 0 & 0 & 0 & 0 & 0 & 0 & 0 & 0 & 0 & 0 & 0 & 0 & 0 & 0 & 0 & 0 & 0 & 0 & 0 & 3 \\
AQ2013Q4 & 0 & 0 & 1 & 0 & 0 & 0 & 0 & 0 & 0 & 0 & 0 & 0 & 0 & 0 & 0 & 0 & 0 & 0 & 0 & 0 & 0 & 0 & 0 & 0 & 0 & 0 & 0 & 0 & 0 & 0 & 1 \\
AQ2014Q1 & 0 & 0 & 0 & 0 & 1 & 0 & 0 & 0 & 0 & 1 & 0 & 0 & 0 & 0 & 0 & 0 & 0 & 0 & 0 & 0 & 0 & 0 & 0 & 0 & 0 & 0 & 0 & 0 & 0 & 0 & 2 \\
AQ2014Q2 & 0 & 0 & 0 & 0 & 0 & 0 & 0 & 0 & 0 & 0 & 0 & 0 & 0 & 0 & 0 & 0 & 0 & 0 & 0 & 0 & 0 & 0 & 0 & 0 & 0 & 0 & 0 & 0 & 0 & 0 & 0 \\
AQ2014Q3 & 0 & 1 & 0 & 0 & 1 & 0 & 0 & 0 & 0 & 0 & 0 & 0 & 0 & 0 & 0 & 0 & 0 & 0 & 0 & 0 & 0 & 0 & 0 & 0 & 0 & 0 & 0 & 0 & 0 & 0 & 2 \\
\bottomrule
\end{tabular}%
}
\end{table}
\end{landscape}

\begin{table}

\caption{Quarterly Run-off Triangle of the number of data breaches that affect more than 500 North Dakota residents}
\centering
\resizebox{\linewidth}{!}{
\begin{tabular}[t]{lrrrrrrrrrrrrr}
\toprule
  & DQ1 & DQ2 & DQ3 & DQ4 & DQ5 & DQ6 & DQ7 & DQ8 & DQ9 & DQ10 & DQ11 & DQ12 & Sum\\
\midrule
AQ2019Q1 & 1 & 2 & 2 & 0 & 0 & 0 & 0 & 0 & 0 & 0 & 0 & 0 & 5\\
AQ2019Q2 & 2 & 1 & 0 & 1 & 0 & 0 & 0 & 0 & 0 & 0 & 0 & 0 & 4\\
AQ2019Q3 & 1 & 0 & 0 & 1 & 1 & 0 & 0 & 1 & 0 & 0 & 0 & 0 & 4\\
AQ2019Q4 & 1 & 0 & 0 & 0 & 1 & 0 & 0 & 1 & 0 & 0 & 0 & 0 & 3\\
AQ2020Q1 & 0 & 2 & 18 & 8 & 1 & 0 & 0 & 0 & 0 & 0 & 0 & 0 & 29\\

AQ2020Q2 & 1 & 5 & 1 & 0 & 1 & 2 & 0 & 0 & 0 & 0 & 0 & 0 & 10\\
AQ2020Q3 & 2 & 1 & 2 & 0 & 1 & 0 & 0 & 0 & 0 & 0 & 0 & 0 & 6\\
AQ2020Q4 & 0 & 3 & 0 & 1 & 0 & 0 & 0 & 0 & 0 & 0 & 0 & 0 & 4\\
AQ2021Q1 & 2 & 3 & 5 & 1 & 0 & 0 & 0 & 0 & 0 & 0 & 0 & 0 & 11\\
AQ2021Q2 & 3 & 3 & 1 & 0 & 0 & 0 & 0 & 0 & 0 & 0 & 0 & 0 & 7\\

AQ2021Q3 & 1 & 3 & 0 & 0 & 0 & 0 & 0 & 0 & 0 & 0 & 0 & 0 & 4\\
AQ2021Q4 & 3 & 0 & 0 & 0 & 0 & 0 & 0 & 0 & 0 & 0 & 0 & 0 & 3\\
Sum & 17 & 23 & 29 & 12 & 5 & 2 & 0 & 2 & 0 & 0 & 0 & 0 & 90\\
\bottomrule
\end{tabular}}
\end{table}

\begin{table}

\caption{Quarterly Run-off Triangle of the number of data breaches that affect between 250 and 499 North Dakota residents}
\centering
\resizebox{\linewidth}{!}{
\begin{tabular}[t]{lrrrrrrrrrrrrr}
\toprule
  & DQ1 & DQ2 & DQ3 & DQ4 & DQ5 & DQ6 & DQ7 & DQ8 & DQ9 & DQ10 & DQ11 & DQ12 & Sum\\
\midrule
AQ2019Q1 & 0 & 0 & 0 & 1 & 0 & 0 & 0 & 0 & 1 & 0 & 0 & 0 & 2\\
AQ2019Q2 & 2 & 0 & 2 & 0 & 0 & 0 & 0 & 0 & 0 & 0 & 0 & 0 & 4\\
AQ2019Q3 & 0 & 1 & 2 & 0 & 0 & 0 & 0 & 0 & 0 & 0 & 0 & 0 & 3\\
AQ2019Q4 & 0 & 0 & 0 & 0 & 0 & 0 & 0 & 0 & 0 & 0 & 0 & 0 & 0\\
AQ2020Q1 & 0 & 0 & 5 & 2 & 2 & 0 & 0 & 0 & 0 & 0 & 0 & 0 & 9\\

AQ2020Q2 & 1 & 1 & 0 & 0 & 0 & 0 & 0 & 0 & 0 & 0 & 0 & 0 & 2\\
AQ2020Q3 & 0 & 0 & 0 & 0 & 0 & 0 & 0 & 0 & 0 & 0 & 0 & 0 & 0\\
AQ2020Q4 & 0 & 2 & 1 & 0 & 0 & 0 & 0 & 0 & 0 & 0 & 0 & 0 & 3\\
AQ2021Q1 & 0 & 1 & 0 & 0 & 0 & 0 & 0 & 0 & 0 & 0 & 0 & 0 & 1\\
AQ2021Q2 & 1 & 0 & 2 & 0 & 0 & 0 & 0 & 0 & 0 & 0 & 0 & 0 & 3\\

AQ2021Q3 & 0 & 0 & 0 & 0 & 0 & 0 & 0 & 0 & 0 & 0 & 0 & 0 & 0\\
AQ2021Q4 & 1 & 0 & 0 & 0 & 0 & 0 & 0 & 0 & 0 & 0 & 0 & 0 & 1\\
Sum & 5 & 5 & 12 & 3 & 2 & 0 & 0 & 0 & 1 & 0 & 0 & 0 & 28\\
\bottomrule
\end{tabular}}
\end{table}

\begin{table}

\caption{Quarterly Run-off Triangle of the number of data breaches that affect more than 500 Delaware residents}
\centering
\resizebox{\linewidth}{!}{
\begin{tabular}[t]{lrrrrrrrrrrrrrrrr}
\toprule
  & DQ1 & DQ2 & DQ3 & DQ4 & DQ5 & DQ6 & DQ7 & DQ8 & DQ9 & DQ10 & DQ11 & DQ12 & DQ13 & DQ14 & DQ15 & Sum\\
\midrule
AQ2018Q2 & 0 & 1 & 1 & 0 & 1 & 0 & 0 & 0 & 0 & 0 & 0 & 0 & 0 & 0 & 0 & 3\\
AQ2018Q3 & 2 & 0 & 0 & 1 & 0 & 1 & 0 & 0 & 0 & 0 & 0 & 0 & 0 & 0 & 0 & 4\\
AQ2018Q4 & 0 & 1 & 1 & 2 & 0 & 0 & 0 & 0 & 0 & 0 & 0 & 0 & 0 & 0 & 0 & 4\\
AQ2019Q1 & 1 & 2 & 3 & 0 & 0 & 0 & 0 & 0 & 0 & 0 & 0 & 0 & 0 & 0 & 0 & 6\\
AQ2019Q2 & 2 & 2 & 0 & 0 & 0 & 0 & 0 & 0 & 0 & 0 & 0 & 0 & 0 & 0 & 0 & 4\\

AQ2019Q3 & 1 & 1 & 0 & 0 & 1 & 0 & 0 & 0 & 0 & 0 & 0 & 0 & 0 & 0 & 0 & 3\\
AQ2019Q4 & 1 & 1 & 0 & 0 & 1 & 1 & 0 & 0 & 0 & 0 & 0 & 0 & 0 & 0 & 0 & 4\\
AQ2020Q1 & 0 & 3 & 4 & 1 & 0 & 0 & 0 & 0 & 0 & 0 & 0 & 0 & 0 & 0 & 0 & 8\\
AQ2020Q2 & 3 & 11 & 5 & 2 & 0 & 0 & 0 & 0 & 0 & 0 & 0 & 0 & 0 & 0 & 0 & 21\\
AQ2020Q3 & 0 & 2 & 2 & 0 & 0 & 1 & 0 & 0 & 0 & 0 & 0 & 0 & 0 & 0 & 0 & 5\\

AQ2020Q4 & 1 & 1 & 1 & 0 & 0 & 0 & 0 & 0 & 0 & 0 & 0 & 0 & 0 & 0 & 0 & 3\\
AQ2021Q1 & 3 & 3 & 0 & 1 & 0 & 0 & 0 & 0 & 0 & 0 & 0 & 0 & 0 & 0 & 0 & 7\\
AQ2021Q2 & 3 & 2 & 3 & 0 & 0 & 0 & 0 & 0 & 0 & 0 & 0 & 0 & 0 & 0 & 0 & 8\\
AQ2021Q3 & 4 & 3 & 0 & 0 & 0 & 0 & 0 & 0 & 0 & 0 & 0 & 0 & 0 & 0 & 0 & 7\\
AQ2021Q4 & 4 & 0 & 0 & 0 & 0 & 0 & 0 & 0 & 0 & 0 & 0 & 0 & 0 & 0 & 0 & 4\\

Sum & 25 & 33 & 20 & 7 & 3 & 3 & 0 & 0 & 0 & 0 & 0 & 0 & 0 & 0 & 0 & 91\\
\bottomrule
\end{tabular}}
\end{table}

\clearpage
\section{GAM Results} \label{app:glm}

Table \ref{tab:glm_mean} presents the coefficients of the variables in the GAM. The table is organised in a specific manner, with variables that are common across states and/or periods appearing first, followed by variables that are unique to individual states (with more numerous data segments listed first), and lastly the dispersion parameter. The variables are presented in the same order as outlined in Section \ref{sec:methodology} (i.e., development period and accident period simplifications, calendar period effects, interactions between accident periods and development periods, and treatments of exceptional observations). Further, the table includes specific naming conventions that are used to identify the variables, which will be explained below.

The variables are identified by a unique name that comprises two components: a \textbf{data segment name} and a \textbf{definition}. The data segment name specifies the segment of data that the variable is applied to, which helps to differentiate it from other segments in the analysis. Naming conventions of data segments and variable definition are provided below. \\

\textbf{Data segment names:}
\begin{itemize}
    \item`IN1' refers to data breaches that affect between 0 and 249 Indiana residents. `IN250' represents breaches that impact between 250 and 499 Indiana residents. `IN500' pertains to breaches that affect more than 500 Indiana residents. And similarly for other states. The only exception is `OR250', which refers to breaches that affect more than 250 Oregon residents. 
    \item `DE500\_ND250' means the same variable applies to both data segments. 
\end{itemize}

\textbf{Definitions:}
\begin{itemize}
    \item The cross ($\times$) denotes the multiplication sign. 
    \item `i, j, c' represent accident quarters (AQ), development quarters (DQ), and calendar quarters (CQ).
    \item indicator functions
    \begin{itemize}
        \item `ind\_j\_2\_5' = $1_{\{2,5\}}(j)$ (as defined in Section \ref{sec:gen_dev_pro}). The first component `ind' refers to `indicator'. The second component `j' represents a specific variable that the indicator variable is associated with. The remaining components `2' and `5' represent specific values that the indicator variable is associated with. 
        \item `ind\_i\_ge\_2020Q1' = $1_{j\ge2020Q1}(j)$
        \item `ind\_j\_le\_4' = $1_{[1,4]}(j)$
        \item `ind\_c\_2017Q4\_2018Q1\_opposite' = $-1_{2017Q4}(c)+1_{2018Q1}(c)$
    \end{itemize}
    \item `max\_0\_and\_j\_minus\_6' = $\max(0,j-6)$
    \item `min\_j\_6' = $\min(j,6)$
    \item `i\textasciicircum{}2' = $i^2$
    \item `IN500\_ND500\_ind\_i\_ge\_2020Q1\_ND500\_ind\_i\_2020Q1' = $1_{\text{IN500, ND500}} \cdot
    1_{i\ge2020Q1}(i) + 1_{\text{ND500}} \cdot 1_{2020Q1}(i)$ 
\end{itemize}

\clearpage
\begin{refltable}{lc}{r}
\caption{GAM Results (mean)}
\label{tab:glm_mean}\\
\toprule
Dependent variable: $\ln(\text{quarterly \# of breaches})$                     &          \\* \midrule
\endfirsthead
\multicolumn{2}{c}%
{{\bfseries Table \thetable\ continued from previous page}} \\
\endhead
\bottomrule
\endfoot
\endlastfoot
Variable name                                                                  & Estimate \\* \midrule
OR250\_IN250\_MT250 $\times$   ind\_j\_1\_2\_5 \textsuperscript{1}                                & -1.5801  \\
IN1\_MT1 $\times$ ind\_j\_2\_5 \textsuperscript{2}                                                & 0.3977   \\
DE500\_ND250 $\times$ ind\_j\_2\_5 \textsuperscript{3}                                            & 0.9916   \\
ND500\_ME250 $\times$ ind\_j\_2\_5 \textsuperscript{4}                                            & 0.8573   \\
IN500\_MT500\_ME500\_WA500\_OR250\_IN250\_MT250   $\times$ ind\_j\_2\_5  \textsuperscript{5}      & 1.2733   \\
IN500\_IN250 $\times$ ind\_j\_5  \textsuperscript{6}                                              & 0.4684   \\
ND500\_WA500\_MT250\_ME250   $\times$ ind\_j\_5 \textsuperscript{7}                               & 0.9916   \\
IN500\_MT500\_ME500\_ND500\_WA\_500\_OR250\_IN250\_MT250\_ME250\_ND250   $\times$ j \textsuperscript{8}                               & -0.6955 \\
IN500\_MT500\_ND500\_WA\_500\_OR250\_IN250\_ME250\_ND250   $\times$ max\_0\_and\_j\_minus\_6 \textsuperscript{9}                      & 0.4698  \\
IN1\_ME1 $\times$   max\_0\_and\_j\_minus\_6 \textsuperscript{10}                                  & -0.1834  \\
WA500\_OR250 $\times$   ind\_i\_ge\_2020Q1  \textsuperscript{11}                                   & 0.4992   \\
MT250\_ME250 $\times$   ind\_i\_ge\_2020Q1 \textsuperscript{12}                                    & -0.5183  \\
IN500\_ND500\_ind\_i\_ge\_2020Q1\_ND500\_ind\_i\_2020Q1 \textsuperscript{13}                       & 1.0772   \\
WA500\_ind\_c\_2017Q2\_WA500\_OR250\_ind\_c\_2018Q4 \textsuperscript{14}                           & 0.5478   \\
IN1 $\times$   ind\_c\_2017Q4\_2018Q1\_opposite \textsuperscript{15}                               & 0.2710   \\
ME1 $\times$   ind\_c\_2018Q2\_2018Q3\_opposite \textsuperscript{16}                               & -0.2371  \\
ME1 $\times$   ind\_c\_2018Q4\_2019Q1\_opposite  \textsuperscript{17}                              & 0.3425   \\
MT1\_ME1 $\times$ ind\_j\_2   $\times$ max\_0\_and\_i\_minus\_2017Q4  \textsuperscript{18}         & 0.0776   \\
CA500 $\times$   ind\_i\_2015Q3\_2015Q4 \textsuperscript{19}                                       & -0.4870  \\
IN1 $\times$   ind\_i\_2015Q2\_2016Q2  \textsuperscript{20}                                        & 0.3814   \\
IN1 $\times$   ind\_i\_2016Q4\_2019Q4 \textsuperscript{21}                                         & -0.2463  \\
ME1 $\times$   ind\_i\_2014Q4\_2016Q1\_2016Q2\_2017Q1\_2017Q2\_2017Q3 \textsuperscript{22}         & 0.4804   \\
IN500\_ind\_j\_5\_ind\_i\_2016Q1\_IN500\_WA500\_OR250\_ind\_j\_5\_ind\_i\_2016Q3\_IN500\_ind\_j\_5\_ind\_i\_2017Q4 \textsuperscript{23} & 1.6916  \\
IN1\_MT1\_ME1 $\times$ ind\_j\_5   $\times$ ind\_i\_2016Q3 \textsuperscript{24}                    & 2.7835   \\
CA500 $\times$ ind\_j\_1                                                       & -1.4722  \\
CA500 $\times$ log\_j\_plus\_1                                                 & -3.1905  \\
CA500 $\times$ log\_j\_plus\_1   $\times$ ind\_j\_le\_4                        & -0.6144  \\
CA500 $\times$ intercept                                                       & 5.7693   \\
CA500 $\times$ i                                                               & 0.0973   \\
CA500 $\times$ i\textasciicircum{}2                                            & -0.0021  \\
CA500 $\times$ ind\_i\_ge\_2020Q1                                              & 0.7959   \\
CA500 $\times$   max\_0\_and\_i\_minus\_2020Q2                                 & 0.0998   \\
CA500 $\times$ ind\_c\_2016Q1                                                  & 0.3363   \\
CA500 $\times$ ind\_c\_2020Q3                                                  & 0.1137   \\
CA500 $\times$ log\_j\_plus\_1   $\times$ ind\_j\_le\_4 $\times$ max\_0\_and\_2014Q3\_minus\_i                     & 0.0549  \\
CA500 $\times$ log\_j\_plus\_1   $\times$ ind\_j\_le\_4 $\times$ j $\times$ max\_0\_and\_i\_minus\_2014Q3          & 0.0135  \\
CA500 $\times$ log\_j\_plus\_1   $\times$ ind\_j\_le\_4 $\times$ j $\times$ max\_0\_and\_i\_minus\_2017Q1          & -0.0097 \\
CA500 $\times$ ind\_j\_1   $\times$ max\_0\_and\_i\_minus\_2017Q1              & -0.0506  \\
CA500 $\times$ ind\_i\_2014Q2                                                  & 0.5007   \\
CA500 $\times$ ind\_i\_2017Q1                                                  & 0.2439   \\
CA500 $\times$ ind\_i\_2021Q2                                                  & -0.4501  \\
CA500 $\times$ ind\_i\_2015Q4   $\times$ ind\_j\_4                             & 1.9590   \\
CA500 $\times$ ind\_i\_2016Q3   $\times$ ind\_j\_5                             & 2.1228   \\
IN1 $\times$ ind\_j\_1\_2\_5                                                   & 0.3970   \\
IN1 $\times$ ind\_j\_5                                                         & -0.6106  \\
IN1 $\times$ min\_j\_6                                                         & -0.7152  \\
IN1 $\times$ intercept                                                         & 5.1813   \\
IN1 $\times$ i                                                                 & -0.0723  \\
IN1 $\times$ i\textasciicircum{}2                                              & -0.0062  \\
IN1 $\times$   max\_0\_and\_i\_minus\_2014Q3                                   & 0.3493   \\
IN1 $\times$ ind\_i\_ge\_2020Q1                                                & 0.0647   \\
IN1 $\times$   max\_0\_and\_i\_minus\_2020Q2                                   & 0.2631   \\
IN1 $\times$ ind\_c\_2014Q4                                                    & 0.3824   \\
IN1 $\times$ ind\_c\_2016Q3                                                    & -0.2559  \\
IN1 $\times$ ind\_c\_2019Q4                                                    & -0.4727  \\
IN1 $\times$ ind\_c\_2020Q4                                                    & 0.2187   \\
IN1 $\times$ ind\_j\_2 $\times$   ind\_i\_ge\_2017Q4                           & 0.3812   \\
IN1 $\times$ ind\_j\_3 $\times$   ind\_i\_ge\_2017Q4                           & 0.9300   \\
IN1 $\times$ ind\_j\_4 $\times$   ind\_i\_ge\_2017Q4                           & 0.7946   \\
IN1 $\times$ ind\_j\_3 $\times$   max\_0\_and\_i\_minus\_2017Q4                & 0.0249   \\
IN1 $\times$ j $\times$   max\_0\_and\_i\_minus\_2017Q4                        & 0.0406   \\
IN1 $\times$   max\_0\_and\_j\_minus\_6 $\times$ max\_0\_and\_i\_minus\_2017Q4 & -0.1309  \\
IN1 $\times$ ind\_i\_2015Q3                                                    & -0.1265  \\
IN1 $\times$ ind\_i\_2016Q1                                                    & 0.6578   \\
IN1 $\times$ ind\_i\_2017Q1                                                    & 0.7881   \\
IN1 $\times$ ind\_i\_2017Q4                                                    & -0.4923  \\
IN1 $\times$ ind\_i\_2018Q1                                                    & 0.1317   \\
IN1 $\times$ ind\_i\_2019Q1                                                    & 0.1186   \\
IN1 $\times$ ind\_i\_2020Q2                                                    & 0.4118   \\
IN1 $\times$ ind\_i\_2016Q1   $\times$ ind\_j\_5                               & 0.7900   \\
MT1 $\times$ ind\_j\_1\_2\_5                                                   & 1.0518   \\
MT1 $\times$ ind\_j\_5                                                         & -1.5313  \\
MT1 $\times$ min\_j\_6                                                         & -0.2827  \\
MT1 $\times$   max\_0\_and\_j\_minus\_6                                        & -0.3223  \\
MT1 $\times$ intercept                                                         & 0.7279   \\
MT1 $\times$ i                                                                 & 0.1443   \\
MT1 $\times$ i\textasciicircum{}2                                              & -0.0029  \\
MT1 $\times$ ind\_i\_ge\_2020Q1                                                & 0.4705   \\
MT1 $\times$ ind\_c\_2019Q2                                                    & 0.1885   \\
MT1 $\times$ ind\_c\_2020Q3                                                    & -0.3137  \\
MT1 $\times$ ind\_j\_3 $\times$   ind\_i\_ge\_2017Q4                           & 1.3183   \\
MT1 $\times$ ind\_j\_4 $\times$   ind\_i\_ge\_2017Q4                           & 0.8188   \\
MT1 $\times$ ind\_j\_5 $\times$   ind\_i\_ge\_2017Q4                           & 1.0009   \\
MT1 $\times$ j $\times$   ind\_i\_ge\_2017Q4                                   & -0.0430  \\
MT1 $\times$ ind\_j\_3 $\times$   max\_0\_and\_i\_minus\_2017Q4                & 0.0731   \\
MT1 $\times$ ind\_j\_4 $\times$   max\_0\_and\_i\_minus\_2017Q4                & 0.0740   \\
MT1 $\times$ ind\_i\_2016Q1                                                    & 0.4164   \\
MT1 $\times$ ind\_i\_2017Q1                                                    & 0.4236   \\
MT1 $\times$ ind\_i\_2016Q1   $\times$ ind\_j\_5                               & 1.5657   \\
MT1 $\times$ ind\_i\_2017Q4   $\times$ ind\_j\_1                               & -1.0562  \\
MT1 $\times$ ind\_i\_2018Q3   $\times$ ind\_j\_5                               & 0.6761   \\
ME1 $\times$ ind\_j\_1\_2\_5                                                   & 0.6670   \\
ME1 $\times$ ind\_j\_2\_5                                                      & 0.1475   \\
ME1 $\times$ ind\_j\_5                                                         & -0.6840  \\
ME1 $\times$ min\_j\_6                                                         & -0.5989  \\
ME1 $\times$ intercept                                                         & 3.2773   \\
ME1 $\times$ i                                                                 & 0.0066   \\
ME1 $\times$ i\textasciicircum{}2                                              & -0.0052  \\
ME1 $\times$   max\_0\_and\_i\_minus\_2014Q3                                   & 0.2252   \\
ME1 $\times$ ind\_i\_ge\_2020Q1                                                & 0.6260   \\
ME1 $\times$ ind\_j\_1 $\times$   ind\_i\_ge\_2017Q4                           & -0.4621  \\
ME1 $\times$ ind\_j\_3 $\times$   ind\_i\_ge\_2017Q4                           & 0.3832   \\
ME1 $\times$ ind\_j\_4 $\times$   ind\_i\_ge\_2017Q4                           & 0.4309   \\
ME1 $\times$ ind\_j\_5 $\times$   ind\_i\_ge\_2017Q4                           & 0.0120   \\
ME1 $\times$ j $\times$   ind\_i\_ge\_2017Q4                                   & 0.1761   \\
ME1 $\times$ ind\_j\_3 $\times$   max\_0\_and\_i\_minus\_2017Q4                & 0.1666   \\
ME1 $\times$ ind\_j\_4 $\times$   max\_0\_and\_i\_minus\_2017Q4                & 0.1186   \\
ME1 $\times$ ind\_j\_5 $\times$   max\_0\_and\_i\_minus\_2017Q4                & 0.0652   \\
ME1 $\times$   max\_0\_and\_j\_minus\_6 $\times$ max\_0\_and\_i\_minus\_2017Q3 & -0.2091  \\
ME1 $\times$ ind\_i\_7                                                         & -0.4358  \\
ME1 $\times$ ind\_i\_2016Q3   $\times$ ind\_j\_4                               & 1.7082   \\
ME1 $\times$ ind\_i\_2016Q1   $\times$ ind\_j\_5                               & 1.0922   \\
WA500 $\times$ ind\_j\_1\_2\_5                                                 & -1.6858  \\
WA500 $\times$ intercept                                                       & 2.5211   \\
WA500 $\times$ i                                                               & -0.0005  \\
WA500 $\times$ i\textasciicircum{}2                                            & 0.0007   \\
WA500 $\times$ ind\_c\_2020Q3                                                  & 0.5364   \\
WA500 $\times$ ind\_i\_2018Q1   $\times$ ind\_j\_5                             & 1.0258   \\
OR250 $\times$ ind\_j\_5                                                       & 0.6630   \\
OR250 $\times$ intercept                                                       & 2.0431   \\
OR250 $\times$ i                                                               & 0.0948   \\
OR250 $\times$ i\textasciicircum{}2                                            & -0.0014  \\
OR250 $\times$ ind\_c\_2020Q2                                                  & -0.6967  \\
OR250 $\times$ ind\_i\_2021Q1                                                  & 0.5014   \\
IN500 $\times$ ind\_j\_1\_2\_5                                                 & -1.4422  \\
IN500 $\times$ intercept                                                       & 0.4537   \\
IN500 $\times$ i                                                               & 0.2020   \\
IN500 $\times$ i\textasciicircum{}2                                            & -0.0039  \\
MT500 $\times$ ind\_j\_1\_2\_5                                                 & -1.8504  \\
MT500 $\times$ ind\_j\_5                                                       & 1.1829   \\
MT500 $\times$ intercept                                                       & 2.9185   \\
MT500 $\times$ i                                                               & -0.1035  \\
MT500 $\times$ i\textasciicircum{}2                                            & 0.0028   \\
ME500 $\times$ ind\_j\_1\_2\_5                                                 & -1.0389  \\
ME500 $\times$   max\_0\_and\_j\_minus\_6                                      & 0.1783   \\
ME500 $\times$ intercept                                                       & 0.2687   \\
ME500 $\times$ i                                                               & 0.0755   \\
ME500 $\times$ i\textasciicircum{}2                                            & -0.0009  \\
ND500 $\times$ ind\_j\_1\_2\_5                                                 & -1.9707  \\
ND500 $\times$ intercept                                                       & 22.9320  \\
ND500 $\times$ i                                                               & -1.2103  \\
ND500 $\times$ i\textasciicircum{}2                                            & 0.0175   \\
DE500 $\times$ ind\_j\_1\_2\_5                                                 & -1.2181  \\
DE500 $\times$ j                                                               & -0.6253  \\
DE500 $\times$   max\_0\_and\_j\_minus\_6                                      & -12.2351 \\
DE500 $\times$ intercept                                                       & 9.9864   \\
DE500 $\times$ i                                                               & -0.5750  \\
DE500 $\times$ i\textasciicircum{}2                                            & 0.0100   \\
DE500 $\times$ ind\_i\_2020Q2                                                  & 1.3953   \\
IN250 $\times$ intercept                                                       & 1.4660   \\
IN250 $\times$ i                                                               & 0.0465   \\
IN250 $\times$ i\textasciicircum{}2                                            & -0.0006  \\
IN250 $\times$ ind\_i\_ge\_2020Q1                                              & 0.6695   \\
IN250 $\times$ ind\_i\_2020Q1                                                  & 0.6447   \\
MT250 $\times$   max\_0\_and\_j\_minus\_6                                      & 0.4992   \\
MT250 $\times$ intercept                                                       & 5.4824   \\
MT250 $\times$ i                                                               & -0.3349  \\
MT250 $\times$ i\textasciicircum{}2                                            & 0.0067   \\
ME250 $\times$ ind\_j\_1\_2\_5                                                 & -1.3984  \\
ME250 $\times$ intercept                                                       & 1.8328   \\
ME250 $\times$ i                                                               & -0.0905  \\
ME250 $\times$ i\textasciicircum{}2                                            & 0.0025   \\
ND250 $\times$ ind\_j\_1\_2\_5                                                 & -2.3052  \\
ND250 $\times$ ind\_j\_5                                                       & 1.3371   \\
ND250 $\times$ intercept                                                       & 50.9165  \\
ND250 $\times$ i                                                               & -2.7139  \\
ND250 $\times$ i\textasciicircum{}2                                            & 0.0359   \\
ND250 $\times$ ind\_i\_ge\_2020Q1                                              & 2.2145   \\
Dispersion parameter                                                           & 1.3250   \\* \bottomrule
Note 1: See the text above this table for variable definitions. \\
Note 2: Periods assigned to have zero weight are OR250\_WA500 $\times$ ind\_i\_2016Q1, \\ CA500 $\times$ ind\_i\_2018Q2\_2018Q3, CA500\_IN1\_MT1\_WA500 $\times$ ind\_i\_2020Q1. \\
\end{refltable}

\clearpage
\section{Model diagnostics} \label{app:diagnostics}

This section provides model diagnostics for the Quarterly Run-off Triangle of the number of data breaches that affect more than 500 California residents. 

Table \ref{tab:Diagnostics_1} - \ref{tab:Diagnostics_3} compare the actual and fitted sum of events by development quarters (DQ), accident quarters (AQ), and calendar quarters (CQ), where Z score = $\frac{\text{Actual} - \text{Fitted}}{\sqrt{\text{Fitted}}}$. Observations in AQ 2018Q2, 2019Q4, 2020Q1 are removed from Table \ref{tab:Diagnostics_1} and \ref{tab:Diagnostics_3} as they have been assigned zero weight in the GAM. 

\begin{figure}[h]
    \centering
    \includegraphics[width=15cm]{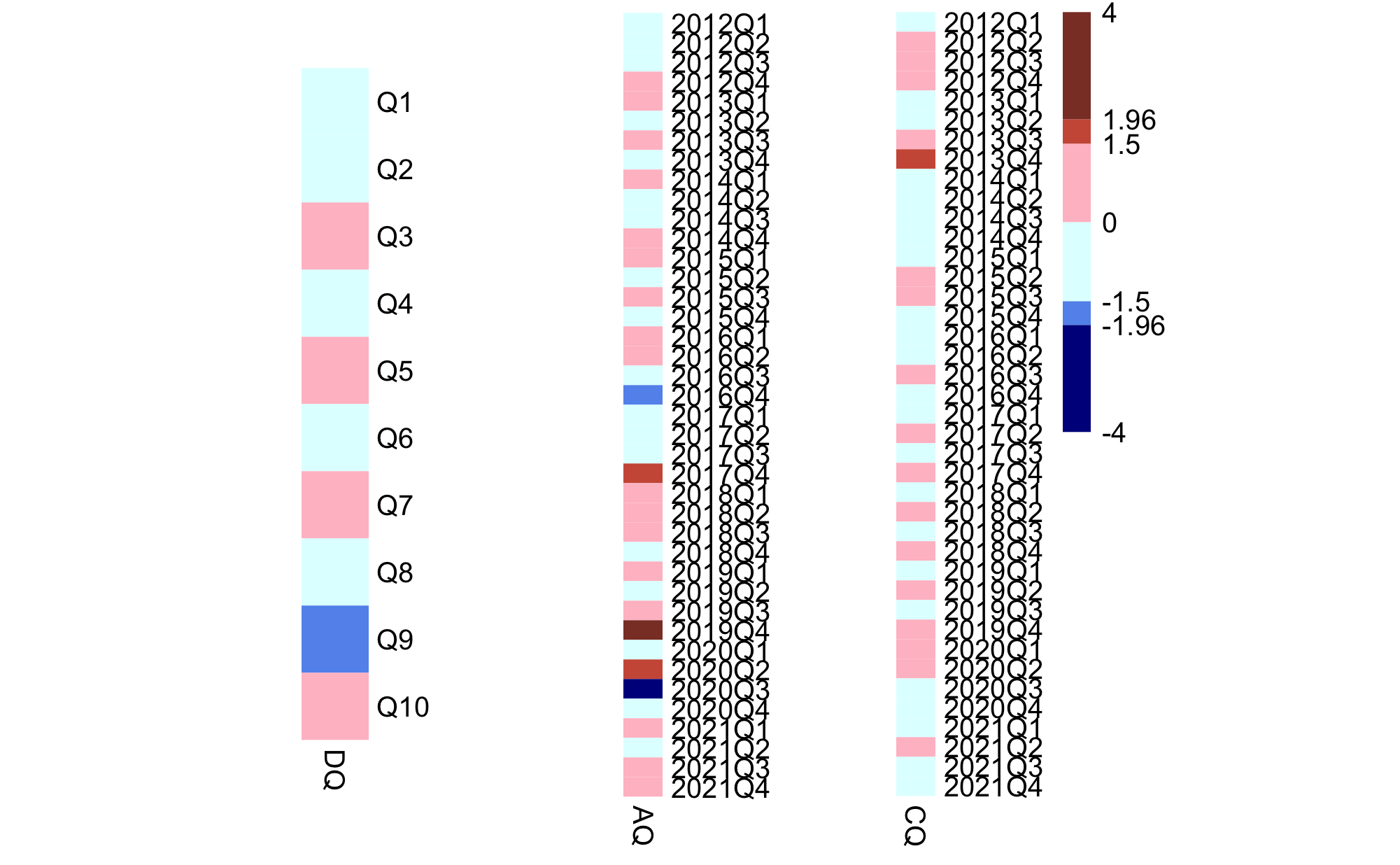}
    \caption{Heatmaps of Z score by development quarters, accident quarters, and calendar quarters}
    \label{fig:hm1}
\end{figure}

\begin{figure}[h]
    \centering
    \includegraphics[width=15cm]{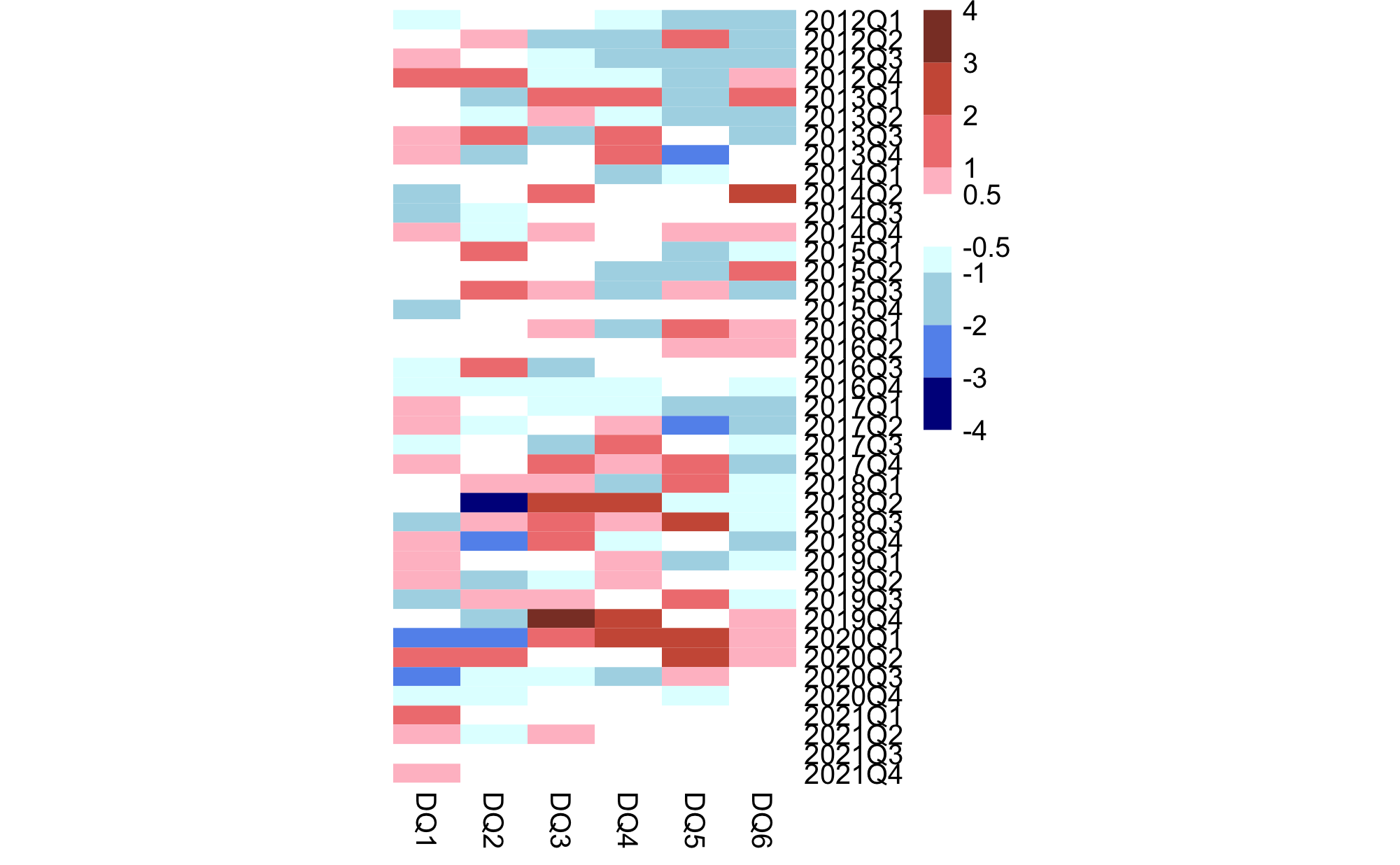}
    \caption{Heatmap of deviance residuals}
    \label{fig:hm2}
\end{figure}

Figure \ref{fig:hm1} visualises the pattern in Z score calculated from Table \ref{tab:Diagnostics_1} - \ref{tab:Diagnostics_3}. Figure \ref{fig:hm2} plots the deviance residuals. These heatmaps reveal no patterns or clusters, which suggest that our modelling is adequate. 

Similar heatmaps are found in all other data segments. In an attempt to reduce the number of pages in the document, they are not presented here and are available upon submission of the data and codes. 

\begin{table}[h]

\caption{Sum of events by development quarters} 
\label{tab:Diagnostics_1}
\centering
\resizebox{\linewidth}{!}{
\begin{tabular}[t]{lrrrrrrrrrr}
\toprule
  & Q1 & Q2 & Q3 & Q4 & Q5 & Q6 & Q7 & Q8 & Q9 & Q10\\
\midrule
Actual Sum & 500 & 714.00 & 294.00 & 161.00 & 123.00 & 48.00 & 35.00 & 21.00 & 9.00 & 14.00\\
Fitted Sum & 500 & 714.79 & 285.37 & 167.89 & 117.73 & 54.28 & 33.13 & 21.39 & 15.16 & 11.13\\
Actual/Fitted & 1 & 1.00 & 1.03 & 0.96 & 1.04 & 0.88 & 1.06 & 0.98 & 0.59 & 1.26\\
Z score & 0 & -0.03 & 0.51 & -0.53 & 0.49 & -0.85 & 0.33 & -0.09 & -1.58 & 0.86\\
\bottomrule
\end{tabular}}
\end{table}

\begin{table}[h]

\caption{Sum of events by accident quarters}
\label{tab:Diagnostics_2}
\centering
\resizebox{\linewidth}{!}{
\begin{tabular}[t]{lrrrrrrrrrrrrrrrrrrrrrrrrrrrrrrrrrrrrrrrr}
\toprule
  & 2012Q1 & 2012Q2 & 2012Q3 & 2012Q4 & 2013Q1 & 2013Q2 & 2013Q3 & 2013Q4 & 2014Q1 & 2014Q2 & 2014Q3 & 2014Q4 & 2015Q1 & 2015Q2 & 2015Q3 & 2015Q4 & 2016Q1 & 2016Q2 & 2016Q3 & 2016Q4 & 2017Q1 & 2017Q2 & 2017Q3 & 2017Q4 & 2018Q1 & 2018Q2 & 2018Q3 & 2018Q4 & 2019Q1 & 2019Q2 & 2019Q3 & 2019Q4 & 2020Q1 & 2020Q2 & 2020Q3 & 2020Q4 & 2021Q1 & 2021Q2 & 2021Q3 & 2021Q4\\
\midrule
Actual Sum & 22.00 & 29.00 & 27.00 & 37.00 & 35.00 & 29.00 & 41.00 & 33.00 & 36.00 & 59 & 31.00 & 46.00 & 47.00 & 42.00 & 33.00 & 42.00 & 62.00 & 56.00 & 78.00 & 47.00 & 78 & 53.00 & 54.00 & 73.00 & 60.00 & 60.00 & 66.00 & 47.00 & 58.00 & 50.00 & 53.00 & 79.00 & 102.00 & 119.00 & 80.00 & 92.00 & 104.00 & 55 & 65.00 & 18.00\\
Fitted Sum & 27.82 & 29.01 & 30.14 & 31.20 & 32.19 & 33.09 & 33.90 & 34.61 & 35.23 & 59 & 36.22 & 39.05 & 42.13 & 44.68 & 29.86 & 45.14 & 57.92 & 54.15 & 79.48 & 58.92 & 78 & 61.03 & 60.73 & 60.20 & 59.44 & 58.47 & 57.29 & 55.91 & 54.34 & 52.60 & 50.72 & 48.85 & 103.72 & 100.17 & 100.32 & 99.55 & 97.80 & 55 & 64.79 & 15.37\\
Actual/Fitted & 0.79 & 1.00 & 0.90 & 1.19 & 1.09 & 0.88 & 1.21 & 0.95 & 1.02 & 1 & 0.86 & 1.18 & 1.12 & 0.94 & 1.11 & 0.93 & 1.07 & 1.03 & 0.98 & 0.80 & 1 & 0.87 & 0.89 & 1.21 & 1.01 & 1.03 & 1.15 & 0.84 & 1.07 & 0.95 & 1.04 & 1.62 & 0.98 & 1.19 & 0.80 & 0.92 & 1.06 & 1 & 1.00 & 1.17\\
Z score & -1.10 & 0.00 & -0.57 & 1.04 & 0.50 & -0.71 & 1.22 & -0.27 & 0.13 & 0 & -0.87 & 1.11 & 0.75 & -0.40 & 0.58 & -0.47 & 0.54 & 0.25 & -0.17 & -1.55 & 0 & -1.03 & -0.86 & 1.65 & 0.07 & 0.20 & 1.15 & -1.19 & 0.50 & -0.36 & 0.32 & 4.31 & -0.17 & 1.88 & -2.03 & -0.76 & 0.63 & 0 & 0.03 & 0.67\\
\bottomrule
\end{tabular}}
\end{table}

\begin{table}[h]

\caption{Sum of events by calendar quarters}
\label{tab:Diagnostics_3}
\centering
\resizebox{\linewidth}{!}{
\begin{tabular}[t]{lrrrrrrrrrrrrrrrrrrrrrrrrrrrrrrrrrrrrrrrr}
\toprule
  & 2012Q1 & 2012Q2 & 2012Q3 & 2012Q4 & 2013Q1 & 2013Q2 & 2013Q3 & 2013Q4 & 2014Q1 & 2014Q2 & 2014Q3 & 2014Q4 & 2015Q1 & 2015Q2 & 2015Q3 & 2015Q4 & 2016Q1 & 2016Q2 & 2016Q3 & 2016Q4 & 2017Q1 & 2017Q2 & 2017Q3 & 2017Q4 & 2018Q1 & 2018Q2 & 2018Q3 & 2018Q4 & 2019Q1 & 2019Q2 & 2019Q3 & 2019Q4 & 2020Q1 & 2020Q2 & 2020Q3 & 2020Q4 & 2021Q1 & 2021Q2 & 2021Q3 & 2021Q4\\
\midrule
Actual Sum & 7.00 & 19.00 & 27.00 & 29.00 & 27.00 & 24.00 & 31.00 & 40.00 & 25.00 & 37.00 & 37.00 & 35.00 & 31.00 & 48.00 & 41.00 & 31.00 & 54 & 43.00 & 63.00 & 51.00 & 60.00 & 68.00 & 75.00 & 65.00 & 54.00 & 50.00 & 36.00 & 55.00 & 47.00 & 59.00 & 45.00 & 45.00 & 24.00 & 34.00 & 71 & 73.00 & 92.00 & 105.00 & 93.00 & 106.00\\
Fitted Sum & 8.46 & 18.78 & 23.39 & 26.01 & 27.85 & 29.26 & 30.39 & 31.32 & 32.08 & 40.30 & 40.43 & 36.59 & 37.08 & 39.84 & 36.30 & 32.73 & 54 & 45.90 & 60.66 & 52.41 & 60.28 & 62.64 & 83.43 & 59.97 & 59.36 & 45.13 & 37.48 & 47.90 & 50.58 & 51.94 & 51.89 & 42.42 & 23.54 & 30.80 & 71 & 81.18 & 94.81 & 97.66 & 93.29 & 107.92\\
Actual/Fitted & 0.83 & 1.01 & 1.15 & 1.11 & 0.97 & 0.82 & 1.02 & 1.28 & 0.78 & 0.92 & 0.92 & 0.96 & 0.84 & 1.20 & 1.13 & 0.95 & 1 & 0.94 & 1.04 & 0.97 & 1.00 & 1.09 & 0.90 & 1.08 & 0.91 & 1.11 & 0.96 & 1.15 & 0.93 & 1.14 & 0.87 & 1.06 & 1.02 & 1.10 & 1 & 0.90 & 0.97 & 1.08 & 1.00 & 0.98\\
Z score & -0.50 & 0.05 & 0.75 & 0.59 & -0.16 & -0.97 & 0.11 & 1.55 & -1.25 & -0.52 & -0.54 & -0.26 & -1.00 & 1.29 & 0.78 & -0.30 & 0 & -0.43 & 0.30 & -0.19 & -0.04 & 0.68 & -0.92 & 0.65 & -0.70 & 0.72 & -0.24 & 1.03 & -0.50 & 0.98 & -0.96 & 0.40 & 0.10 & 0.58 & 0 & -0.91 & -0.29 & 0.74 & -0.03 & -0.18\\
\bottomrule
\end{tabular}}
\end{table}

\clearpage
\section{Frequency trend} \label{app:freq}

\begin{figure}[htp]
    \centering
    \includegraphics[width=10cm]{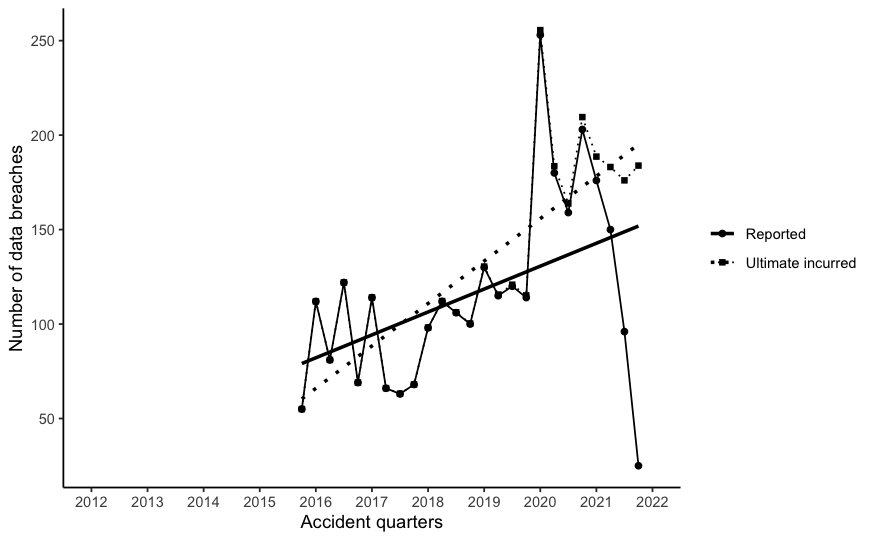}
    \caption{Reported versus Ultimate incurred breaches (AQ 2015Q4 - 2021Q4, MT(0-249))}
    \label{fig:MT_freq}
\end{figure}

\begin{figure}[htp]
    \centering
    \includegraphics[width=10cm]{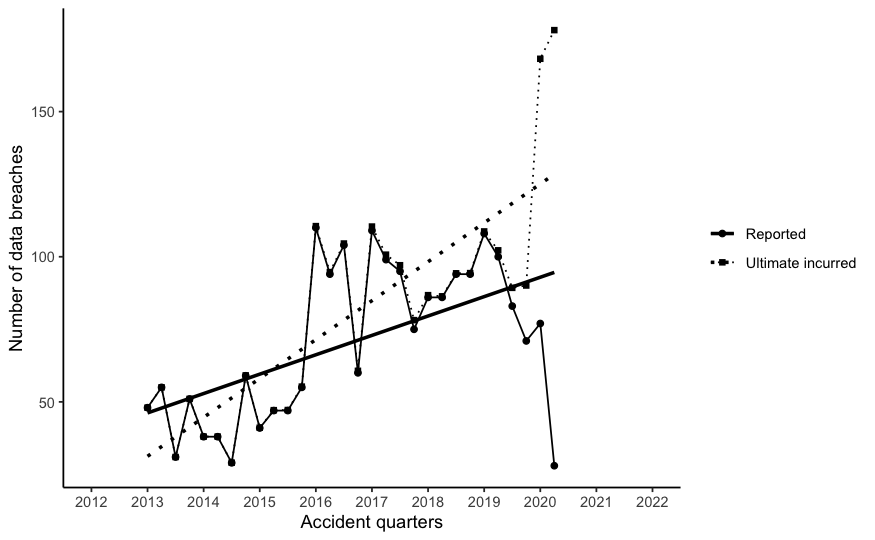}
    \caption{Reported versus Ultimate incurred breaches (AQ 2013Q1 - 2020Q2, ME(0-249))}
    \label{fig:ME_freq}
\end{figure}

\newpage
\section{Interactions between development periods and accident periods} \label{app:DP}

\subsection{CA(\texorpdfstring{$>$499}{\texttwoinferior})} \label{app:CA_DP}

\begin{figure}[htp]
    \centering
    \includegraphics[width=10cm]{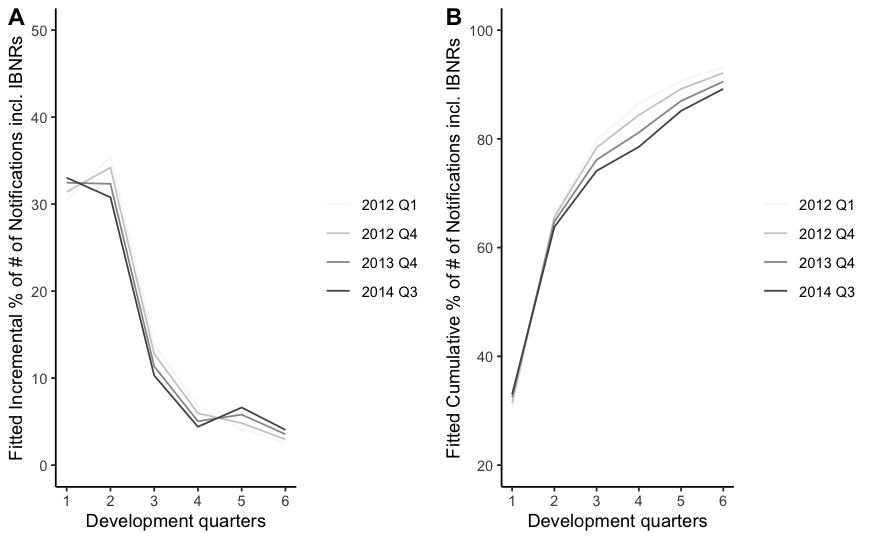}
    \caption{Development pattern trend (AQ 2012Q1 - AQ 2014Q3, CA($>$499))}
    \label{fig:CA_DP1}
\end{figure}

\begin{figure}[htp]
    \centering
    \includegraphics[width=10cm]{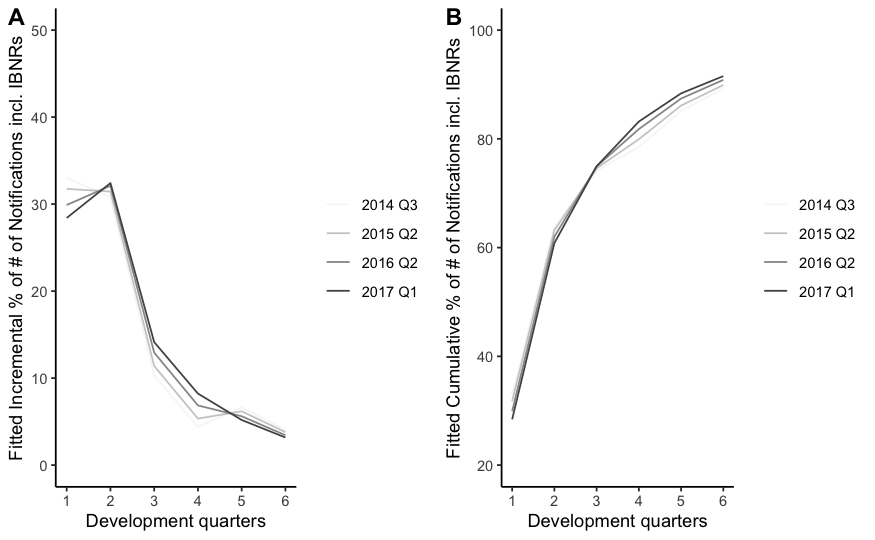}
    \caption{Development pattern trend (AQ 2014Q3 - AQ 2017Q1, CA($>$499))}
    \label{fig:CA_DP2}
\end{figure}

\begin{figure}[htp]
    \centering
    \includegraphics[width=10cm]{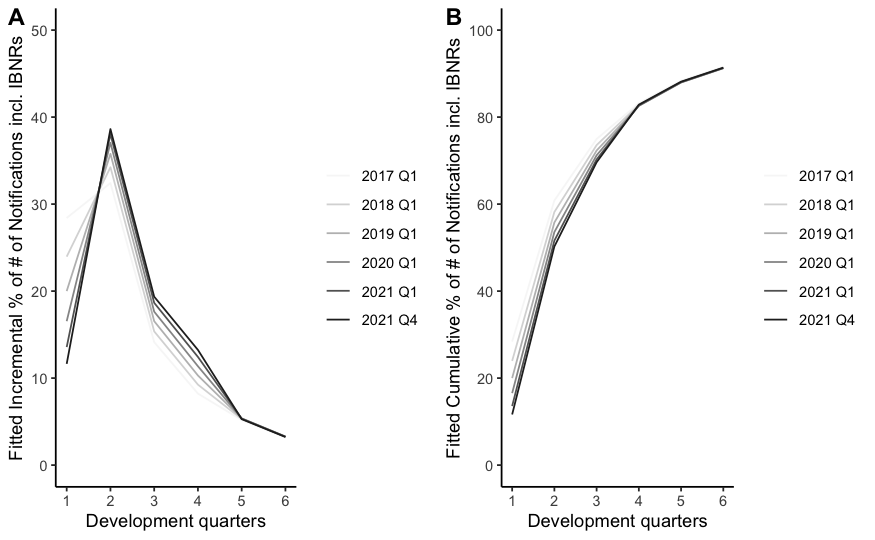}
    \caption{Development pattern trend (AQ 2017Q1 - AQ 2021Q4, CA($>$499))}
    \label{fig:CA_DP3_dup}
\end{figure}

Figures \ref{fig:CA_DP1} - \ref{fig:CA_DP3_dup} show the trends in development pattern of breaches that affect more than 500 California residents between 2012Q1 and 2021Q4. On average, around 80$\%$ of breaches are disclosed within the first four quarters of occurrence, and 90$\%$ within six quarters. The change in development pattern manifests itself in the first four DQs. Fitted incremental and cumulative percentage of breaches reported in DQ 1-4 are plotted respectively, based on the total number of breaches including IBNRs.

There are two change-points: 2014Q3 and 2017Q1. From 2012Q1 to 2014Q3, the incremental percentage of breaches reported in DQ 1 becomes larger, and smaller in DQ 2-4 respectively. The cumulative percentage of breaches starting from DQ 2 becomes smaller, and the cumulative percentage of breaches reported within one year from occurrence decreases from 86$\%$ to 79$\%$.

From 2014Q3 to 2017Q1, the changes are reversed. The incremental percentage of breaches reported in DQ 1 becomes smaller, and larger in DQ 2-4 respectively. However, while the cumulative percentage in DQ 1 and DQ 2 becomes lower, it becomes larger in DQ 3 and DQ 4. The cumulative percentage of breaches reported within one year from occurrence increases from 79$\%$ to 83$\%$.

Although the period between 2017Q1 and 2021Q4 continues the same pattern of change as the period between 2014Q3 to 2017Q1, that is, the incremental percentage of breaches reported in DQ 1 becomes smaller, and larger in DQ 2-4 respectively, the magnitude of change is greater. The percentage of breaches reported in DQ 1 decreases from 28$\%$ to 12$\%$. The cumulative percentage in all DQs before DQ 4 becomes lower, and finally in DQ 4, catches up to the level observed for breaches occurred in 2017Q1. The cumulative percentage of breaches reported within one year from occurrence keeps around 83$\%$.

\newpage
\subsection{IN(0-249)} \label{app:IN_DP}

\begin{figure}[htp]
    \centering
    \includegraphics[width=10cm]{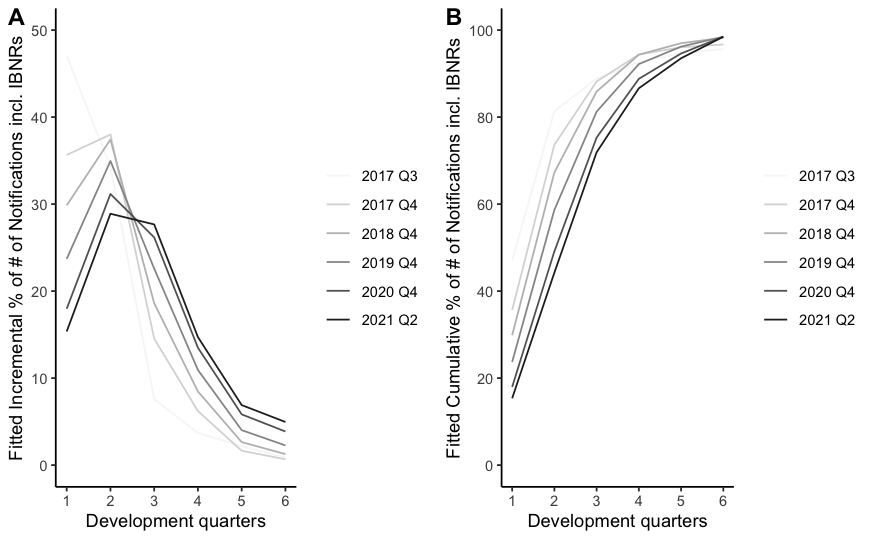}
    \caption{Development pattern trend (AQ 2014Q1 - AQ 2021Q2, IN (0-249))}
    \label{fig:IN_DP_dup}
\end{figure}

Figure \ref{fig:IN_DP_dup} shows the trend in development pattern of breaches that affect between 0 and 249 Indiana residents between 2014Q1 and 2021Q2. On average, around 90$\%$ of breaches are disclosed within four quarters of occurrence, and almost all breaches are reported within six quarters. The change in development pattern is observed in the first six DQs. Fitted incremental and cumulative percentage of breaches reported in DQ 1-6 are plotted respectively, based on the total number of breaches including IBNRs.

In the period between 2014Q1 and 2017Q3, data breaches are subject to a constant development pattern and 2017Q4 is the point of change. From 2017Q4 onward, the incremental percentage of breaches reported in DQ 1 and DQ 2 becomes smaller, and larger in all subsequent DQs. The cumulative percentage in all DQs before DQ 6 is consistently lower than before, until which it catches up to the level observed for breaches occurred in 2017Q3. Cumulative percentage of breaches reported within one year from occurrence decreases from 93$\%$ to 87$\%$, and cumulative percentage within six quarters increases from 96$\%$ to 98$\%$, when comparing 2017Q3 and 2021Q2.

\newpage
\subsection{MT(0-249)} \label{app:MT_DP}

\begin{figure}[htp]
    \centering
    \includegraphics[width=10cm]{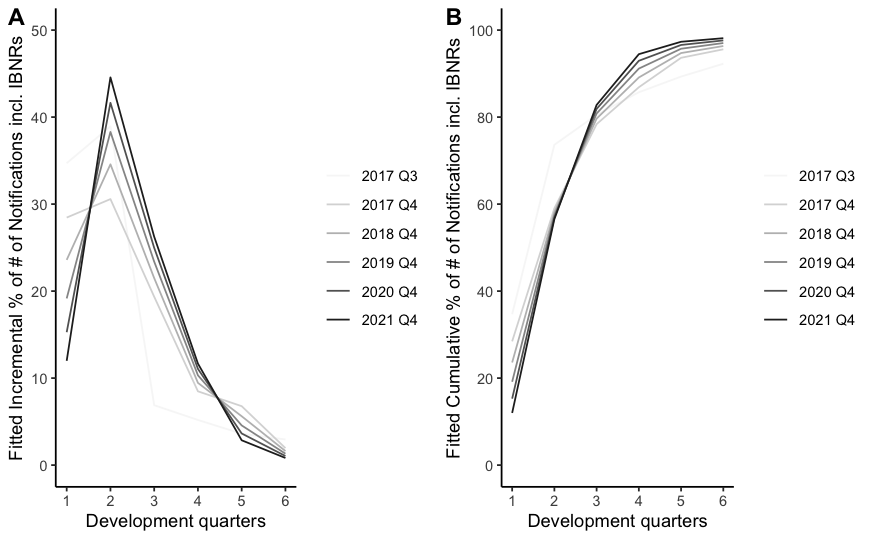}
    \caption{Development pattern trend (AQ 2015Q4 - AQ 2021Q4, MT (0-249))}
    \label{fig:MT_DP_dup}
\end{figure}

Figure \ref{fig:MT_DP_dup} shows the trend in development pattern of breaches that affect between 0 and 249 Montana residents between 2015Q4 and 2021Q4. On average, around 90$\%$ of breaches are disclosed within four quarters of occurrence, and almost all breaches are reported within six quarters. The change in development pattern is observed in the first six DQs. Fitted incremental and cumulative percentage of breaches reported in DQ 1-6 are plotted respectively, based on the total number of breaches including IBNRs.

In the period between 2015Q4 and 2017Q3, data breaches are subject to a constant development pattern and 2017Q4 is the point of change. From 2017Q4 onward, the incremental percentage of breaches reported in DQ 1 becomes smaller, larger in all DQs between DQ 2 and DQ 4, and smaller in DQ 5 and DQ 6. The cumulative percentage is lower in the first two DQs, but higher in all later DQs. Cumulative percentage of breaches reported within one year from occurrence increases from 86$\%$ to 94$\%$, and cumulative percentage within six quarters increases from 92$\%$ to 98$\%$, when comparing 2015Q4 and 2021Q2.

\newpage
\subsection{ME(0-249)} \label{app:ME_DP}

\begin{figure}[htp]
    \centering
    \includegraphics[width=10cm]{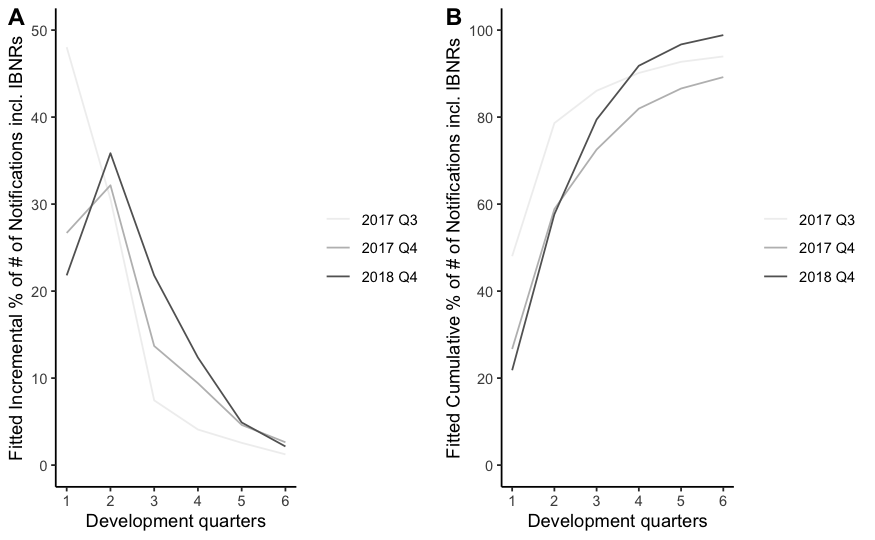}
    \caption{Development pattern trend (AQ 2013Q1 - AQ 2018Q4, ME (0-249))}
    \label{fig:ME_DP1}
\end{figure}

\begin{figure}[htp]
    \centering
    \includegraphics[width=10cm]{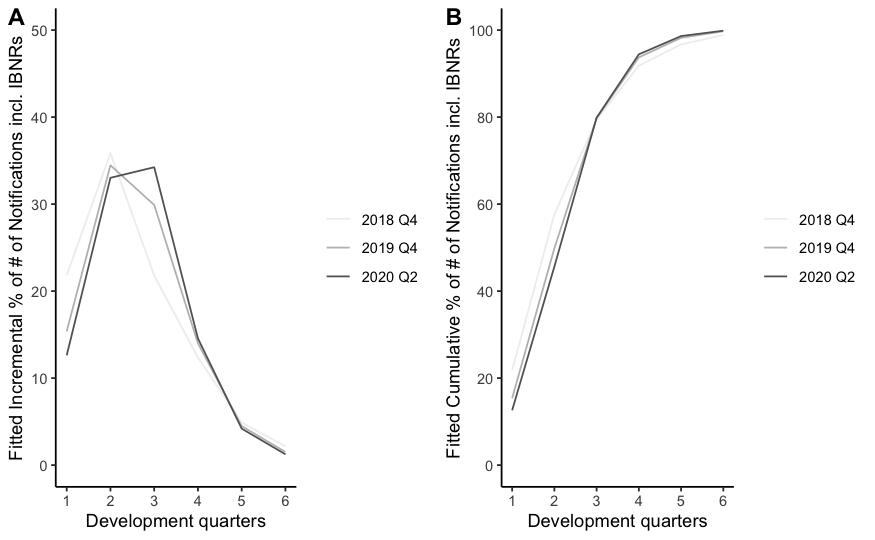}
    \caption{Development pattern trend (AQ 2018Q4 - AQ 2020Q2, ME (0-249))}
    \label{fig:ME_DP2}
\end{figure}

Figures \ref{fig:ME_DP1} and \ref{fig:ME_DP2} show the trends in development pattern of breaches that affect between 0 and 249 Maine residents between 2013Q1 and 2020Q2. On average, around 90$\%$ of breaches are disclosed within four quarters of occurrence, and almost all breaches are reported within six quarters. The change in development pattern is observed in the first six DQs. Fitted incremental and cumulative percentage of breaches reported in DQ 1-6 are plotted respectively, based on the total number of breaches including IBNRs.

In the period between 2013Q1 and 2017Q3, data breaches are subject to a constant development pattern. Data breaches occurred in 2017Q4 are more delayed: the cumulative percentage at all DQs is lower. Then, between 2018Q1 and 2018Q4, the delay is shortening. The cumulative percentage at DQ 1 and DQ 2 is indistinguishable from that of breaches occurred in 2018Q1, but the cumulative percentage at all subsequent DQs is higher.

From 2018Q4 onward, the incremental percentage of breaches reported in DQ 1 and DQ 2 becomes smaller, larger in DQ 3 and DQ 4, and smaller in DQ 5 and DQ 6. The cumulative percentage is lower in the first two DQs, but higher in all later DQs. Cumulative percentage of breaches reported within one year from occurrence increases from 90$\%$ to 94$\%$, and cumulative percentage within six quarters increases from 94$\%$ to 100$\%$, when comparing 2013Q1 and 2020Q2.

\newpage
\section{Average reporting delay} \label{app:AD}

\subsection{MT(0-249)}

\begin{figure}[htp]
    \centering
    \includegraphics[width=10cm]{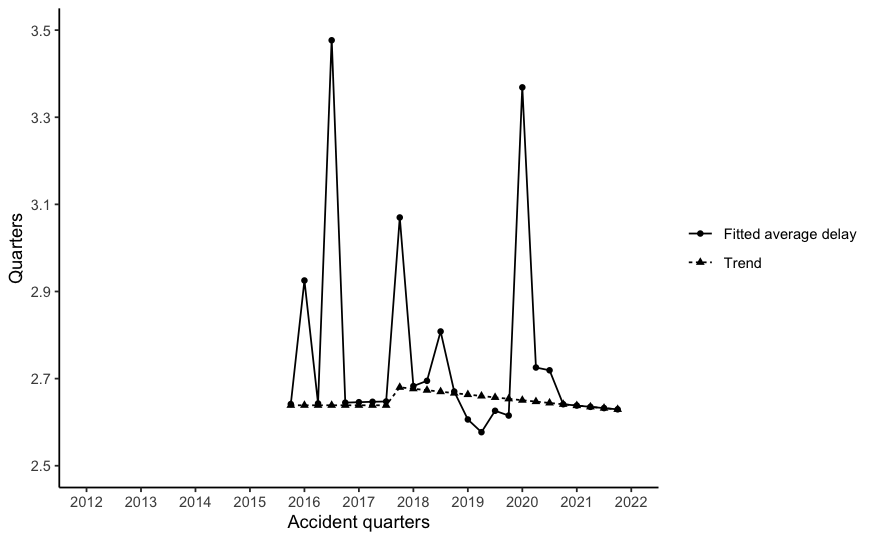}
    \caption{Fitted average delay and its trend - MT(0-249)}
    \label{fig:MT_AD_dup}
\end{figure}

Shown in Figure \ref{fig:MT_AD_dup}, the average time to report data breaches is 2.64 quarters between 2013Q1 and 2017Q3, increases to 2.68 quarters in 2017Q4, and gradually decreases to the same level as before. The average reporting delay is fairly constant over time.

\subsection{ME(0-249)}

\begin{figure}[htp]
    \centering
    \includegraphics[width=10cm]{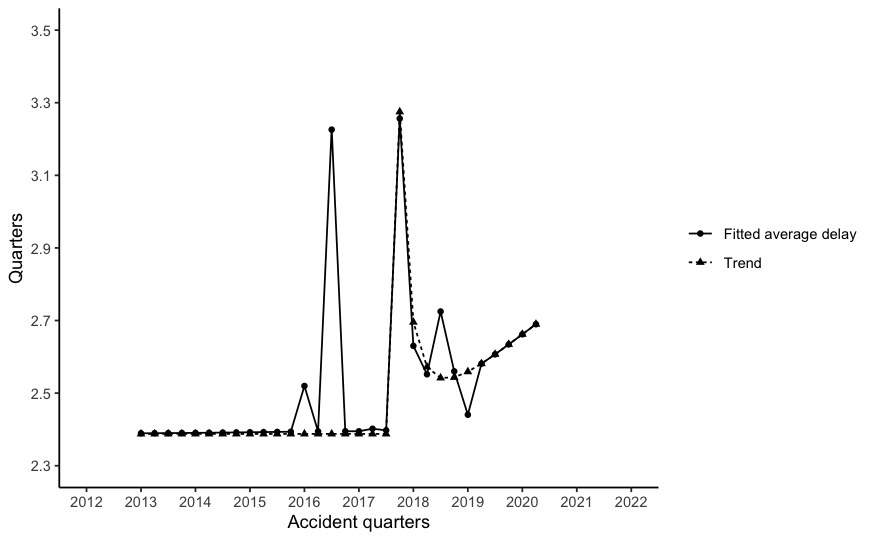}
    \caption{Fitted average delay and its trend - ME(0-249)}
    \label{fig:ME_AD}
\end{figure}

Shown in Figure \ref{fig:ME_AD}, the average time to report data breaches is 2.39 quarters between 2015Q4 and 2017Q3, increases to 2.7 quarters in 2018Q1, decreases at a decreasing rate until 2018Q4 to 2.54 quarters, and increases slightly to 2.69 quarters in 2020Q2.

\end{document}